\documentclass{elsarticle}
\usepackage{placeins}
\usepackage{amsmath,amssymb}
\usepackage{color}
\usepackage{fullpage}
\usepackage{booktabs}
\usepackage[english]{babel}
\usepackage{graphicx}
\usepackage{xcolor}
\usepackage[unicode]{hyperref}
\usepackage{float}

\usepackage{caption}
\usepackage{subcaption}
\usepackage{xspace}
\usepackage{mathtools}
\usepackage{setspace}
\usepackage{tabularx}
\newcommand{\eg}{\emph{e.g.}\xspace}
\newcommand{\ie}{\emph{i.e.}\xspace}

\renewcommand{\vec}[1]{\boldsymbol{#1}}     
\newcommand{\ten}[1]{\boldsymbol{#1}}       
\newcommand{\mat}[1]{\mathsf{#1}}          
\newcommand{\params}{\boldsymbol{\mu}}      
\newcommand{\paramspace}{\boldsymbol{\Theta}} 

\newcommand{\wi}{\mathrm{Wi}}

\newcommand{\stdof}[1]{\mathrm{std}(#1)}

\newcommand{\subs}{\mathrm{s}}              
\newcommand{\subp}{\mathrm{p}}                 
\newcommand{\subexp}{\mathrm{exp}}

\begin{document}
\begin{frontmatter}

\title{Bayesian inference in active microrheology \\ with wall and particle-particle interactions}

\author[1]{Parajal Rai\corref{cor1}}
\ead{p.rai@tue.nl}
\author[1]{Michelle M.A. Spanjaards}
\author[1]{Ye Wang}
\author[1]{Patrick D. Anderson}
\author[1]{Nick Jaensson}
\cortext[cor1]{Corresponding author.}
\address[1]{Department of Mechanical Engineering,
            Eindhoven University of Technology,
            5600 MB Eindhoven, The Netherlands}

\begin{abstract}
Active microrheology infers rheological properties from the motion of a force-driven
probe. In small samples, however, the probe is often close to walls or other probes,
and the resulting hydrodynamic interactions bias the inferred parameters. We present a
Bayesian framework in which these interactions are built into simplified analytical
models for Newtonian and linear viscoelastic fluids and test it on noisy synthetic
data from finite-element simulations of Newtonian and Oldroyd-B fluids. Accounting for
the interactions substantially improves the accuracy of the inferred parameters. Near a
wall, oblique forcing allows the material
parameters and the wall distance to be identified jointly from a single experiment. In
the nonlinear viscoelastic regime, posterior predictive checks over multiple force
levels reveal the inadequacy of the linear model. Finally, when the probe size is
comparable to the heterogeneity length scale, the framework distinguishes spatial
variations in the modulus from measurement noise.
\end{abstract}


\end{frontmatter}

\section{Introduction}
\label{sec:introduction}

Complex fluids are structurally inhomogeneous at nanometer-to-micrometer
scales owing to suspended particles, droplets, polymers, or surfactants,
and, as a consequence, exhibit nontrivial rheological behavior with broad
scientific and technological relevance
\cite{osterhold2000rheological,amorim2021insights,ahuja2018rheological,
fischer2011rheology,lee2009thixotropic}.
Many biological materials share this complexity, including synovial fluid,
mucus, saliva, blood plasma, and cytoplasmic mixtures. Because the
mechanical properties of these fluids often reflect the physiological
state of the host, the measurement of their response to external stimuli
can be used for diagnostic purposes. Examples include elevated
extracellular-matrix stiffness \cite{fan2024matrix} at cancer metastatic sites and fibrotic
remodeling in cardiac tissue following infarction \cite{li2018extracellular},
abnormally viscous mucus in cystic fibrosis \cite{lai2009micro}, and
reduced viscoelasticity of synovial fluid in arthritis caused by
hyaluronic-acid degradation \cite{schurz1987rheology,fam2007rheological}.
In all of these cases, the diagnostically relevant information lies in the
local mechanical properties, motivating measurement techniques that can
resolve them.

Conventional bulk rheometers measure properties averaged over a large
sample volume and therefore cannot resolve mechanical heterogeneity at the
cellular or subcellular length scales relevant to many diagnostic
applications
\cite{cox2011remodeling,roeder2004local,wullkopf2018cancer,martin2024scale}.
Microrheology addresses this limitation by probing at the microscale and
in small sample volumes. Passive methods infer viscoelasticity from
thermal fluctuations
\cite{squires2010fluid,burkholder2020nonlinear,zia2018active}, whereas
active methods drive the probe using magnetic fields or optical tweezers
\cite{bausch1999measurement,rich2011size,yao2009microrheology,
robertson2018optical,radiom2021magnetic}. Active microrheology is
particularly suited to stiffer or more viscous media, where thermal
fluctuations are too small to induce measurable Brownian displacement
\cite{xia2018microrheology,mao2022passive}. Material properties are
inferred by fitting measured probe displacements to simplified analytical models
based on classical relations such as Stokes drag or generalized
Stokes-Einstein relations, which typically assume motion in an unbounded
fluid. In practice, however, the small sample volumes that make
microrheology attractive also place the probe close to solid boundaries and, when multiple probes are present, sufficiently close for probe-probe hydrodynamic coupling. Both effects introduce
hydrodynamic interactions that alter the drag on the probe and systematically
bias the inferred properties when they are unaccounted for
\cite{geonzon2021effect}. Although experimental
validations are often performed far from walls
\cite{wilson2009passive,besseris1999rotational,radiom2021magnetic}, real
biological samples cannot avoid solid boundaries, especially when the
sample volume is small or when measurements are performed adjacent to
tissue.

Uncertainty and noise in microrheological data typically arise from variability in applied forces, probe surface properties, optical aberrations, post-processing, and thermal fluctuations. These 
variations are most commonly handled by ordinary least-squares fitting, which accounts only for uncorrelated Gaussian noise and does not
capture the correlated model bias introduced by unmodeled
effects such as wall proximity and probe-probe interactions
\cite{rappel2020tutorial,paul2021bayesian}. Explicitly tracking the probe-to-wall distance, which has been attempted \cite{gong2013active}, places stringent demands on optical resolution and data quality, introducing additional uncertainty. A rigorous statistical
framework that jointly models these correlated effects and propagates all
sources of uncertainty into the inferred parameters is therefore needed.
Bayesian inference offers a natural framework and has already proved useful
for identifying viscoelastic material parameters, interpreting
optical-tweezer microrheology data, and calibrating rheological models in
complex flow geometries
\cite{rappel2020tutorial,paul2021bayesian,rinkens2023uncertainty, rinkens2026bayesian}. 
These studies indicate that reliable parameter estimates often require inferring
the constitutive parameters jointly with the measurement-noise level and
the model's structural bias so that unmodeled physics is not absorbed into 
the material properties.

Building on this, we develop a data-analysis framework that combines
simplified analytical models with Bayesian inference to identify
the material parameters of Newtonian and viscoelastic fluids from the
displacement of a force-controlled sphere. The central idea is that the
hydrodynamic interactions with a nearby wall and with neighboring probes
need not be eliminated from the experiment: they are built into the
forward model, with the probe-to-wall separation inferred jointly with
the constitutive parameters, and the probe-probe interactions accounted
for through a known drag correction. We also use the framework to examine how material heterogeneity on length scales comparable to the particle size affects the inferred response.

The paper is organized as follows.
Section~\ref{sec:bayes} introduces the Bayesian inverse problem and the
observation model used to separate measurement noise from model bias.
Section~\ref{sec:governing_models} describes the full-order finite-element
model (FOM) used to generate reference trajectories, and Section~\ref{sec:rom} states the simplified analytical models used in the inference. Section~\ref{sec:bayesian_inference} then
specifies the prior distributions, likelihood, and Markov chain Monte
Carlo (MCMC) sampler. With these ingredients in place,
Section~\ref{sec:numerical_experiments} reports numerical experiments
on synthetically generated data for Newtonian and viscoelastic fluids.

\section{Bayesian inference framework}
\label{sec:bayes}

A central objective in microrheology is to make reliable inferences
about the rheological properties of a fluid from the motion of an
embedded probe. 
Bayesian inference provides a formulation in which the unknown
parameters are modeled as random variables described by probability
distributions, and the discrepancy between the model and the data can be
explicitly decomposed into experimental noise and model bias, which are often
referred to as \emph{aleatoric} and \emph{epistemic} uncertainty, respectively.
In this work, the Bayesian framework is used exclusively for parameter
identification, \ie, we do not use it for rheological model selection
(as done in, \eg,~\cite{freund2015quantitative,rinkens2026bayesian}).
We develop simplified analytical models for Newtonian and linear
viscoelastic fluids, accounting for wall and particle-particle interactions,
and the inference returns the posterior density of their parameters
given the observed trajectory. The forward models are constructed in
Section~\ref{sec:rom}, and the prior distributions, likelihood, and MCMC sampler are specified in
Section~\ref{sec:bayesian_inference}.


We consider the optical tracking of a particle, which yields its displacement in the plane orthogonal to the optical axis. For clarity, we present the
observation model for a single direction in that plane, taking the
$x$-direction as an example.
A measured trajectory of $N$ probe displacements is then
written as the observation vector 
$\vec{x}_{\mathrm{obs}} = [x_{\mathrm{obs},1},\dots,
x_{\mathrm{obs},N}] \in \mathbb{R}^{N}$, where
$x_{\mathrm{obs},i} \in \mathbb{R}$ is the displacement at time $t_{i}$.
Let ${d}_{i}(\params) \in \mathbb{R}$ denote the prediction of the forward model at $t_{i}$ for the parameter vector
$\params \in \paramspace$, where $\paramspace$ is the parameter space. The forward model is the deterministic
mapping $\vec{d}: \paramspace \to \mathbb{R}^{N}$ that returns the predicted
displacements at the $N$ measurement times, with
$\vec{d}(\params) = [{d}_{1}(\params),\dots,
{d}_{N}(\params)] \in \mathbb{R}^{N}$.
Following the Kennedy and O'Hagan formulation, the observations can be
decomposed as~\cite{kennedy2001calibration}
\begin{equation}
\vec{x}_{\mathrm{obs}}
\;=\;
\vec{d}(\params)
+
\vec{\varepsilon}_{\mathrm{exp}}
+
\vec{\varepsilon}_{\mathrm{bias}},
\label{eq:error_decomposition}
\end{equation}
where $\vec{\varepsilon}_{\mathrm{exp}} \in \mathbb{R}^{N}$ denotes
the experimental noise and
$\vec{\varepsilon}_{\mathrm{bias}} \in \mathbb{R}^{N}$ the
model bias (also called model error or model discrepancy).  
Because we use synthetic data, the model bias can be quantified directly
by comparing the forward model with the full-order solution. We
therefore drop the bias term when it is known to be small, but retain
both contributions when it is large.

Given the observation model~\eqref{eq:error_decomposition}, the
parameter identification problem is posed in the Bayesian setting.
Bayes' theorem expresses the probability density of $\params$ given
the observed trajectory $\vec{x}_{\mathrm{obs}}$, referred to as the
posterior, as
\begin{equation}
p(\params \mid \vec{x}_{\mathrm{obs}})
\;=\;
\frac{p(\vec{x}_{\mathrm{obs}} \mid \params)\,p(\params)}
     {p(\vec{x}_{\mathrm{obs}})}
\;=\;
\frac{\mathcal{L}(\params \mid \vec{x}_{\mathrm{obs}})\,p(\params)}
     {p(\vec{x}_{\mathrm{obs}})},
\label{eq:bayes_intro}
\end{equation}
where
$\mathcal{L}(\params \mid \vec{x}_{\mathrm{obs}})
 \equiv p(\vec{x}_{\mathrm{obs}} \mid \params)$
is the likelihood, $p(\params)$ the prior, and
\begin{equation}
p(\vec{x}_{\mathrm{obs}})
\;=\;
\int_{\paramspace}
\mathcal{L}(\params \mid \vec{x}_{\mathrm{obs}})\,p(\params)\,
\mathrm{d}\params,
\label{eq:evidence}
\end{equation}
the evidence.

The prior $p(\params)$ encodes information available before the
trajectory is observed, such as positivity constraints, expert knowledge, 
independent measurements, or physical intuition regarding the individual 
parameters.
The likelihood $\mathcal{L}(\params \mid \vec{x}_{\mathrm{obs}})$
quantifies how plausibly the model with parameter vector $\params$ explains the observed
trajectory. 
The evidence $p(\vec{x}_{\mathrm{obs}})$ normalizes the posterior over
$\paramspace$, but its evaluation requires an integral over the entire
parameter domain, which becomes computationally prohibitive in high
dimensions. The posterior is therefore characterized using Markov
chain Monte Carlo sampling, which generates samples from
$p(\params \mid \vec{x}_{\mathrm{obs}})$ using only ratios of posterior
densities, in which the evidence cancels.
The posterior combines the prior with the information contributed by
the data through the likelihood. 

A sharp, informative prior dominates
the posterior when the data are sparse or noisy. In contrast, a broad prior
is overwhelmed by the likelihood when the dataset is large or strongly
informative. From the posterior, one extracts posterior means as point
estimates, credible intervals describing the spread of plausible
parameter values, and the correlation structure between inferred
parameters. For the nonlinear forward models considered here, the
posterior admits no closed form and is characterized numerically, as
detailed in Section~\ref{sec:bayesian_inference}.

\section{Full-order model}
\label{sec:governing_models}

This section presents the full-order model used to generate the
high-fidelity particle trajectories that serve as synthetic ground
truth for Bayesian inference. We first specify the particle-wall geometry and the imposed forcing, then introduce the governing equations for incompressible creeping flow around a
translating rigid sphere. We next describe the Newtonian and viscoelastic
constitutive models, as well as the boundary conditions that close the force-driven
particle problem. Finally, we outline the finite element solution procedure
for the coupled velocity-pressure-conformation system, explaining how the
particle displacement is obtained from the computed rigid-body velocity.

\begin{figure}[!ht]
    \centering

    \begin{subfigure}[t]{0.48\linewidth}
        \centering
        \includegraphics[width=\linewidth]{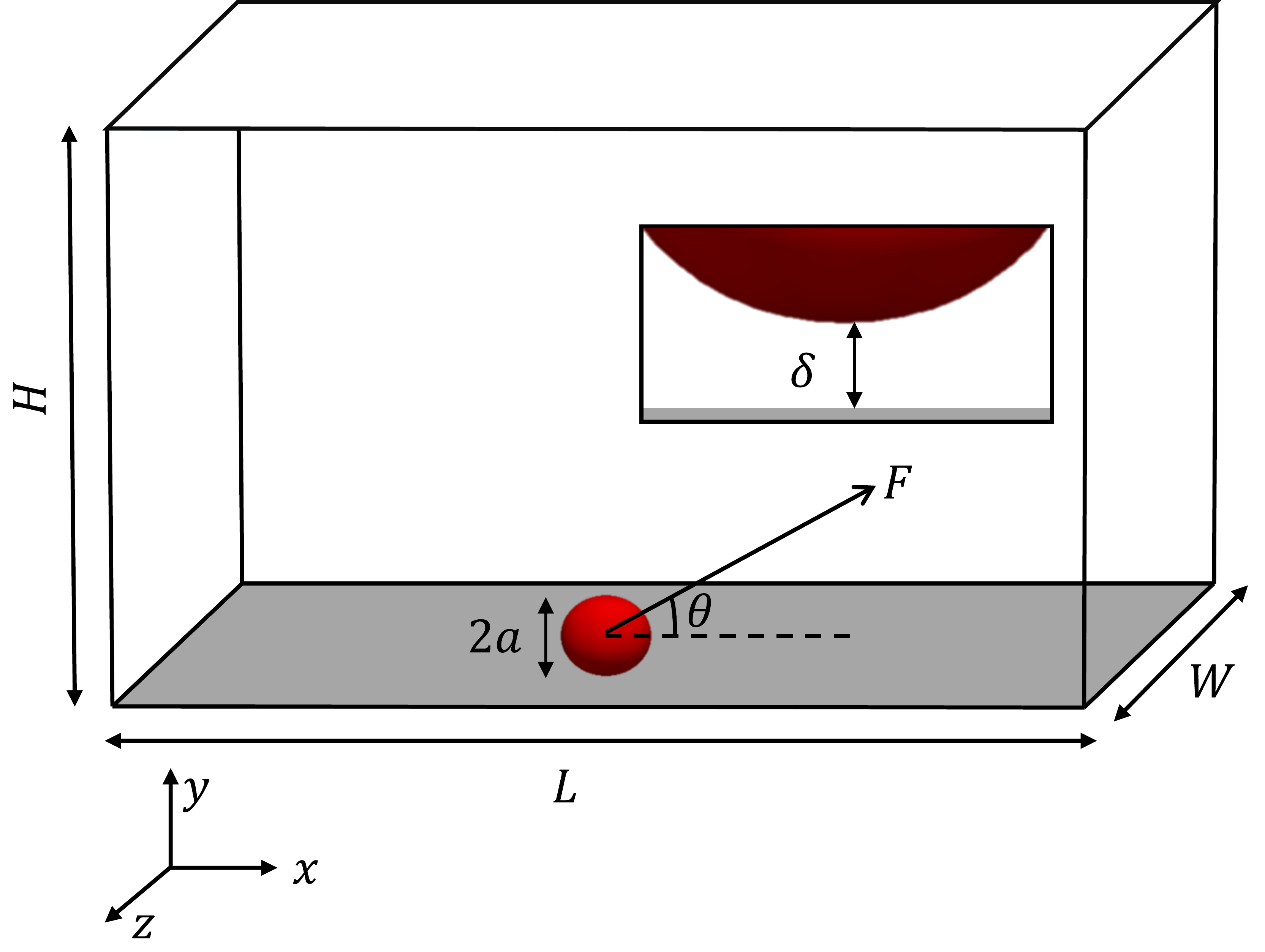}
        \caption{\emph{Case~I}: particle close to a single planar wall.}
        \label{fig:single_wall}
    \end{subfigure}
    \hfill
    \begin{subfigure}[t]{0.48\linewidth}
        \centering
        \includegraphics[width=\linewidth]{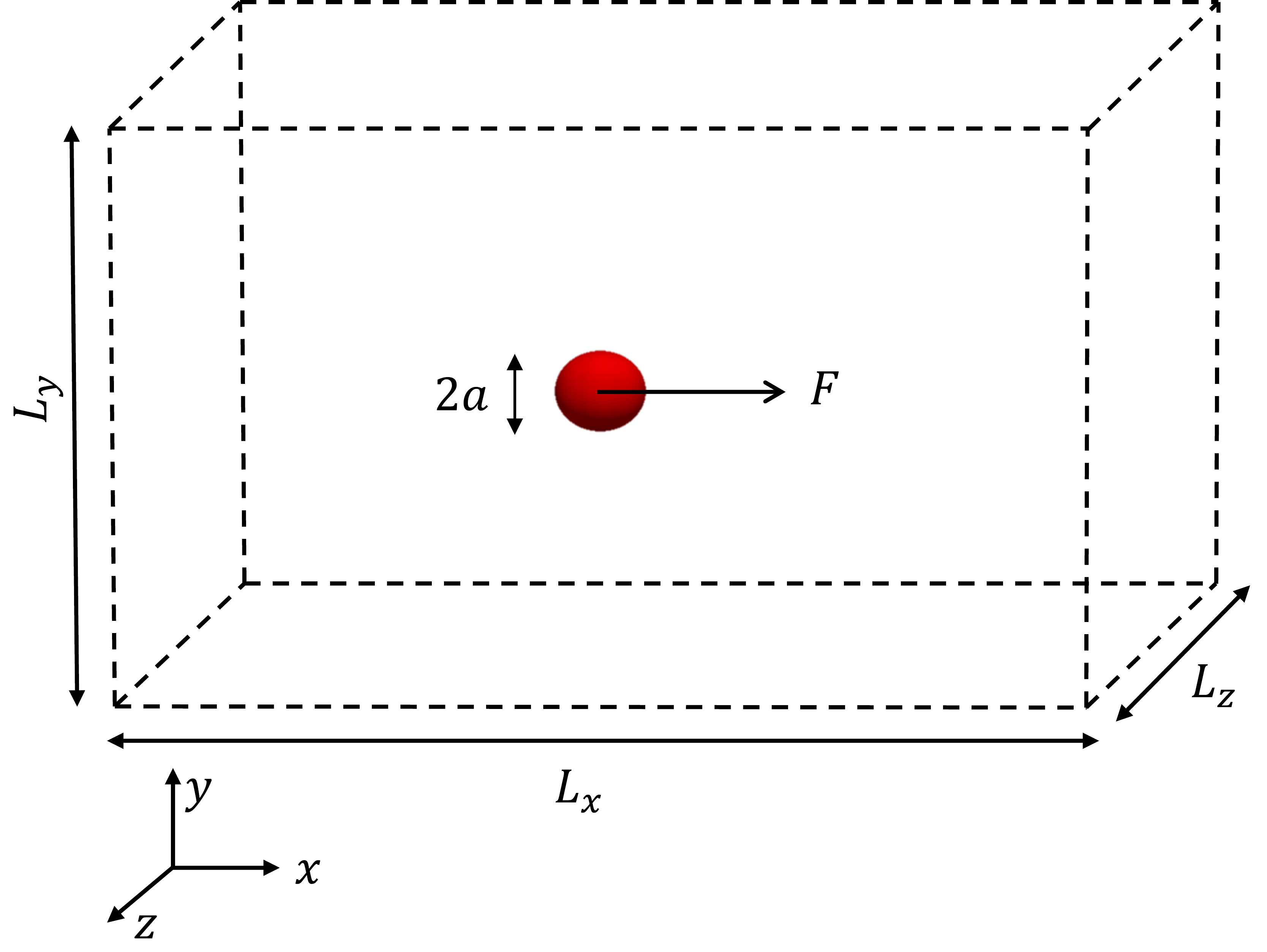}
        \caption{\emph{Case~II}: particle in a triperiodic domain.}
        \label{fig:periodic_domain}
    \end{subfigure}
\caption{Schematics of the configurations considered in the full-order model. 
(a) A rigid sphere of radius $a$ translating near a single planar wall,
where $\delta$ denotes the particle-to-wall separation. The inset above the particle provides a magnified view of the particle-wall distance. (b) The corresponding particle suspended in a triperiodic domain (in this work $L_x = L_y = L_z=L$).}
    \label{fig:domain_comparison}
\end{figure}

\subsection{Problem definition}
\label{subsec:problem_def}
We consider a single non-Brownian, inertialess, rigid spherical particle
of radius $a$ translating through a complex fluid, driven by a constant
external force $\vec{F}$ of magnitude $F=\|\vec{F}\|$ applied at a fixed
angle $\theta$ relative to the $x$-axis,
\begin{equation}
\vec{F} = \bigl(F\cos\theta,\; F\sin\theta,\; 0\bigr),
\label{eq:force_angle}
\end{equation}
so that $\theta=0^{\circ}$ corresponds to forcing in the $x$-direction and
$\theta=90^{\circ}$ to forcing in the $y$-direction.
The suspending fluid is modeled as either Newtonian or viscoelastic. We study two configurations, shown in Figure~\ref{fig:domain_comparison},
together with the unbounded domain to which both configurations approach as the limiting case.

\emph{Case~I}:
In the first configuration (Figure~\ref{fig:single_wall}), the particle
translates near a planar rigid wall with a normal in the $y$-direction, 
initially placed at a
wall-normal separation $\delta_0$ from it. The resulting particle motion
quantifies the wall-induced mobility correction as a function of
$\delta/a$ and $\theta$. 
No-slip conditions are imposed on all walls of the bounding box, and the
domain dimensions are chosen sufficiently large compared with
the particle radius so that the finite computational box approximates a
particle moving near a single wall in an otherwise unbounded domain. We verify this through the domain-size convergence study reported in \ref{sec:mesh_conv}.

\emph{Case~II}: In the second 
configuration (Figure~\ref{fig:periodic_domain}), periodic boundary conditions 
are imposed on the outer faces of the domain so that the problem describes
a periodic array of particles. In this case, the angle $\theta$ is set to 
zero, and particle motion is purely in the $x$-direction. 
The domain size is chosen such that the particle interacts with its own
periodic images; the separation between the particle
and its nearest image is set by the domain dimensions $L_x$, $L_y$, and $L_z$.
In this work, we only consider cubic triperiodic domains with periodic spacing $L$:
$L_x = L_y = L_z=L$.

\emph{Case~III (isolated)}: Both configurations reduce to this unbounded case in the limits
$a \ll \delta \ll H$ with $L, W \gg a$ (\emph{Case~I}) and $L \gg a$ (\emph{Case~II}). 
For numerical efficiency, we approximate this unbounded problem on a 2D axisymmetric 
mesh (see \ref{sec:axisymmetric}).

\subsection{Governing equations}
\label{sec:governing_equations}
Let $\Omega$ denote the total domain containing both the fluid and the
particle, and let $P \subset \Omega$ denote the particle domain, so that
$\Omega \setminus P$ is the fluid domain. Assuming incompressible,
inertialess flow, the momentum and mass balances read
\begin{align}
-\nabla \cdot \ten{\sigma} &= \vec{0}
\quad \text{in } \Omega \setminus P,
\label{eq:momentum} \\
\nabla \cdot \vec{u} &= 0
\quad \text{in } \Omega \setminus P,
\label{eq:continuity}
\end{align}
where $\vec{u}$ is the fluid velocity and $\ten{\sigma}$ is the Cauchy
stress. For the viscoelastic (Oldroyd-B) fluid, we express $\ten{\sigma}$
directly through the conformation tensor $\ten{c}$,
\begin{equation}
\ten{\sigma} = -p\,\ten{I} + 2\eta_{\mathrm{s}}\,\ten{D}
             + G\,(\ten{c} - \ten{I}),
\label{eq:sigma_c}
\end{equation}
where $p$ is the pressure, $\ten{I}$ the unit tensor, $\eta_{\mathrm{s}}$
the solvent viscosity, $\ten{D} = \tfrac{1}{2}\bigl(\nabla\vec{u} +
(\nabla\vec{u})^{\mathrm{T}}\bigr)$ the rate-of-deformation tensor, and
$G=\eta_{\mathrm{p}}/\lambda$ the polymer modulus, with $\eta_{\mathrm{p}}>0$
the polymer viscosity and $\lambda>0$ the relaxation time. The conformation
tensor evolves according to
\begin{equation}
\lambda\,\overset{\nabla}{\ten{c}}
+ (\ten{c} - \ten{I}) = \ten{0}
\quad \text{in } \Omega \setminus P,
\qquad
\overset{\nabla}{\ten{c}}
= \frac{\mathrm{D}\ten{c}}{\mathrm{D}t}
  - (\nabla\vec{u})^{\mathrm{T}}\cdot\ten{c}
  - \ten{c}\cdot\nabla\vec{u},
\label{eq:conformation}
\end{equation}
where $\overset{\nabla}{(\cdot)}$ is the upper-convected derivative and
$\mathrm{D}/\mathrm{D}t$ is the material derivative. A Newtonian fluid 
with viscosity $\eta_{\mathrm{s}}$ is
recovered as $G \to 0$, for which the polymeric
stress vanishes. Together with the boundary and initial conditions below,
Equations~\eqref{eq:momentum}, \eqref{eq:continuity}, \eqref{eq:sigma_c}, and
\eqref{eq:conformation} define the coupled problem for $\vec{u}$, $p$, and
$\ten{c}$.

\subsection{Boundary conditions}
The system is closed by conditions on the particle surface and the outer
boundaries. On the particle surface, the fluid velocity equals the rigid-body
velocity of the particle,
\begin{equation}
\vec{u} = \vec{U} + \vec{\omega}\times\vec{r}
\quad \text{on } \partial P,
\label{eq:no_slip_particle}
\end{equation}
where $\vec{U}$ and $\vec{\omega}$ are the translational and angular
velocities, and $\vec{r}$ is the position vector measured from the particle
center. The motion is force-controlled: $\vec{F}$ is prescribed by
Equation~\eqref{eq:force_angle} and the applied torque is zero, while $\vec{U}$ and
$\vec{\omega}$ are unknowns that follow from the inertialess force and torque
balances,
\begin{align}
\vec{F} &= \int_{\partial P} \ten{\sigma}\cdot\vec{n}\,\mathrm{d}S,
\label{eq:force_balance} \\
\vec{0} &= \int_{\partial P} \vec{r} \times (\ten{\sigma}\cdot\vec{n})\,\mathrm{d}S,
\label{eq:torque_balance}
\end{align}
where $\vec{n}$ is the outward unit normal on $\partial P$. On the outer
boundaries, \emph{Case~I} imposes no-slip conditions on the rigid walls,
whereas \emph{Case~II} imposes triperiodic conditions on $\partial\Omega$,
representing an infinite periodic array (similar to \cite{egelmeers2024numerical,egelmeers2025direct}). For \emph{Case~II}, the velocity is constrained to have zero net
flow rate in the $x$-direction, so that the applied
force does not drive a net flow through the periodic domain. In both cases, the
pressure is defined up to a constant, which is removed by fixing it at one reference
point.

Let $ \vec{X}=(X, Y, Z) $ denote the position of the particle center. The particle motion is governed by \begin{equation} \frac{\mathrm{d}X}{\mathrm{d}t}=U_x, \qquad \frac{\mathrm{d}Y}{\mathrm{d}t}=U_y, \qquad \frac{\mathrm{d}Z}{\mathrm{d}t}=U_z, \end{equation} where $U_x$, $U_y$, and $U_z$ are the Cartesian components of the particle translational velocity.

\subsection{Initial conditions}
At $t=0$, the fluid is in a
stress-free state, so the conformation tensor equals the identity,
\begin{equation}
\ten{c}(t=0) = \ten{c}_0 = \ten{I}.
\label{eq:c_initial}
\end{equation}
The particle starts from its prescribed initial position
$\bigl(X(0), Y(0), Z(0)\bigr) = \bigl(X_0, Y_0, Z_0\bigr)$.

\subsection{Numerical method}
\label{sec:numerical_method}
In this section, we give a concise outline of the numerical approach used in the finite-element model. We refer the reader to earlier works for details of the numerical implementation \cite{egelmeers2024numerical,egelmeers2025direct,jaensson2016direct}. The coupled velocity-pressure-conformation problem is solved by the finite
element method with tetrahedral $P_2$-$P_1$-$P_1$ elements for the velocity,
pressure, and conformation fields. The quadratic-velocity/linear-pressure pair
satisfies the Ladyzhenskaya-Babu\v{s}ka-Brezzi (LBB) condition. Boundary-fitted meshes are generated with
Gmsh~\cite{geuzaine2009gmsh} and updated in time to conform to the moving
particle surface while the outer boundaries remain fixed. Because the mesh
follows the particle but is not fully Lagrangian, the equations are cast in an
arbitrary Lagrangian-Eulerian (ALE) frame~\cite{hu2001direct}. The domain is remeshed with the solution projected
onto the new mesh whenever the mesh deformation becomes too large.   To maintain
stability at high Weissenberg numbers, the conformation tensor is advanced in
its log-representation~\cite{fattal2004constitutive,hulsen2005flow}, with its
convective term stabilized by the streamline-upwind/Petrov-Galerkin (SUPG)
method~\cite{brooks1982streamline} and the momentum balance by the DEVSS-G
formulation~\cite{guenette1995new,bogaerds2002stability}. At each time step, the
system is solved for $\vec{u}$, $p$, $\ten{c}$, and the rigid-body velocities
$\vec{U}$ and $\vec{\omega}$ subject to the force
balance~\eqref{eq:force_balance}, after which the particle center is advanced by
a second-order Adams-Bashforth scheme with a first-order Euler startup. For
\emph{Case~I}, the force has no $z$-component, making the problem symmetric about
the plane through the particle center normal to $z$, so only half of the domain
is simulated with a symmetry condition on that plane. We do not implement this symmetric mesh for \emph{Case~II} due to the use of a heterogeneous microstructure (Section \ref{sec:heterogeneous_inference}), which makes the problem non-symmetric.

\section{Simplified analytical model}
\label{sec:rom}

This section develops the simplified analytical model (SAM) used as the
forward model in the Bayesian inference. The model relates the applied force
to the particle motion through Stokes mobility, with corrections for wall and
particle-particle interactions. It is then validated against the full-order model (FOM) of Section~\ref{sec:governing_models}. 

Throughout this section, the particle center starts at the origin, $(X_0,\, Y_0,\, Z_0)=(0,0,0)$, where $Y$ denotes the wall-normal displacement measured from the particle's initial position, taken as positive in the direction away from the wall. The instantaneous particle-wall gap is $\delta(t)=\delta_0+Y(t)$, with $\delta_0=\delta(0)>0$ denoting the initial gap. 
\subsection{Newtonian wall effects}
\label{sec:wall_corrections}

\begin{figure}
    \centering
    \includegraphics[width=0.66\linewidth]{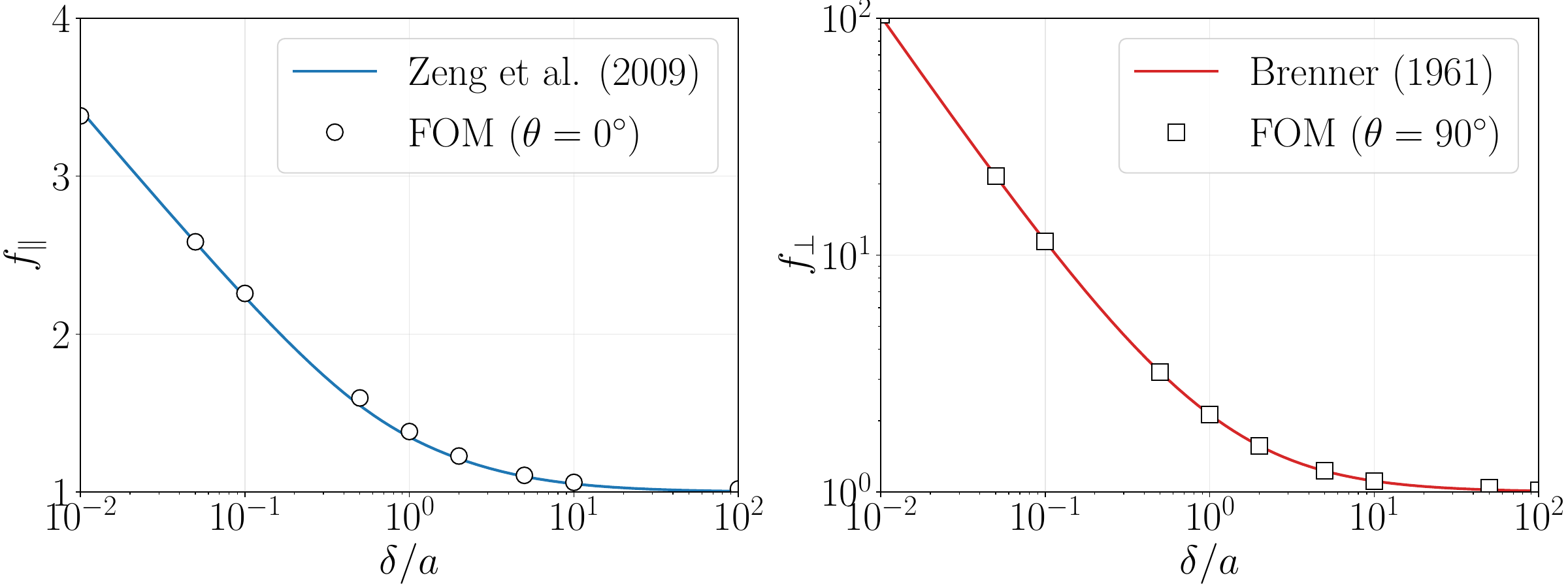}
    \caption{Wall-correction factors $f_{\parallel}$ and $f_{\perp}$ compared
    with FOM reference data.}
    \label{fig:wall_correction}
\end{figure}

In an unbounded creeping flow, the drag on a sphere translating with velocity
$\vec{U}$ is given by Stokes' law,
$\vec{F}=6\pi\eta_{\mathrm{s}}a\vec{U}$. A nearby wall increases the drag and
makes it anisotropic, so the resistance is different for motion parallel and
normal to the wall. In the configuration considered here (\emph{Case~I}), symmetry
keeps the particle in the $x$-$y$ plane, and the SAM tracks only $X(t)$ and
$Y(t)$. The wall-corrected drag is written as
\begin{equation}
\vec{F}
= 6\pi\,\eta_{\mathrm{s}}\,a\,
  \bigl(f_{\parallel}\,U_{\parallel}\,\vec{e}_x
  + f_{\perp}\,U_{\perp}\,\vec{e}_y\bigr),
\label{eq:modified_stokes_drag}
\end{equation}
where $U_{\parallel}$ and $U_{\perp}$ are the wall-parallel and wall-normal
velocity components. For perpendicular
motion, we use Brenner's exact solution~\cite{brenner1961slow},
\begin{equation}
f_{\perp}
= \dfrac{4}{3}\sinh\alpha
\sum_{n=1}^{\infty}
\dfrac{n(n+1)}{(2n-1)(2n+3)}
\left[
\dfrac{2\sinh((2n+1)\alpha) + (2n+1)\sinh(2\alpha)}
     {4\sinh^{2}((n+\frac{1}{2})\alpha) - (2n+1)^{2}\sinh^{2}\alpha}
- 1\right],
\label{eq:brenner}
\end{equation}
with $\alpha=\operatorname{arccosh}((\delta+a)/a)$. For numerical efficiency, 
the series is truncated at $N_{\mathrm{trunc}}=100$ terms, 
which is sufficient for the smallest gaps
considered here (Figure~\ref{fig:wall_correction}).
For parallel motion, we use the Zeng
interpolant~\cite{zeng2009forces,loth2023fluid},
\begin{equation}
f_{\parallel}\approx
1.028 - \dfrac{0.07\,a^{2}}{a^{2}+\delta^{2}}
- \dfrac{8}{15}\ln\!\left(\dfrac{135\,\delta}{135\,a + 128\,\delta}\right).
\label{eq:f_parallel_zeng}
\end{equation}
This expression interpolates smoothly between the lubrication solution at small
$\delta$ and the far-field solution at large $\delta$. It strictly applies to
a non-rotating sphere, whereas the FOM allows rotation. For the values of
$\delta$ considered here, however, the drag on a torque-free sphere differs
from that on a non-rotating sphere by only a few percent~\cite{goldman1967slow}.
This difference is small compared with typical microrheology uncertainties.
Both correction factors agree with the FOM reference data across the tested
wall separations (Figure~\ref{fig:wall_correction}). The comparisons shown here use the converged discretization M3/D2 identified in \ref{sec:mesh_conv}.

\begin{figure}[!ht]
    \centering
    \includegraphics[width=0.95\linewidth]{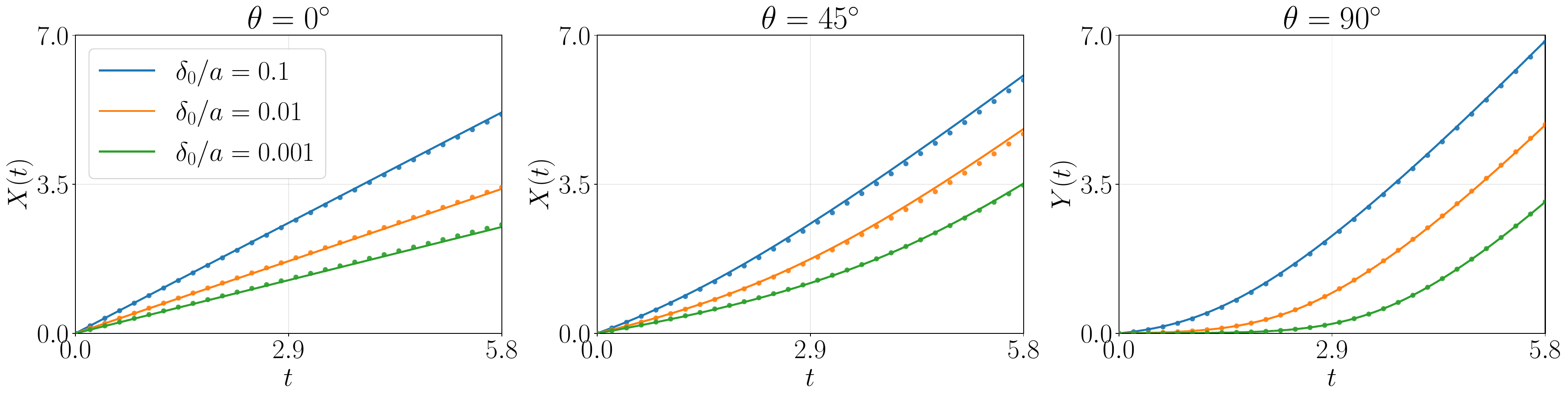}
\caption{Simplified analytical model (solid lines)
versus FOM displacements (dots) for initial wall distances
$\delta_0/a \in \{0.1,\,0.01,\,0.001\}$ and forcing angles
$\theta \in \{0^\circ,\,45^\circ,\,90^\circ\}$.}
    \label{fig:newtonian_displacement_3x3}
\end{figure}

With the wall factors evaluated at the instantaneous gap $\delta(t)$, a
constant force gives
\begin{equation}
\dfrac{\mathrm{d}X}{\mathrm{d}t}
= \dfrac{F\cos\theta}{6\pi\,a\,\eta_{\mathrm{s}}\,f_{\parallel}(\delta(t),a)},
\qquad
\dfrac{\mathrm{d}Y}{\mathrm{d}t}
= \dfrac{F\sin\theta}{6\pi\,a\,\eta_{\mathrm{s}}\,f_{\perp}(\delta(t),a)},
\label{eq:newtonian_displacement}
\end{equation}
where $\mathrm{d}X/\mathrm{d}t$ and $\mathrm{d}Y/\mathrm{d}t$ are the particle
velocities in the $x$- and $y$-directions.
We solve this nonlinear system numerically using
SciPy's \texttt{solve\_ivp} with adaptive Runge-Kutta time stepping \cite{2020SciPy-NMeth}.
As shown in Figure~\ref{fig:newtonian_displacement_3x3}, the resulting trajectories predicted by the SAM agree closely with the FOM across all tested initial gaps and forcing angles for a force magnitude $F=12\pi$, with particle radius $a=1$ and solvent viscosity $\eta_{\mathrm{s}}=1$.

\subsection{Newtonian particle-particle effects}
\label{sec:hasimoto}

We next consider a periodic array of particles moving through a Newtonian
fluid, with forcing in the $x$-direction (\emph{Case~II}). Hasimoto~\cite{hasimoto1959periodic}
showed that the drag in such a periodic lattice is increased relative to the
unbounded Stokes drag by the factor $1/Q$, where $Q<1$ is a mobility ratio,
\begin{equation}
\vec{F} = \dfrac{6\pi\,\eta_{\mathrm{s}}\,a\,\vec{U}}{Q}.
\label{eq:hasimoto_drag}
\end{equation}
For a simple cubic lattice of spacing $L$, measured in units of $a$, this ratio is
\begin{equation}
Q(L) = 1 - \dfrac{2.8373}{L} + \dfrac{4\pi}{3}\dfrac{1}{L^{3}} + \cdots.
\label{eq:hasimoto_Q}
\end{equation}
This correction forms the SAM for \emph{Case~II}. Table~\ref{tab:hasimoto} verifies it
against FOM data with $F=6\pi$, $\eta_{\mathrm{s}}=1$, and $a=1$. The Hasimoto-corrected velocity matches the FOM data with an error of at most $0.5\%$ for all values of $L$ considered.

\begin{table}[!ht]
\centering
\begin{tabular}{cccc}
\toprule
$L$ & $U \, \text{(FOM)} $ & $U \, \text{(SAM)} $ & rel.\ error \\
\midrule
5  & 0.463722 & 0.466050 & $0.50\%$ \\
10 & 0.720247 & 0.720458 & $0.029\%$ \\
20 & 0.858431 & 0.858658 & $0.026\%$ \\
40 & 0.928885 & 0.929132 & $0.027\%$ \\
60 & 0.952005 & 0.952731 & $0.076\%$ \\
\bottomrule
\end{tabular}
\caption{Comparison between the FOM velocity and the Hasimoto-corrected SAM
velocity for a periodic array of particles.}
\label{tab:hasimoto}
\end{table}

\subsection{Viscoelasticity}
\label{sec:ve_rom}

For viscoelastic fluids, the model is extended to a creep-recovery experiment.
A constant force is applied over $0\le t\le t_0$ and then removed:
\begin{equation}
\label{eq:ve_forcing}
F_x(t)=
\begin{cases}
F\cos\theta, & 0\le t\le t_0,\\
0, & t>t_0,
\end{cases}
\qquad
F_y(t)=
\begin{cases}
F\sin\theta, & 0\le t\le t_0,\\
0, & t>t_0.
\end{cases}
\end{equation}
We consider small deformations, so the SAM is restricted to the linear regime.
The unbounded model is derived first. Wall and particle-particle corrections
are then added as mobility modifications.

As a starting point for the viscoelastic SAM, we use the standard Maxwell
expression for the scalar polymeric stress $\tau_{\mathrm{p}}(t)$
\cite{morrison2001understanding},
\begin{equation}
\lambda\,\dfrac{\mathrm{d}\tau_{\mathrm{p}}}{\mathrm{d}t}
+ \tau_{\mathrm{p}} = \eta_{\mathrm{p}}\,\dfrac{U}{a}.
\label{eq:scalar_tau_oldroyd_b}
\end{equation}
Here $U/a$ is used as a scalar shear-rate scale. Although the configurations
considered here may involve different geometric length scales (such as the gap distance), this simplified
choice gives accurate predictions when wall or particle-particle interactions are
included through the Newtonian mobility corrections, as shown later in this section.

The time-dependent polymeric drag is written as
$F_{\mathrm{p}}=C\,\tau_{\mathrm{p}}$. Matching the steady-state Stokes drag
$6\pi a\,\eta_{\mathrm{p}}U$ gives $C=6\pi a^{2}$. Substitution into Equation~\eqref{eq:scalar_tau_oldroyd_b} yields
\begin{equation}
\lambda\,\dfrac{\mathrm{d}F_{\mathrm{p}}}{\mathrm{d}t} + F_{\mathrm{p}}
= 6\pi a\,\eta_{\mathrm{p}}\,U,
\label{eq:Fp_relax}
\end{equation}
which is a first-order ODE for the polymeric drag. Equation~\eqref{eq:Fp_relax}
is the linear, first-order mobility limit of the weakly nonlinear
force-velocity relation derived for a single-mode Maxwell model
in~\cite{joens2022unsteady}. For a constant applied velocity, its solution
reproduces their linear start-up force. 

In the absence of inertia, the applied force $F(t)$ is balanced by the solvent
and polymeric drag,
\begin{equation}
\underbrace{6\pi a\,\eta_{\mathrm{s}}\,U}_{\text{solvent drag}}
+ \underbrace{F_{\mathrm{p}}}_{\text{polymer drag}}
= F(t).
\label{eq:ve_balance_unbounded}
\end{equation}
The wall-bounded viscoelastic model is obtained by applying the Newtonian
wall-correction factors to the mobility. In the linear regime, the polymeric
stress is assumed to be weakly affected by the boundary, so the same
hydrodynamic correction is used for the solvent and polymeric drag. The factors
$f_{\parallel}$ and $f_{\perp}$ describe motion parallel and normal to the
wall, respectively, and are evaluated at the instantaneous gap
$\delta(t)=\delta_0+Y(t)$. 
With $F_{\mathrm{p},x}$ and $F_{\mathrm{p},y}$ denoting the wall-corrected
polymeric drag components, the wall-bounded system is
\begin{equation}
\label{eq:oldroyd_b_force_ode}
\begin{aligned}
\dfrac{\mathrm{d}X}{\mathrm{d}t}
&=
\dfrac{1}{6\pi a\eta_{\mathrm{s}}}
\dfrac{F_x(t) - F_{\mathrm{p},x}}{f_{\parallel}\!\left(\delta(t),a\right)},
&\qquad
\dfrac{\mathrm{d}Y}{\mathrm{d}t}
&=
\dfrac{1}{6\pi a\eta_{\mathrm{s}}}
\dfrac{F_y(t) - F_{\mathrm{p},y}}{f_{\perp}\!\left(\delta(t),a\right)},
\\[0.5em]
\dfrac{\mathrm{d}F_{\mathrm{p},x}}{\mathrm{d}t}
&=
-\dfrac{F_{\mathrm{p},x}}{\lambda}
+
\dfrac{6\pi a\eta_{\mathrm{p}}f_{\parallel}\!\left(\delta(t),a\right)}{\lambda}
\dfrac{\mathrm{d}X}{\mathrm{d}t},
&\qquad
\dfrac{\mathrm{d}F_{\mathrm{p},y}}{\mathrm{d}t}
&=
-\dfrac{F_{\mathrm{p},y}}{\lambda}
+
\dfrac{6\pi a\eta_{\mathrm{p}}f_{\perp}\!\left(\delta(t),a\right)}{\lambda}
\dfrac{\mathrm{d}Y}{\mathrm{d}t},
\end{aligned}
\end{equation}
with initial conditions
\begin{equation}
\label{eq:ve_wall_initial_conditions}
X(0)=X_0,
\qquad
Y(0)=Y_0,
\qquad
F_{\mathrm{p},x}(0)=0,
\qquad
F_{\mathrm{p},y}(0)=0,
\end{equation}
representing a stress-free initial state of the fluid. Because $\delta(t)=\delta_0+Y(t)$, the wall factors change as the particle
moves under oblique forcing ($\theta\neq0$). This couples the $x$- and
$y$-dynamics: motion in $y$ changes the gap, and the changing gap modifies the
mobility in both directions.  We solve this nonlinear system numerically using
SciPy's \texttt{solve\_ivp} with adaptive Runge-Kutta time stepping \cite{2020SciPy-NMeth}.

For $\theta=0$ and for the particle-particle interaction problem in Section~\ref{sec:ve_pp},
the correction coefficients are constant. The system then decouples and has an analytical solution (obtained by, for example, a Laplace transform similar to \cite{furst2017microrheology}), written in terms of
the scaled relaxation time
$\tau=\lambda\eta_{\mathrm{s}}/(\eta_{\mathrm{s}}+\eta_{\mathrm{p}})$ as follows:
\begin{subequations}\label{eq:analytic_creep}
\begin{align}
  X(t) &= \frac{F}{6\pi a (\eta_{\mathrm{s}} + \eta_{\mathrm{p}})f_{\parallel}\!\left(\delta_0,a\right)}\, t
        + \frac{F\,\lambda\,\eta_{\mathrm{p}}}{6\pi a (\eta_{\mathrm{s}} + \eta_{\mathrm{p}})^2 f_{\parallel}\!\left(\delta_0,a\right)}
          \left(1 - e^{-t/\tau}\right),
        && 0 \le t \le t_0,
        \label{eq:analytic_loading} \\[6pt]
  X(t) &= X(t_0)
        - \frac{F\,\lambda\,\eta_{\mathrm{p}}}{6\pi a (\eta_{\mathrm{s}} + \eta_{\mathrm{p}})^2 f_{\parallel}\!\left(\delta_0,a\right)}
          \left(1 - e^{-t_0/\tau}\right)
          \left(1 - e^{-(t-t_0)/\tau}\right),
        && t > t_0.
        \label{eq:analytic_recovery}
\end{align}
\end{subequations}

Equation~\eqref{eq:analytic_loading} describes loading, and
Equation~\eqref{eq:analytic_recovery} describes recovery after the force is removed
at $t=t_0$.

\begin{figure}[!ht]
    \centering
    \includegraphics[width=\linewidth]{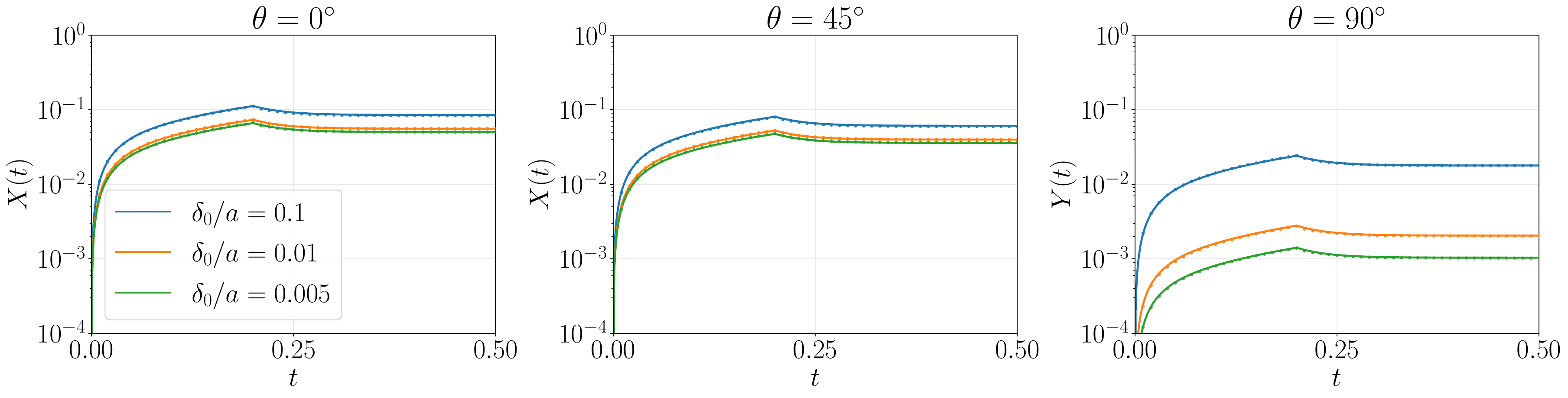}
\caption{Simplified analytical model (solid lines)
versus FOM particle displacements (dots) for wall distances
$\delta_0/a \in \{0.1,\,0.01,\,0.005\}$ and forcing angles
$\theta \in \{0^\circ,\,45^\circ,\,90^\circ\}$.}
\label{fig:viscoelastic_displacement}
\end{figure}

We consider the parameter set
$(\eta_{\mathrm{s}},\eta_{\mathrm{p}},\lambda)
= (0.5,\,0.9,\,0.1)$,
with forcing magnitude \(F=8\pi\) and force-removal time \(t_0=0.2\).
The wall-corrected SAM is compared with the FOM in
Figure~\ref{fig:viscoelastic_displacement} for initial particle-wall gaps
$\delta_0/a \in \{0.1,\,0.01,\,0.005\}$
and forcing angles
$\theta \in \{0^\circ,\,45^\circ,\,90^\circ\}$.
These cases test the model where
the corrections are strongest: small gaps produce large resistance factors,
and oblique forcing makes those factors vary in time. Across all nine cases,
the SAM trajectories nearly coincide with the FOM displacements without
adjusting the constitutive parameters. Thus, using a common wall factor for the
solvent and polymeric drag introduces no visible error in the linear regime.

\FloatBarrier
\subsubsection{Particle-particle effects}
\label{sec:ve_pp}

The interaction between the probe and its periodic images is incorporated as a
mobility rescaling, following the same idea used for wall effects. From
Section~\ref{sec:hasimoto}, the mobility is multiplied by $Q(L)$. This is
equivalent to setting $f_{\parallel}=f_{\perp}=1/Q$ in
system~\eqref{eq:oldroyd_b_force_ode}.

\begin{figure}[!ht]
    \centering
    \includegraphics[width=0.63\linewidth]{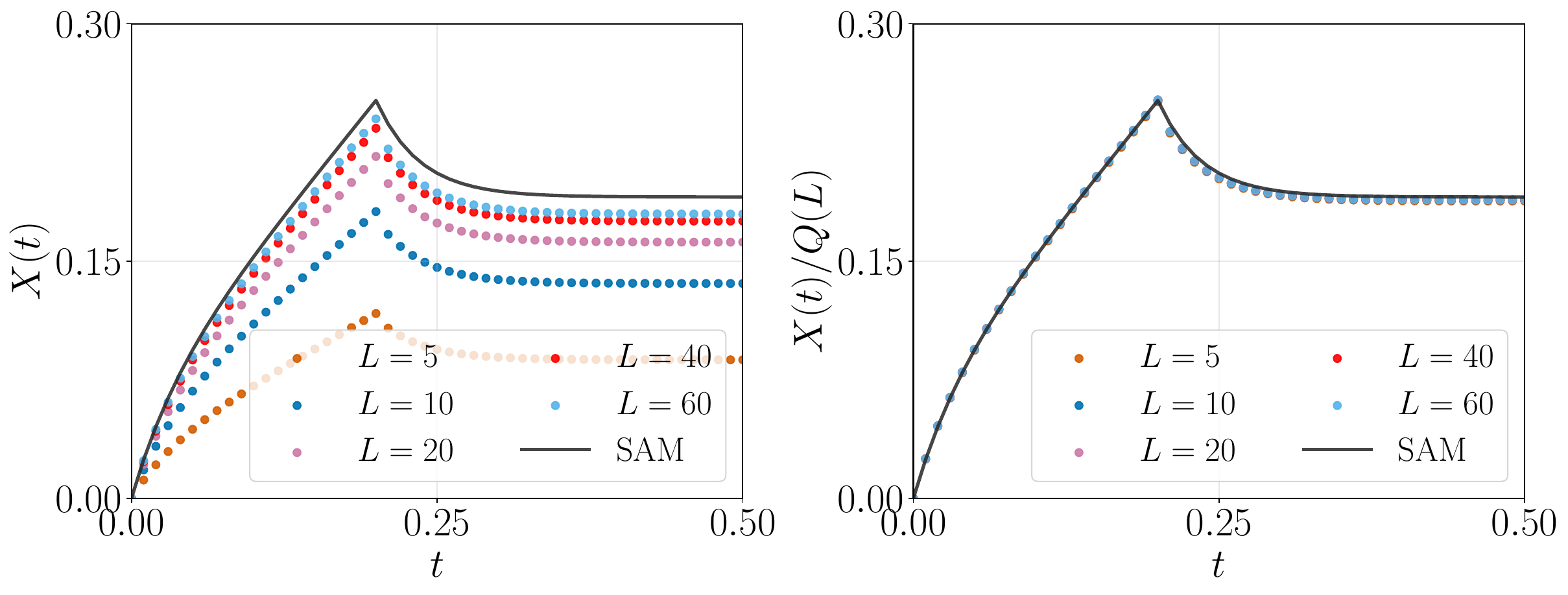}
    \caption{FOM particle displacements $X(t)$ compared to the unbounded SAM (left) and the
    Hasimoto-corrected particle displacement $X(t)/Q(L)$ (right) for box sizes
    $L\in\{5,\,10,\,20,\,40,\,60\}$. Dots show the FOM data and the solid line shows the unbounded SAM prediction.}
    \label{fig:hasimoto_ve}
\end{figure}

To assess the periodic-image correction, we consider the parameter set $(\eta_{\mathrm{s}},\eta_{\mathrm{p}},\lambda) = (0.5,\,0.9,\,0.1)$, with forcing magnitude $F=8\pi$ and force-removal time $t_0=0.2$, and compare the resulting SAM and FOM displacements in Figure~\ref{fig:hasimoto_ve}. The FOM results use mesh M3, defined in \ref{sec:conv_periodic}. The left panel shows
the FOM displacements for periodic box sizes $L\in\{5,\,10,\,20,\,40,\,60\}$.
Smaller boxes increase the interaction with neighboring images, so the curves
fall below the unbounded response. The right panel shows the same trajectories
divided by $Q(L)$. The corrected curves collapse onto the unbounded SAM prediction for all box sizes, including the transient build-up during creep and
the elastic recoil during recovery. The same scalar correction that recovers
the Newtonian mobility therefore also recovers the viscoelastic creep dynamics
without changing the constitutive parameters.

\FloatBarrier

\section{Inference procedure}
\label{sec:bayesian_inference}

The forward models of Section~\ref{sec:rom} define the deterministic map from
the physical parameter vector $\params$ to the predicted particle trajectory
$\vec{d}(\params)$. For a Newtonian fluid,
$\params=[\eta_{\mathrm{s}},\,\delta_0]$, while for a viscoelastic fluid,
$\params=[\eta_{\mathrm{s}},\,\eta_{\mathrm{p}},\,\lambda,\,\delta_0]$.
The gap parameter $\delta_0$ is included only for \emph{Case~I}. The general Bayesian
framework was introduced in
Section~\ref{sec:bayes}. We now specify the priors, the likelihood,
and the MCMC sampler.

\subsection{Prior distributions}
\label{subsec:priors}

The prior encodes what is known before the trajectory is observed. For the
physical parameters, we assume positivity and broad, physically reasonable
ranges. Each component of $\params$ is transformed with a base-10 logarithm:
\begin{equation}
 \hat{\params}
= \bigl[\log_{10}\eta_{\mathrm{s}},\,\log_{10}\delta_0\bigr],   
\end{equation}
for the Newtonian problem, and
\begin{equation}
    \hat{\params}
= \bigl[\log_{10}\eta_{\mathrm{s}},\,\log_{10}\eta_{\mathrm{p}},
\,\log_{10}\lambda,\,\log_{10}\delta_0\bigr],
\end{equation}
for the viscoelastic problem. We place a uniform prior on these transformed
variables (with bounds specified in Section \ref{sec:numerical_experiments}), allowing the inference to span orders of magnitude.

The remaining unknowns describe the observation error. Following Kennedy and
O'Hagan~\cite{kennedy2001calibration}, we separate independent measurement
noise from correlated model discrepancy:
\begin{equation}
\vec{\varepsilon}_{\mathrm{exp}}
\sim \mathcal{N}\!\left(\vec{0},\,\sigma_{\mathrm{exp}}^{2}\,\mat{I}\right),
\qquad
\vec{\varepsilon}_{\mathrm{bias}}
\sim \mathcal{N}\!\left(\vec{0},\,\sigma_{\mathrm{bias}}^{2}\,\mat{K}\right).
\label{eq:exp_noise_prior}
\end{equation}
Here $\sigma_{\mathrm{exp}}>0$ is the measurement-noise standard deviation, and
$\sigma_{\mathrm{bias}}>0$ is the amplitude of the correlated discrepancy. The
discrepancy is modeled as a zero-mean Gaussian process with the stationary
exponential kernel~\cite{brynjarsdottir2014,rasmussen2006gpml}
\begin{equation}
\mat{K}_{ij}
= \exp\!\left(-\dfrac{|t_{i} - t_{j}|}{\ell_{\mathrm{bias}}}\right),
\label{eq:kernel}
\end{equation}
where the parameter $\ell_{\mathrm{bias}}$ is the correlation timescale of the
discrepancy, in the same time units as $t_i$.

When model bias is inferred, $\ell_{\mathrm{bias}}$ is fixed to a prescribed value (its influence is studied in Section~\ref{subsec:wall_newtonian}) and
$\sigma_{\mathrm{exp}}$ and $\sigma_{\mathrm{bias}}$ are inferred jointly with
$\params$. We employ exponential priors for these parameters:
\begin{equation}
\sigma_{\mathrm{exp}}\sim \mathcal{E}(\beta),
\qquad
\sigma_{\mathrm{bias}}\sim \mathcal{E}(\beta),
\label{eq:priors_noise}
\end{equation}
where $\beta$ is the rate parameter of the exponential distribution. These
priors favor small noise amplitudes while still allowing larger values when
supported by the data. Case-specific numerical values are reported in
Section~\ref{sec:numerical_experiments}.

\subsection{Likelihood function}
\label{subsec:likelihood}

Because both error terms are Gaussian, their sum is also Gaussian:
$\vec{\varepsilon}=\vec{\varepsilon}_{\mathrm{exp}}
+\vec{\varepsilon}_{\mathrm{bias}}$. Its covariance is
\begin{equation}
\mat{\Sigma}
= \sigma_{\mathrm{exp}}^{2}\,\mat{I}
+ \sigma_{\mathrm{bias}}^{2}\,\mat{K}
\,\in\, \mathbb{R}^{N\times N},
\label{eq:covariance}
\end{equation}
where $\mat{I}$ is the identity matrix and $\mat{K}$ is the correlation matrix
from Equation~\eqref{eq:kernel}. Both have size $N\times N$ for a trajectory with
$N$ observations, denoted by
$\vec{x}_{\mathrm{obs}}=[x_{\mathrm{obs},1},\dots,x_{\mathrm{obs},N}]$.

Writing the physical parameters as $10^{\hat{\params}}$, the residual
against the data is
\begin{equation}
\vec{r}
= \vec{x}_{\mathrm{obs}} - \vec{d}(10^{\hat{\params}}),
\qquad
\vec{r},\,\vec{x}_{\mathrm{obs}},\,\vec{d}\in \mathbb{R}^{N}.
\label{eq:residual}
\end{equation}
The likelihood is the density of the combined error evaluated at this residual:
\begin{equation}
\mathcal{L}\bigl(\hat{\params},\,\sigma_{\mathrm{exp}},\,\sigma_{\mathrm{bias}}
\mid \vec{x}_{\mathrm{obs}}\bigr)
= \dfrac{1}{(2\pi)^{N/2}\,\det(\Sigma)^{1/2}}
  \exp\!\left(-\tfrac{1}{2}\,\vec{r}^{\top}\mat{\Sigma}^{-1}\vec{r}\right).
\label{eq:likelihood}
\end{equation}
The corresponding log-likelihood is
\begin{equation}
\log \mathcal{L}\bigl(\hat{\params},\,\sigma_{\mathrm{exp}},\,\sigma_{\mathrm{bias}}
\mid \vec{x}_{\mathrm{obs}}\bigr)
= -\tfrac{1}{2}\,\vec{r}^{\top}\mat{\Sigma}^{-1}\vec{r}
  -\tfrac{1}{2}\,\log\left(\det(\Sigma)\right)
  -\tfrac{N}{2}\log(2\pi).
\label{eq:loglikelihood}
\end{equation}
In most cases, we perform inference on a single particle trajectory in a single coordinate direction. However, we also consider multiple trajectories (Section~\ref{subsec:nonlinear}) and multiple coordinate directions (Section~\ref{subsec:wall_ve}). In these cases, the likelihood is straightforwardly extended to include the corresponding outputs of the SAM.

\subsection{Posterior sampling}
\label{subsec:posterior}

Bayes' theorem (Equation \eqref{eq:bayes_intro}) specifies the posterior. The inferred
quantities are the transformed physical parameters $\hat{\params}$ and the
error hyperparameters $\sigma_{\mathrm{exp}}$ and $\sigma_{\mathrm{bias}}$.
The evidence $p(\vec{x}_{\mathrm{obs}})$ only adds a constant, so the target
density is
\begin{equation}
\log p\bigl(\hat{\params},\sigma_{\mathrm{exp}},\sigma_{\mathrm{bias}}
      \mid \vec{x}_{\mathrm{obs}}\bigr)
= \log \mathcal{L}\bigl(\hat{\params},\,\sigma_{\mathrm{exp}},\,\sigma_{\mathrm{bias}}
\mid \vec{x}_{\mathrm{obs}}\bigr)
+ \log p\bigl(\hat{\params},\sigma_{\mathrm{exp}},\sigma_{\mathrm{bias}}\bigr)
+ \mathrm{const},
\label{eq:log_posterior}
\end{equation}
where the constant $-\log p(\vec{x}_{\mathrm{obs}})$ is omitted in practice. Because the
forward models are nonlinear, and the covariance contains inferred
hyperparameters, this posterior has no closed form. We therefore sample it
using Markov chain Monte Carlo. Specifically, we use the Affine Invariant
Stretch Move sampler of Goodman and Weare~\cite{goodman2010ensemble},
implemented in \texttt{emcee}~\cite{foremanmackey2013emcee}. Multiple walkers
explore the posterior in parallel, which helps reduce sensitivity to their
starting positions. 

In all experiments, we use $N_{\mathrm{w}} = 2N_{\mathrm{dim}}$ walkers and $10{,}000$ iterations each. Since the material parameters
span several orders of magnitude, prior-initialized walkers often start where
the likelihood is nearly flat (\eg, if the true viscosity value is $\mathcal{O}(1)$, proposals
of $90$ and $100$ are nearly indistinguishable), causing poor mixing. Therefore, we run a short warm-up chain of ($10\%$ of the full number of iterations) from the prior and initialize the walkers near the highest-posterior states encountered by adding Gaussian perturbations of 10\% of each parameter's marginal standard deviation. Convergence is assessed
using trace plots and the Gelman-Rubin
statistic~\cite{gelman1992rubin,brooks1998convergence}. An analysis of this MCMC sampler for use in flow problems, specifically for the identification of constitutive parameters, can be found in \cite{rinkens2025sampler,rinkens2026sampler}.
The first $30\%$ of samples are discarded as
burn-in, allowing the walkers to move from their prior initialization to the
typical set. We accept a chain when
the walkers mix around a common stationary region, $\hat{R}$ is close to one, 
and in practice $\hat{R}\lesssim1.1$. We additionally verify stability by checking that the posterior means, credible
intervals, and correlations remain stable as the chain length is increased.
Posterior distributions for the material parameters are obtained by applying
$\params=10^{\hat{\params}}$ to each sample. The hyperparameters are sampled in physical units and require no transformation.

In plotting the results, we generally show the mean of the posterior, as well as the 95\% credible interval, where the latter is computed as mean $\pm 1.96$ posterior standard deviations.
Finally, we note that we use plots of the posterior predictive
distribution (PPD) to check if the model and uncertainties are correctly identified visually.
The PPD is obtained by propagating samples from the calibrated posterior
through the SAM and adding the observation noise. If the model-bias term
is included, each draw additionally includes a realization of the Gaussian-process
discrepancy, so the band reflects both the measurement noise and the estimated structural error. 

\section{Numerical experiments}
\label{sec:numerical_experiments}

We assess the inference procedure on a sequence of numerical experiments
of increasing complexity, using finite-element solutions as the ground
truth data throughout. The experiments progress from an isolated particle to a wall-bounded particle, and finally to particle-particle interactions in
periodic suspensions. Similarly, the constitutive equations vary from Newtonian to linear viscoelastic behavior and then into regimes where
 linear viscoelasticity begins to break down. Table~\ref{tab:experiment_design}
summarizes the forcing, geometry, and true parameters for each case.
\begin{table}[!ht]
\centering
\small
\setlength{\tabcolsep}{4pt}
\begin{tabularx}{\textwidth}{@{}l l l l X@{}}
\toprule
Case & Fluid & Forcing $F$, $\theta$ & Geometry ($L, \delta_0/a$) & True parameters \\
\midrule
\emph{Case~III} (\S\ref{subsec:isolated})
 & Newtonian & $12\pi$, $0^{\circ}$ & unbounded & $\eta_{\mathrm{s}}=1$ \\
\emph{Case~III} (\S\ref{subsec:ve_isolated})
 & Oldroyd-B & $8\pi$, $0^{\circ}$ & unbounded
 & $(\eta_{\mathrm{s}},\eta_{\mathrm{p}},\lambda)=(0.5,0.9,0.1)$ \\
\emph{Case~III} (\S\ref{subsec:nonlinear})
 & Oldroyd-B & $1024\pi$, $0^{\circ}$ & unbounded
 & as above \\
\emph{Case~III} (\S\ref{subsec:nonlinear})
 & Oldroyd-B & $\{16,32,64\}\pi$, $0^{\circ}$ & unbounded
 & as above \\
\emph{Case~III} (\S\ref{subsec:nonlinear})
 & Oldroyd-B & $\{256,512,1024\}\pi$, $0^{\circ}$ & unbounded
 & as above \\
\emph{Case~I} (\S\ref{subsec:wall_newtonian})
 & Newtonian & $12\pi$, $\{0^{\circ},45^{\circ}, 90^{\circ}\}$ & $\delta_0/a\in\{0.001,0.01,0.1\}$
 & $\eta_{\mathrm{s}}=1$ \\
\emph{Case~I} (\S\ref{subsec:wall_ve})
 & Oldroyd-B & $8\pi$, $45^{\circ}$ & $\delta_0/a\in\{0.005,0.01,0.1\}$
 & $(\eta_{\mathrm{s}},\eta_{\mathrm{p}},\lambda)=(0.5,0.9,0.1)$ \\
\emph{Case~II} (\S\ref{subsec:particle_particle})
 & Oldroyd-B & $8\pi$, $0^{\circ}$ & periodic ($L$)
 & $(\eta_{\mathrm{s}},\eta_{\mathrm{p}},\lambda)=(0.5,0.9,0.1)$\\
\bottomrule
\end{tabularx}
\caption{Overview of the numerical experiments, listing for each case the
fluid type, the applied force magnitude and angle, the geometry, and the true
material parameters. The probe radius is $a=1$ throughout. Section references
point to where each experiment is presented.}
\label{tab:experiment_design}
\end{table}

\FloatBarrier

\subsection{Isolated particle}
\label{subsec:isolated}
\subsubsection{Newtonian fluid}
We begin with an isolated particle of radius $a=1$ driven by a constant
force $F=12\pi$ at an angle $\theta=0^\circ$ in an unbounded domain.  
The fluid is Newtonian with solvent viscosity $\eta_{\mathrm{s}}=1$. 
Although trivial, this problem verifies our implementation and provides a clean baseline before wall effects are introduced.

\begin{figure}[!ht]
    \centering
    \includegraphics[width=0.93\linewidth]{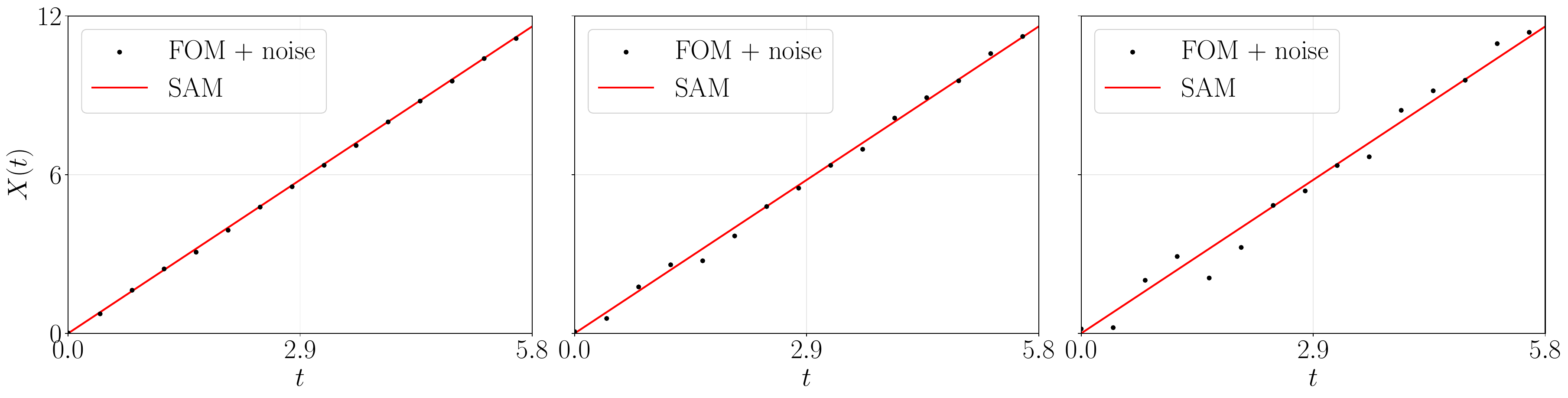}
\caption{Particle displacement for a Newtonian fluid with synthetic Gaussian noise
added to the FOM data. The noise is zero-mean with standard deviation
$\sigma_{\mathrm{exp}}$ chosen as a percentage of the maximum displacement
$\Delta X_{\max}$: $0.5\%$ in the left panel, $2\%$ in the middle
panel, and $5\%$ in the right panel. The red curve denotes the SAM, and the black dots denote the noisy FOM data.}
  \label{fig:sigma_noise_newtonian}
\end{figure}

The particle trajectory is simulated using the FOM until $t=5.8$, and we assume 
that 14 uniformly spaced data points were acquired ($N=14$). 
To mimic experimental data, we add uncorrelated zero-mean Gaussian noise 
to these data points.
To test the sensitivity of the inference to the noise level, we use standard deviations of
$0.5\%$, $2\%$, $5\%$, $10\%$ and $20\%$ of the maximum displacement  $\Delta X_{\max}=\max_t|X(t)|$. The representative trajectories are shown in Figure~\ref{fig:sigma_noise_newtonian} and are compared to the SAM predictions using the ground truth parameter values.
As Figure~\ref{fig:sigma_noise_newtonian}
shows, and as expected for this case, the SAM (red curve) closely matches the FOM
trajectories (black dots), so there is no systematic model discrepancy to
account for. We therefore drop the model-bias term for this case and infer only
$\eta_{\mathrm{s}}$ and the experimental-noise scale $\sigma_{\mathrm{exp}}$.

\begin{table}[!ht]
\centering
\begin{tabular}{ll}
\toprule
Parameter & Prior \\
\midrule
$\hat{\eta}_{\mathrm{s}}$ & $\mathcal{U}(-2,\,2)$ \\
$\sigma_{\mathrm{exp}}$      & $\mathcal{E}(0.9)$ \\
\bottomrule
\end{tabular}
\caption{Prior distributions used for the Newtonian inference cases.}
\label{tab:prior_distributions_newtonian}
\end{table}
The solvent viscosity is assigned a broad, uninformative log-uniform prior, $\hat{\eta}_{\mathrm{s}}=\log_{10}\eta_{\mathrm{s}}\sim\mathcal{U}(-2,2)$, placing
equal weight on each decade of $\eta_{\mathrm{s}}$ over the range $10^{-2}$ to $10^{2}$. 
The exponential prior on $\sigma_{\mathrm{exp}}$ is parameterized by its rate
$\beta$, so that its mean is $1/\beta$. We set this mean of the exponential prior to $10\%$ of the
maximum displacement, giving $\beta = 0.9$ for this case
(Table~\ref{tab:prior_distributions_newtonian}). 
 
The posterior is summarized in Table~\ref{tab:noise_newtonian}, which shows that the
inferred viscosity is robust: even at $20\%$ noise, the posterior mean of
$\eta_{\mathrm{s}}$ stays within about $4\%$ of the true value. 
The inferred experimental-noise parameter, $\mathrm{mean}(\sigma_{\subexp})$, increases nearly linearly with the imposed noise level, while its posterior uncertainty grows correspondingly.
This baseline confirms that, in the absence of model discrepancy, the inference procedure attributes the added noise to $\sigma_{\subexp}$ rather than biasing the viscosity estimate.

\begin{table}[!ht] \centering \begin{tabular}{c cc cc} \toprule Noise & $\text{mean}({\eta}_{\subs})$ & $\stdof{{\eta}_{\subs}}$ & $\text{mean}({\sigma}_{\subexp})$ & $\stdof{{\sigma}_{\subexp}}$ \\ \midrule $0.5\%$ & $1.006$ & $2.24\times 10^{-3}$ & $5.46\times 10^{-2}$ & $1.14\times 10^{-2}$ \\ $2\%$ & $1.002$ & $8.75\times 10^{-3}$ & $2.19\times 10^{-1}$ & $4.67\times 10^{-2}$ \\ $5\%$ & $9.95\times 10^{-1}$ & $2.07\times 10^{-2}$ & $5.35\times 10^{-1}$ & $1.09\times 10^{-1}$ \\ $10\%$ & $9.83\times 10^{-1}$ & $4.24\times 10^{-2}$ & $1.07$ & $2.11\times 10^{-1}$ \\ $20\%$ & $9.66\times 10^{-1}$ & $7.95\times 10^{-2}$ & $2.07$ & $3.84\times 10^{-1}$ \\ \bottomrule \end{tabular} \caption{Results for varying prescribed noise percentage on inference for the Newtonian unbounded model. } \label{tab:noise_newtonian} \end{table}

\FloatBarrier

\subsubsection{Viscoelastic fluid}
\label{subsec:ve_isolated}
Having established that the unbounded Newtonian case yields a well-identified
posterior for a single parameter, we now consider a
viscoelastic fluid. In contrast to the linear Newtonian displacement, a viscoelastic
medium exhibits memory, so the response contains both viscous and elastic
contributions through additional material parameters (\eg, relaxation time). 

\begin{figure}[!ht]
    \centering
    \includegraphics[width=0.95\linewidth]{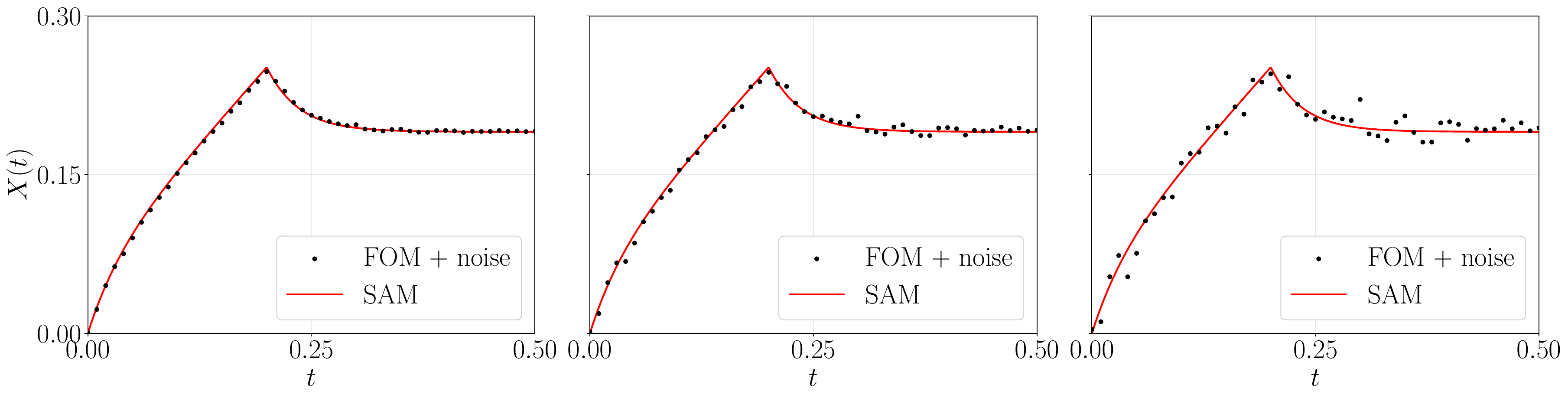}
\caption{Viscoelastic displacement response with synthetic Gaussian noise
added to the FOM data. The noise is zero-mean with standard deviation
$\sigma_{\mathrm{exp}}$ chosen as a percentage of the maximum displacement
$\Delta X_{\max}$: $0.5\%$ in the left panel, $2\%$ in the middle
panel, and $5\%$ in the right panel. The red curve denotes the SAM, and the black dots denote the noisy FOM data.}
  \label{fig:sigma_noise_viscoelastic}
\end{figure}

We consider the displacement of the particle with radius $a = 1$  driven by a constant force $F = 8\pi$ at
$\theta = 0^{\circ}$ over $0 \le t \le t_{0} = 0.2$. The force is then removed, and the simulation continues to $t = 0.5$ to capture the full creep-recovery response. We consider 51 uniformly spaced data points ($N=51$) for the inference.  The fluid is Oldroyd-B with true parameters
$\eta_{\mathrm{s}} = 0.5$, $\eta_{\mathrm{p}} = 0.9$, and
$\lambda = 0.1$. We also estimate the Weissenberg number as $\wi = \lambda\, X(t_0)/(t_0\, a)$,
using the maximum displacement reached at the end of forcing, $t_0 = 0.2$. With
$X(t_0) \approx 0.25$, $\lambda = 0.1$, and $a = 1$, this gives $\wi = 0.125$,
confirming that the experiment lies in the linear viscoelastic regime. As in the Newtonian case, we test the sensitivity of the inference to
measurement noise by adding zero-mean Gaussian noise to the FOM trajectory, with standard deviations of $0.5\%$, $2\%$, $5\%$, $10\%$ and $20\%$ of the maximum
displacement $\Delta X_{\max}$ (Figure~\ref{fig:sigma_noise_viscoelastic}). Similarly, for this case, the SAM using the ground-truth rheological parameters (red curve) closely matches the
FOM trajectories (black dots), so there is no systematic model discrepancy to
account for. We therefore drop the model-bias term for this case and infer the
three material parameters $\eta_{\text{s}}$, $\eta_{\text{p}}$, and $\lambda$ together
with the experimental-noise scale $\sigma_{\subexp}$. The broad log-uniform priors for the physical parameters and the exponential prior for the hyperparameter are listed in
Table~\ref{tab:prior_distributions_viscoelastic}.

The resulting posteriors reported in
Table~\ref{tab:noise_viscoelastic} are accurate up to
about $10\%$ noise, where $(\eta_{\mathrm{s}}, \eta_{\mathrm{p}}, \lambda)$ are still recovered close to their true values with a small spread. At $20\%$,
however, the inference degrades sharply: the inferred parameters drift well away from their true values, and
the estimates are no longer accurate. This contrast reflects the greater complexity of the creep-recovery
response relative to the Newtonian case.

\begin{table}[!ht]
\centering
\begin{tabular}{ll}
\toprule
Parameter & Prior \\
\midrule
$\hat{\eta}_{\mathrm{s}}$ & $\mathcal{U}(-2,\,2)$ \\
$\hat{\eta}_{\mathrm{p}}$ & $\mathcal{U}(-2,\,2)$ \\
$\hat{\lambda}$ & $\mathcal{U}(-2,\,2)$ \\
$\sigma_{\mathrm{exp}}$           & $\mathcal{E}(39)$ \\
\bottomrule
\end{tabular}
\caption{Prior distributions used for the viscoelastic inference cases.}
\label{tab:prior_distributions_viscoelastic}
\end{table}

\begin{table}[!ht] 
\centering \scriptsize 
\resizebox{\textwidth}{!}
{ \begin{tabular}{c cc cc cc cc} \toprule Noise & $\text{mean}({\eta}_{\subs})$ & $\stdof{{\eta}_{\subs}}$ & $\text{mean}({\eta}_{\subp})$ & $\stdof{{\eta}_{\subp}}$ & ${\text{mean}(\lambda})$ & $\stdof{{\lambda}}$ & ${\text{mean}(\sigma}_{\subexp})$ & $\stdof{{\sigma}_{\subexp}}$ \\ \midrule $0.5\%$ & $0.52$ & $5.80\times 10^{-3}$ & $0.87$ & $5.22\times 10^{-3}$ & $0.10$ & $1.23\times 10^{-3}$ & $1.29\times 10^{-3}$ & $1.38\times 10^{-4}$ \\ $2\%$ & $0.53$ & $1.70\times 10^{-2}$ & $0.85$ & $1.54\times 10^{-2}$ & $0.10$ & $3.81\times 10^{-3}$ & $3.90\times 10^{-3}$ & $3.99\times 10^{-4}$ \\ $5\%$ & $0.55$ & $4.43\times 10^{-2}$ & $0.83$ & $4.02 \times 10^{-2}$ & $0.10$ & $9.81 \times 10^{-3}$ & $9.70\times 10^{-3}$ & $1.02\times 10^{-3}$ \\ $10\%$ & $0.58$ & $8.74\times 10^{-2}$ & $0.78$ & $7.92 \times 10^{-2}$ & $0.10$ & $2.26 \times 10^{-2}$ & $1.93\times 10^{-2}$ & $2.10\times 10^{-3}$ \\ $20\%$ & $0.67$ & $2.61\times 10^{-1}$ & $3.25$ & $1.13 \times 10^{1}$ & $2.90$ & $1.16 \times 10^{1}$ & $3.92\times 10^{-2}$ & $3.87\times 10^{-3}$ \\ \bottomrule \end{tabular} } \caption{Results for varying prescribed noise percentage on inference for the unbounded creep-recovery model.} \label{tab:noise_viscoelastic} \end{table}

\begin{figure}[!ht]
    \centering
    \includegraphics[width=0.8\linewidth]{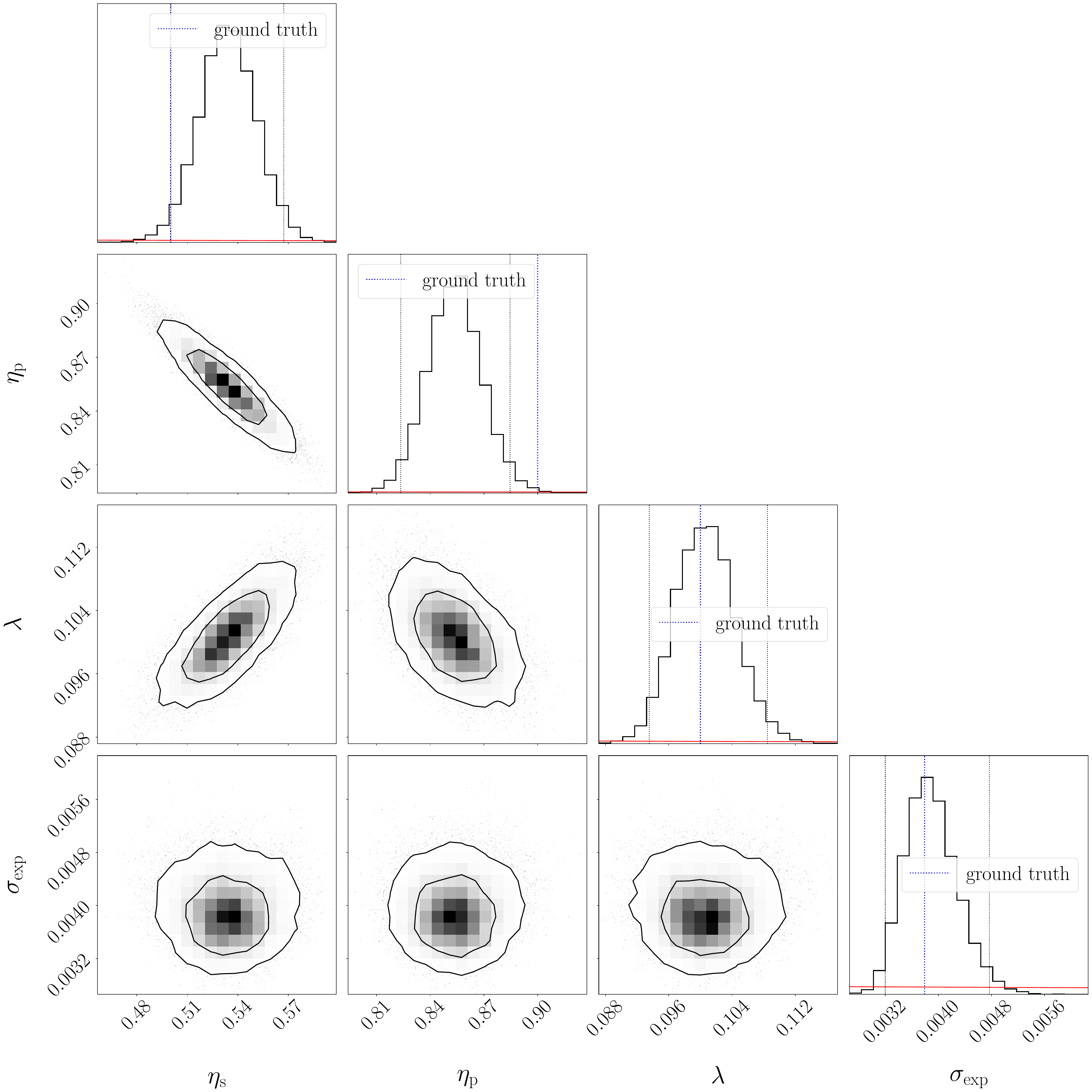}
\caption{Corner plot for the isolated-particle creep-recovery case with true
parameters $\eta_{\mathrm{s}} = 0.5$, $\eta_{\mathrm{p}} = 0.9$,
$\lambda = 0.1$. Diagonal panels show the 1D marginal posteriors, with red curves the priors and blue dotted lines the ground truth. Dashed vertical lines mark the $95\%$ credible interval. Off-diagonal panels show the 2D joint
posteriors, with contours enclosing the $68\%$ and $95\%$ credible regions. The
synthetic noise standard deviation is $2\%$ of the maximum displacement at
$t = t_0 = 0.2$.}
        \label{viscoelastic:case1-posterior}
\end{figure}
In Figure~\ref{viscoelastic:case1-posterior}, we show the corner plot (generated
using \cite{corner}) of the posterior for the case of $2\%$ noise. The 1D
marginals are approximately Gaussian and substantially narrower than the
priors, indicating an informative update from the data. All parameters are strongly
correlated, as expected from the analytical solution
in Equation~\eqref{eq:analytic_creep}, where the slope of the loading phase depends on the sum $\eta_{\mathrm{s}}+\eta_{\mathrm{p}}$, indicating the negative $\eta_{\mathrm{s}}$-$\eta_{\mathrm{p}}$ correlation. However, the magnitude of the recovery phase depends on the product $\lambda\eta_{\mathrm{p}}$, giving the
negative $\eta_{\mathrm{p}}$-$\lambda$ correlation, and combined together, these imply the
positive $\eta_{\mathrm{s}}$-$\lambda$ correlation. The relaxation time $\lambda$ is inferred accurately, with the posterior mean
close to the ground truth. The ground truth for $\eta_{\mathrm{s}}$ and
$\eta_{\mathrm{p}}$ falls slightly outside their credible intervals, which we attribute to their strong negative correlation, and therefore the inference selects a slightly shifted
$(\eta_{\mathrm{s}}, \eta_{\mathrm{p}})$ pair along this degeneracy without
affecting the overall creep response.
\FloatBarrier

\subsubsection{Nonlinear viscoelastic}
\label{subsec:nonlinear}

\begin{figure}[!ht]
    \centering

    \begin{subfigure}{0.32\linewidth}
        \centering
        \includegraphics[width=\linewidth]{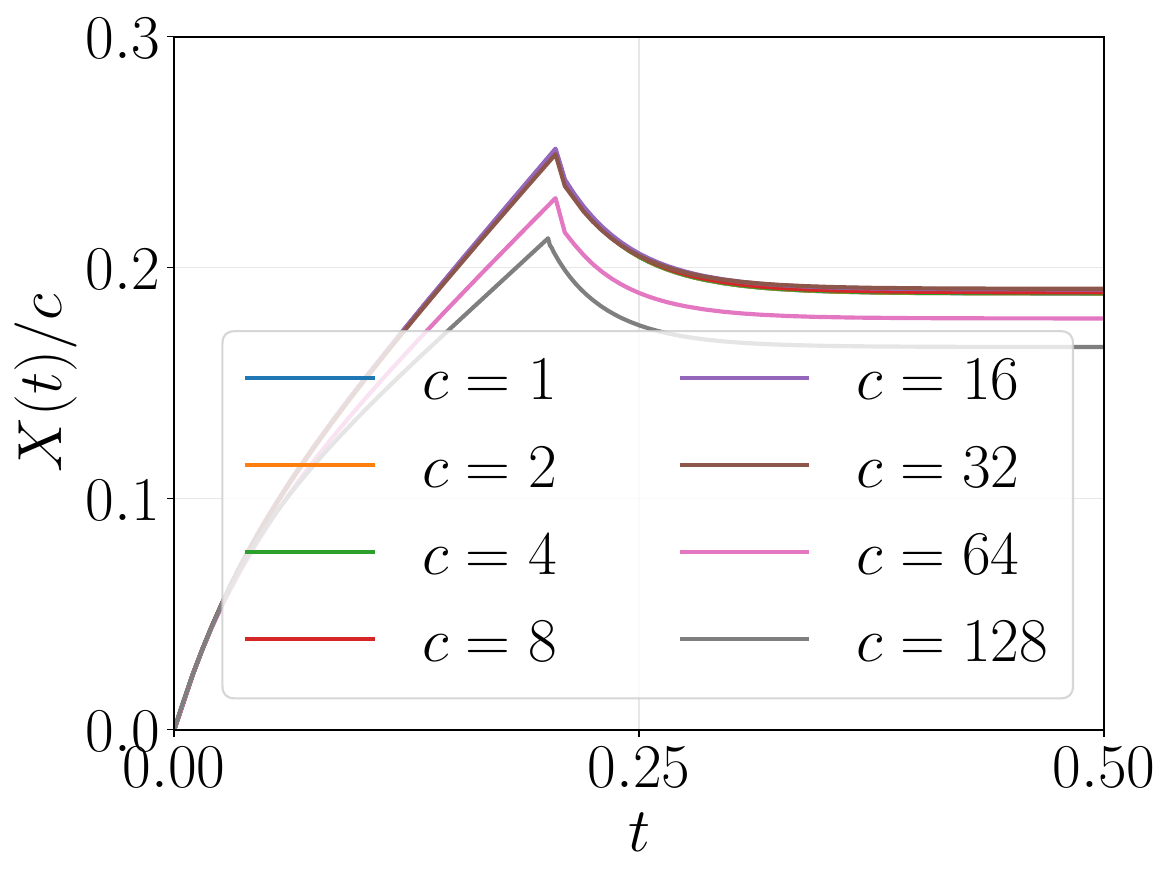}
        \caption{ FOM response.}
        \label{fig:non_linear_analysis}
    \end{subfigure}
    \begin{subfigure}{0.33\linewidth}
        \centering
        \includegraphics[width=\linewidth]
{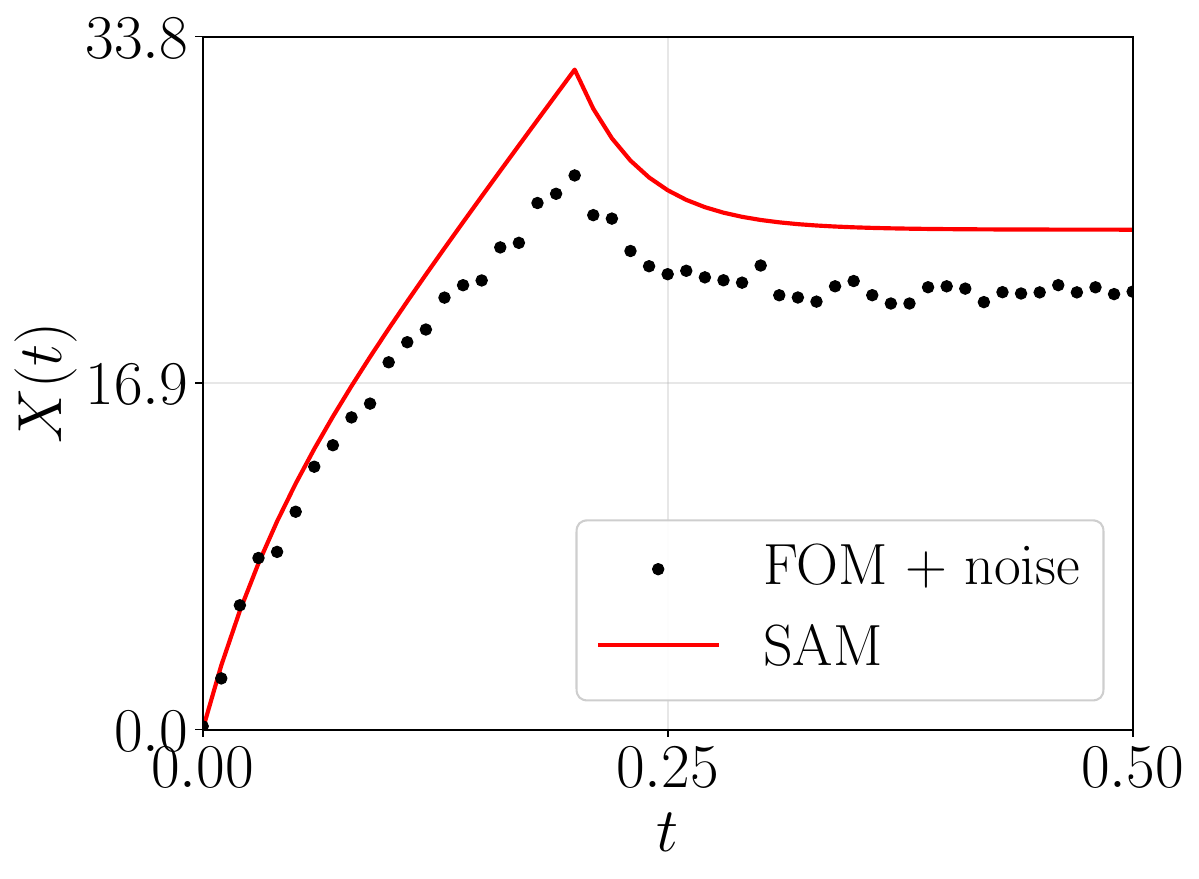}
        \caption{FOM and SAM.}
        \label{fig:non-lin-displacement}
    \end{subfigure}
    \begin{subfigure}{0.32\linewidth}
        \centering
        \includegraphics[width=\linewidth]{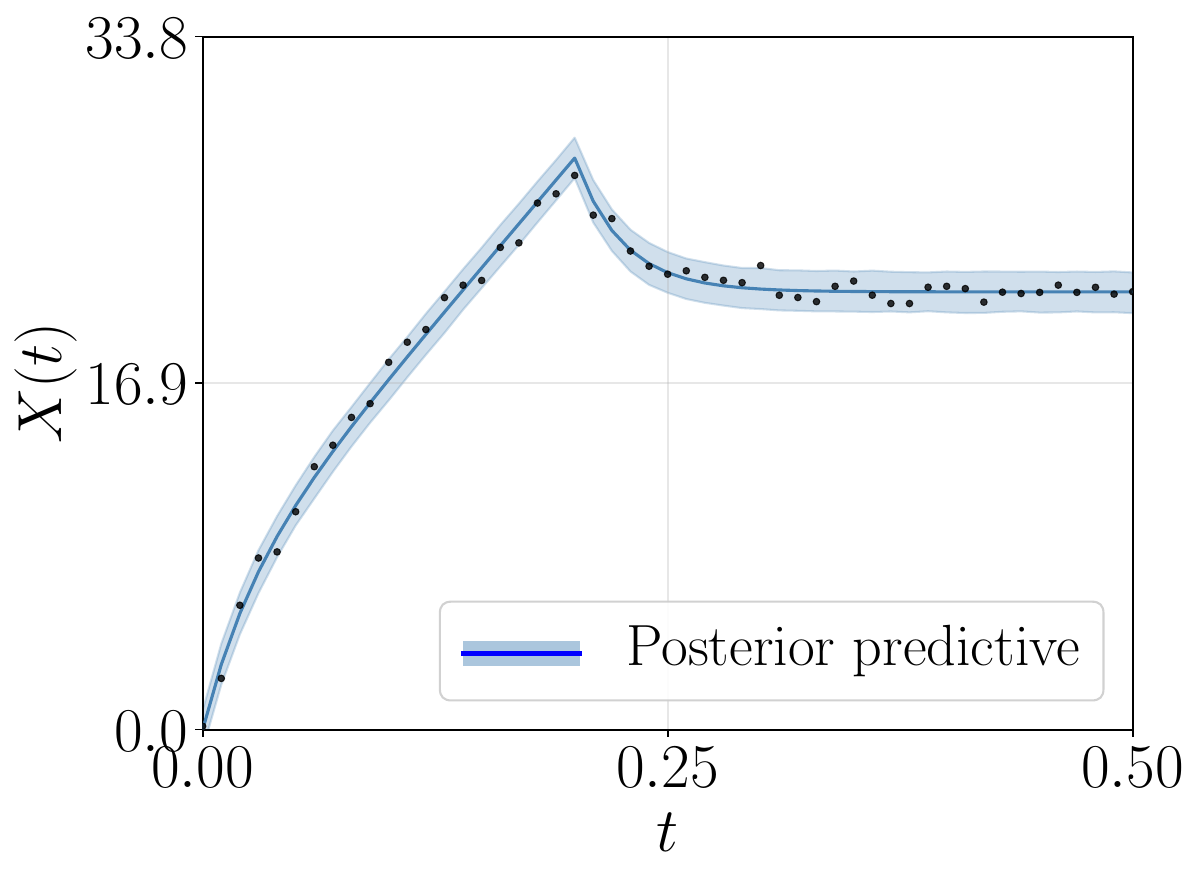}
        \caption{Posterior predictive.}
        \label{fig:nonlinear_pp_nobias}
    \end{subfigure}
\caption{(a) FOM response for different forcing amplitudes, $F = 8\pi c$. (b) High-Weissenberg-number creep test at $F = 1024\pi$, comparing the FOM and SAM using the ground-truth constitutive parameters. (c) Posterior predictive response of the SAM after parameter inference.}
    \label{fig:nonlinear_single_predictive}
\end{figure}

To investigate where the linearity assumptions break down, we vary the force
magnitude on the particle. The FOM data are generated from $t=0$ to $t=0.5$ with the force removed at
$t_0=0.2$, sampled at $N=51$ equally spaced points. The FOM results in
Figure~\ref{fig:non_linear_analysis} use forces that are scaled relative to
the baseline creep case, with $F=8\pi c$ (so that $c=1$ corresponds to
$F=8\pi$) and the displacement scaled by $c$. Up to approximately $c=32$
($F=256\pi$), the scaled trajectories nearly collapse, indicating an approximately linear response. For larger forcing, the collapse is lost, marking the onset of nonlinear viscoelastic effects. We therefore consider a
strongly nonlinear case with $F=1024\pi$ ($c=128$) and $\theta=0^\circ$, for which
$\wi\gg 1$ and the linear creep model is not expected to be
quantitatively accurate. With the Weissenberg number defined as in Section~\ref{subsec:ve_isolated} and with $X(t_0) = 27.22$, $\lambda = 0.1$, and $a = 1$, this gives $\wi = 13.61$, confirming that the experimental data lie in the nonlinear viscoelastic regime.

To mimic experimental data, we
add uncorrelated zero-mean Gaussian noise with a standard deviation of $2\%$ of the
maximum displacement $\Delta X_{\max}$, and we use the (linear) SAM to perform inference on the nonlinear case of $F=1024 \pi$. The priors are similar to the linear viscoelastic case (Table \ref{tab:prior_nonlinear_viscoelastic}), except that the exponential prior is scaled separately for each force level.
\begin{table}[!ht]
\centering
\begin{tabular}{ll}
\toprule
Parameter & Prior \\
\midrule
$\hat{\eta}_{\mathrm{s}}$ & $\mathcal{U}(-2,\,2)$ \\
$\hat{\eta}_{\mathrm{p}}$ & $\mathcal{U}(-2,\,2)$ \\
$\hat{\lambda}$ & $\mathcal{U}(-2,\,2)$ \\
$\sigma_{\mathrm{exp}}$           & $\mathcal{E}\big((0.1\,\Delta X_{\max})^{-1}\big)$ \\
\bottomrule
\end{tabular}
\caption{Prior distributions used for the nonlinear viscoelastic inference cases.}
\label{tab:prior_nonlinear_viscoelastic}
\end{table}
Figure~\ref{fig:non-lin-displacement} compares the FOM trajectory with the SAM
evaluated at the ground-truth parameters in this high-$\wi$ regime, revealing a
clear structural discrepancy. For a single trajectory, however, this error can be
largely hidden by the posterior predictive fit (Figure~\ref{fig:nonlinear_pp_nobias}),
which shows the $95\%$ band is narrow and closely follows the FOM data, giving a misleadingly
confident fit. As Table~\ref{tab:nonlinear_inference_nobias} shows, this agreement
is achieved only by shifting the material parameters away from their true values
$(\eta_{\subs},\eta_{\subp},\lambda) = (0.5,\,0.9,\,0.1)$, most visibly in
$\eta_{\subp}$. The inference selects the wrong parameters to compensate for the
missing nonlinear physics, so the SAM reproduces the observations while
being structurally incorrect.
\begin{table}[!ht]
\centering
\scriptsize
\resizebox{\textwidth}{!}{%
\begin{tabular}{cc cc cc cc}
\toprule
$\text{mean}({\eta}_{\subs})$ & $\stdof{{\eta}_{\subs}}$
& $\text{mean}({\eta}_{\subp})$ & $\stdof{{\eta}_{\subp}}$
& ${\text{mean}(\lambda})$ & $\stdof{{\lambda}}$
& ${\text{mean}(\sigma}_{\subexp})$ & $\stdof{{\sigma}_{\subexp}}$ \\
\midrule
$0.47$ & $2.39\times 10^{-2}$
& $1.12$ & $2.23\times 10^{-2}$
& $0.087$ & $3.20\times 10^{-3}$
& $0.49$ & $5.02\times 10^{-2}$ \\
\bottomrule
\end{tabular}%
}
\caption{Posterior means and standard deviations of the inferred parameters for the nonlinear viscoelastic model with \(2\%\) measurement noise and an applied force of \(F = 1024\pi\) (single trajectory).}
\label{tab:nonlinear_inference_nobias}
\end{table}

The single-trajectory fit is therefore not a reliable diagnostic of
model adequacy. To expose the model bias, we fit several force levels jointly, hypothesizing that a biased parameter set cannot compensate for the
missing nonlinearities at all forces simultaneously. Although the standard deviation of the added noise varies per trajectory, we infer a single value $\sigma_{\subexp}$ for simplicity. We compare a
lower-force set, $F\in\{16\pi,32\pi,64\pi\}$, which stays in the
linear-response regime, with a high-force set,
$F\in\{1024\pi,512\pi,256\pi\}$, which enters the high-$\wi$ regime.

\begin{figure}[!ht]
    \centering
    \begin{subfigure}{0.34\linewidth}
        \centering
        \includegraphics[width=\linewidth]
        {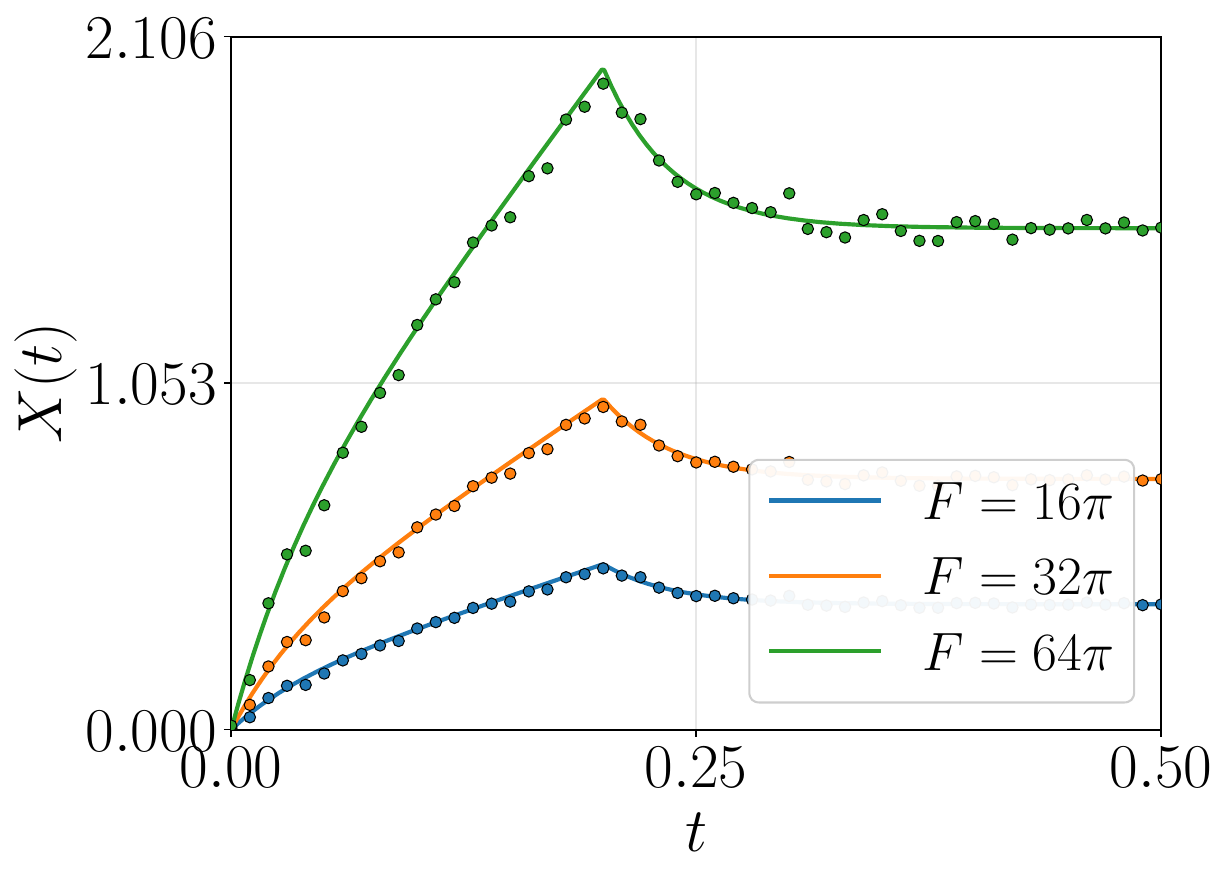}
        \caption{Linear-regime displacement data.}
        \label{fig:linmix-data-noise}
    \end{subfigure}
    \hspace{0.4cm}
    \begin{subfigure}{0.34\linewidth}
        \centering
        \includegraphics[width=\linewidth]
        {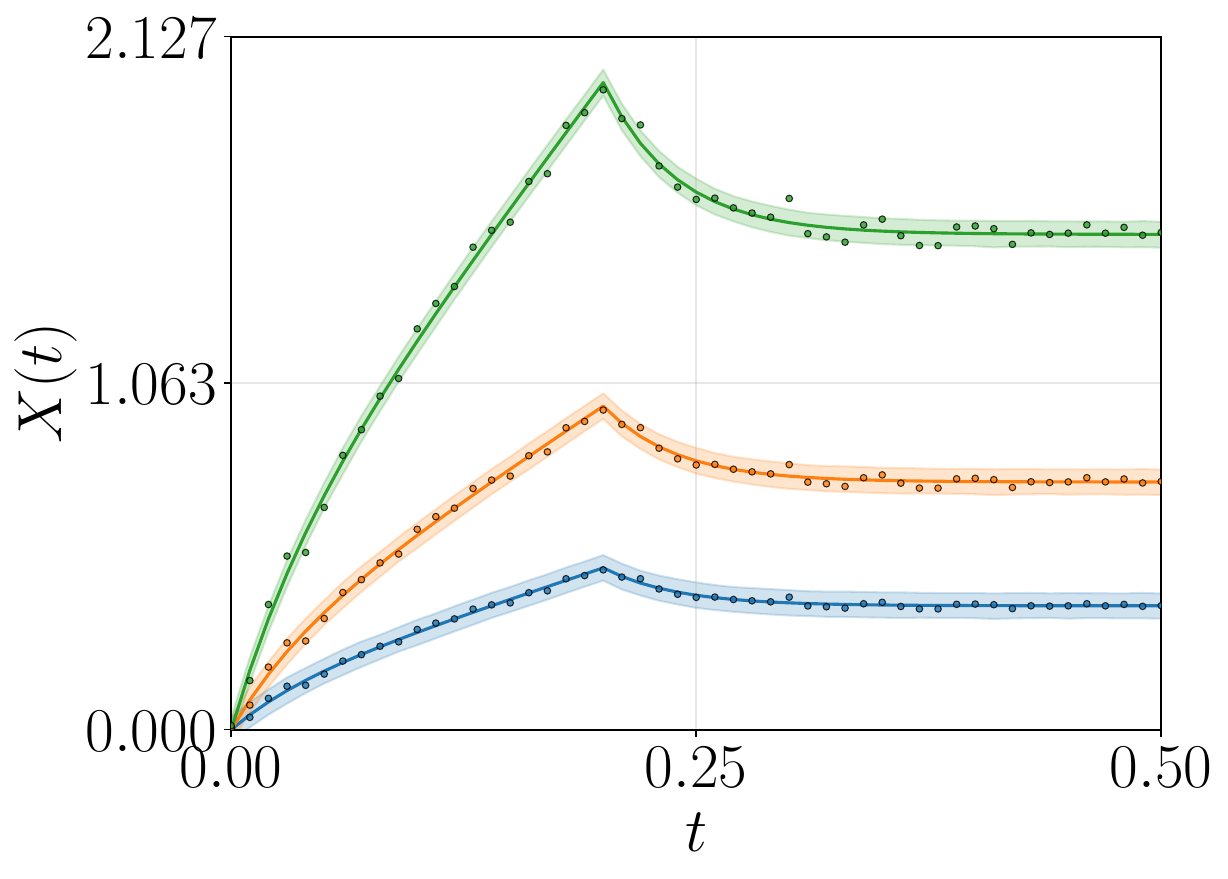}
        \caption{Linear-regime posterior predictive.}
        \label{fig:linmix-posterior-predictive}
    \end{subfigure}
    

    \begin{subfigure}{0.34\linewidth}
        \centering
        \includegraphics[width=\linewidth]
        {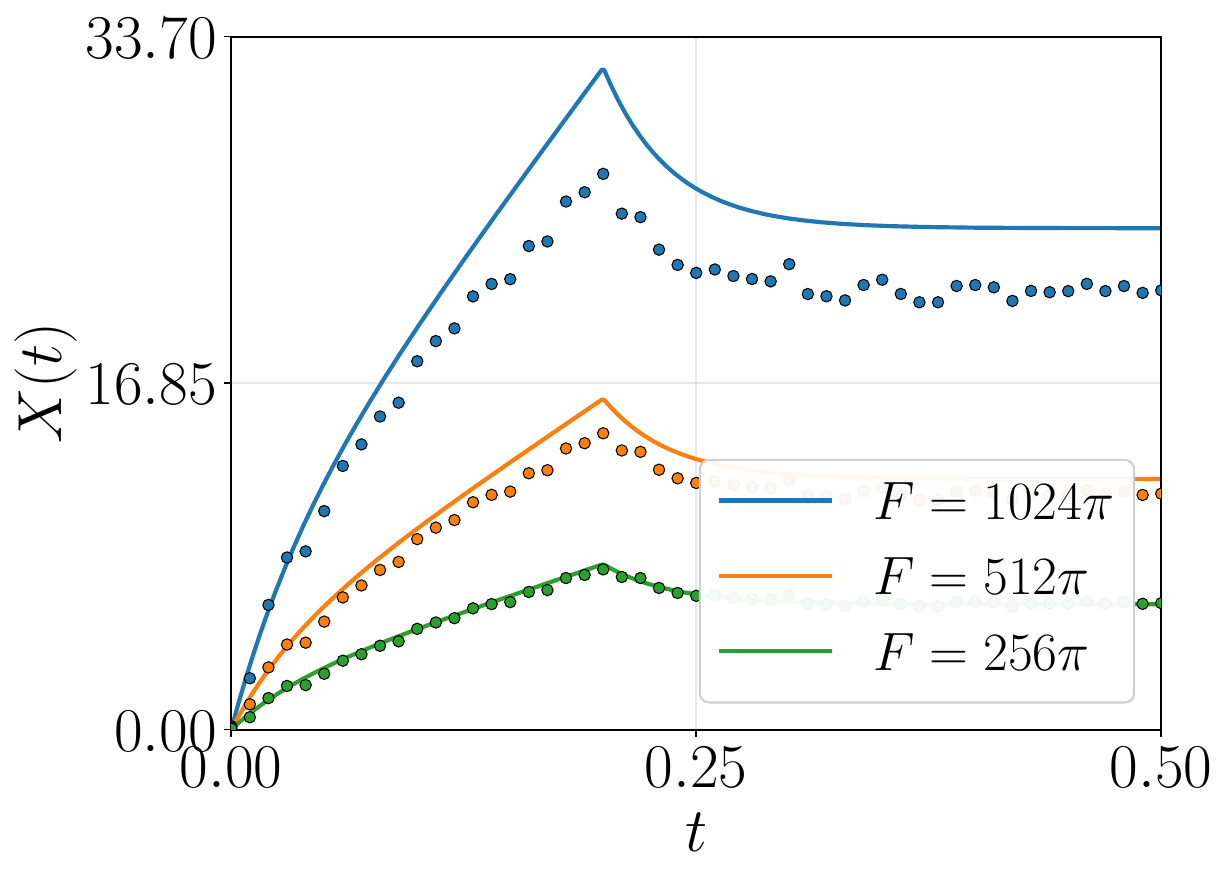}
        \caption{High-$\wi$ displacement data.}
        \label{fig:nonlinmix-data-noise}
    \end{subfigure}
    \hspace{0.4cm}
    \begin{subfigure}{0.34\linewidth}
        \centering
        \includegraphics[width=\linewidth]
        {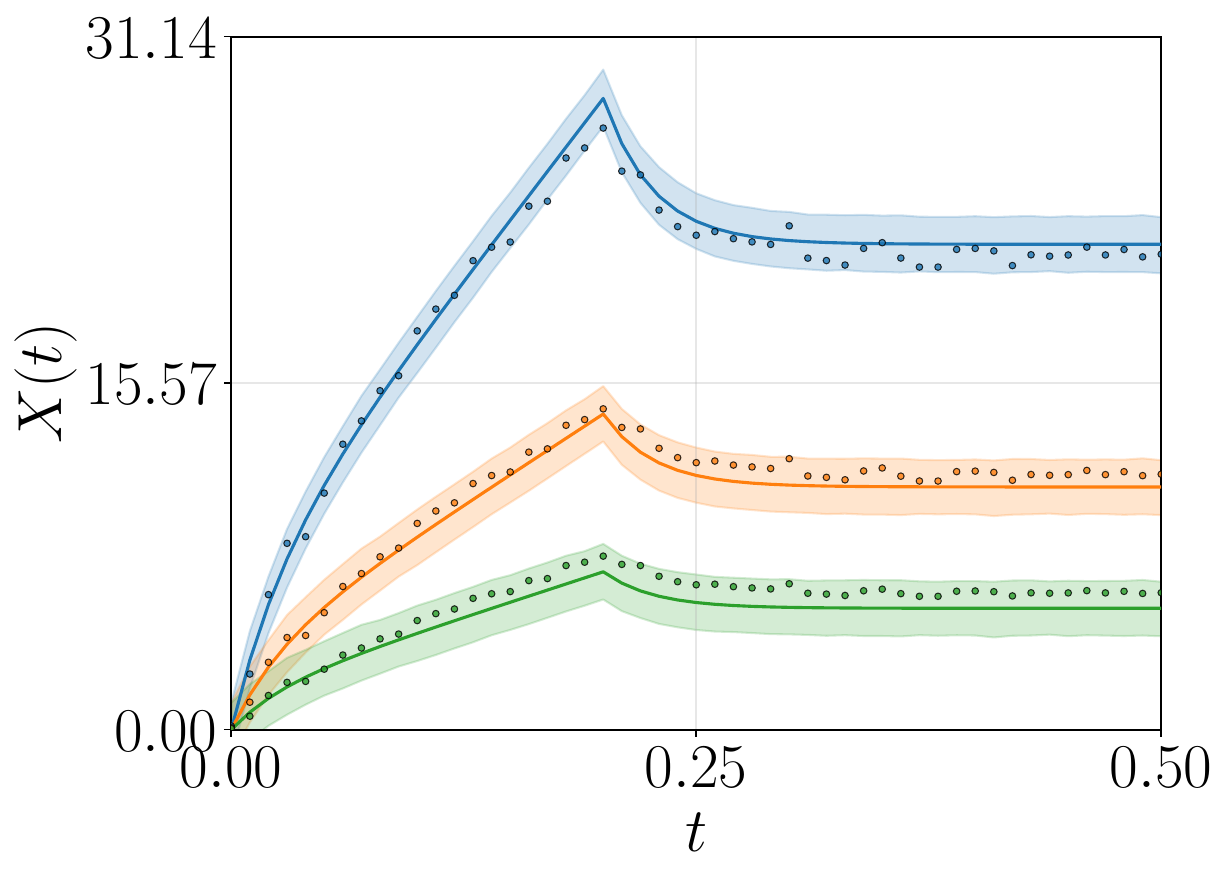}
        \caption{High-$\wi$ posterior predictive.}
        \label{fig:nonlinmix-posterior-predictive}
    \end{subfigure}
    \caption{Joint posterior predictive checks for mixed creep-recovery
    data. The left panels show the displacement curves, and the right
    panels show the posterior predictive bands. The upper panels are the low-force set
    $F \in \{16\pi,\,32\pi,\,64\pi\}$ and the lower panels are the high-force set
    $F \in \{1024\pi,\,512\pi,\,256\pi\}$.}
    \label{fig:predictive_posterior-nonlin1}
\end{figure}

The multi-trajectory inference results are presented in Figure~\ref{fig:predictive_posterior-nonlin1}, where we show the corresponding posterior predictive plots. For the lower-force set, the
SAM describes all three trajectories well. For the high-force set, systematic discrepancies grow with $F$, especially near peak displacement and
during recovery. The statistics are summarized in Table~\ref{tab:nonlin_viscoelastic_multiple}, where it can be observed that
the increase in the inferred discrepancy amplitude \(\sigma_{\mathrm{exp}}\) for the nonlinear dataset reflects the systematic mismatch of the linear viscoelastic SAM at high \(\wi\). Although the inferred parameter values for the joint fit are similar to the fit on a single trajectory, the posterior predictive plot reveals the unrepresented nonlinear physics in the forward model.

\begin{table}[!ht]
\centering
\scriptsize
\resizebox{\textwidth}{!}{%
\begin{tabular}{c cc cc cc cc}
\toprule
Case
& $\text{mean}({\eta}_{\subs})$ & $\stdof{{\eta}_{\subs}}$
& $\text{mean}({\eta}_{\subp})$ & $\stdof{{\eta}_{\subp}}$
& ${\text{mean}(\lambda})$ & $\stdof{{\lambda}}$
& ${\text{mean}(\sigma}_{\subexp})$ & $\stdof{{\sigma}_{\subexp}}$ \\
\midrule
Linear
& $0.54$ & $9.94\times 10^{-3}$
& $0.86$ & $8.96\times 10^{-3}$
& $0.100$ & $2.10 \times 10^{-3}$
& $1.96 \times 10^{-2}$ & $1.14\times 10^{-3}$ \\
\midrule
Nonlinear 
& $0.48$ & $2.63\times 10^{-2}$
& $1.08$ & $2.47\times 10^{-2}$
& $0.087$ & $3.64 \times 10^{-3}$
& $6.30 \times 10^{-1}$ & $3.68\times 10^{-2}$ \\
\bottomrule
\end{tabular}%
}
\caption{Comparison of the posterior means and standard deviations of the inferred viscoelastic parameters obtained from the low-force linear-regime and high-force nonlinear creep-recovery datasets with \(2\%\) measurement noise (multiple trajectories). }
\label{tab:nonlin_viscoelastic_multiple}
\end{table}

\FloatBarrier
\subsection{Wall effects}

\subsubsection{Newtonian fluid}
\label{subsec:wall_newtonian}

We now introduce a wall and first consider wall-parallel motion. In the
wall-corrected Stokes relation, the resistance depends on the product
$\eta_{\mathrm{s}} f_{\parallel}$, so increasing the viscosity
or decreasing the wall distance reduces the particle mobility in the same
way. From a single wall-parallel trajectory, these two effects cannot be
separated: many pairs $(\eta_{\mathrm{s}},\delta_0)$ produce the same
displacement slope. 
Breaking this degeneracy from a purely parallel trajectory would
require external information, namely a tight prior on either $\eta_{\mathrm{s}}$
or $\delta_0$, for example, by rheological measurements or by additional observations of the gap distance. 

Alternatively, this degeneracy is broken when the forcing is oblique. In that case, the particle does not remain at a fixed wall distance: its normal motion changes $\delta$ over time, causing
$f_{\parallel}$ to vary along the trajectory. This coupled parallel-normal motion provides extra information that is not present in a purely wall-parallel trajectory, allowing $\eta_{\mathrm{s}}$ and $\delta_0$ to be uniquely identified.
\begin{table}[!ht]
\centering
\begin{tabular}{ll}
\toprule
Parameter & Prior \\
\midrule
$\hat{\eta}_{\mathrm{s}}$ & $\mathcal{U}(-2,\,2)$ \\
${\hat{\delta}_0}$ & $\mathcal{U}(-4,\,2)$  \\
$\sigma_{\mathrm{exp}}$      & $\mathcal{E}(2)$ \\
\bottomrule
\end{tabular}
\caption{Prior distributions used for the wall-bounded Newtonian inference cases.}
\label{tab:wall_eta_delta}
\end{table} 

\begin{figure}[!ht] \centering \includegraphics[width=0.6\linewidth]{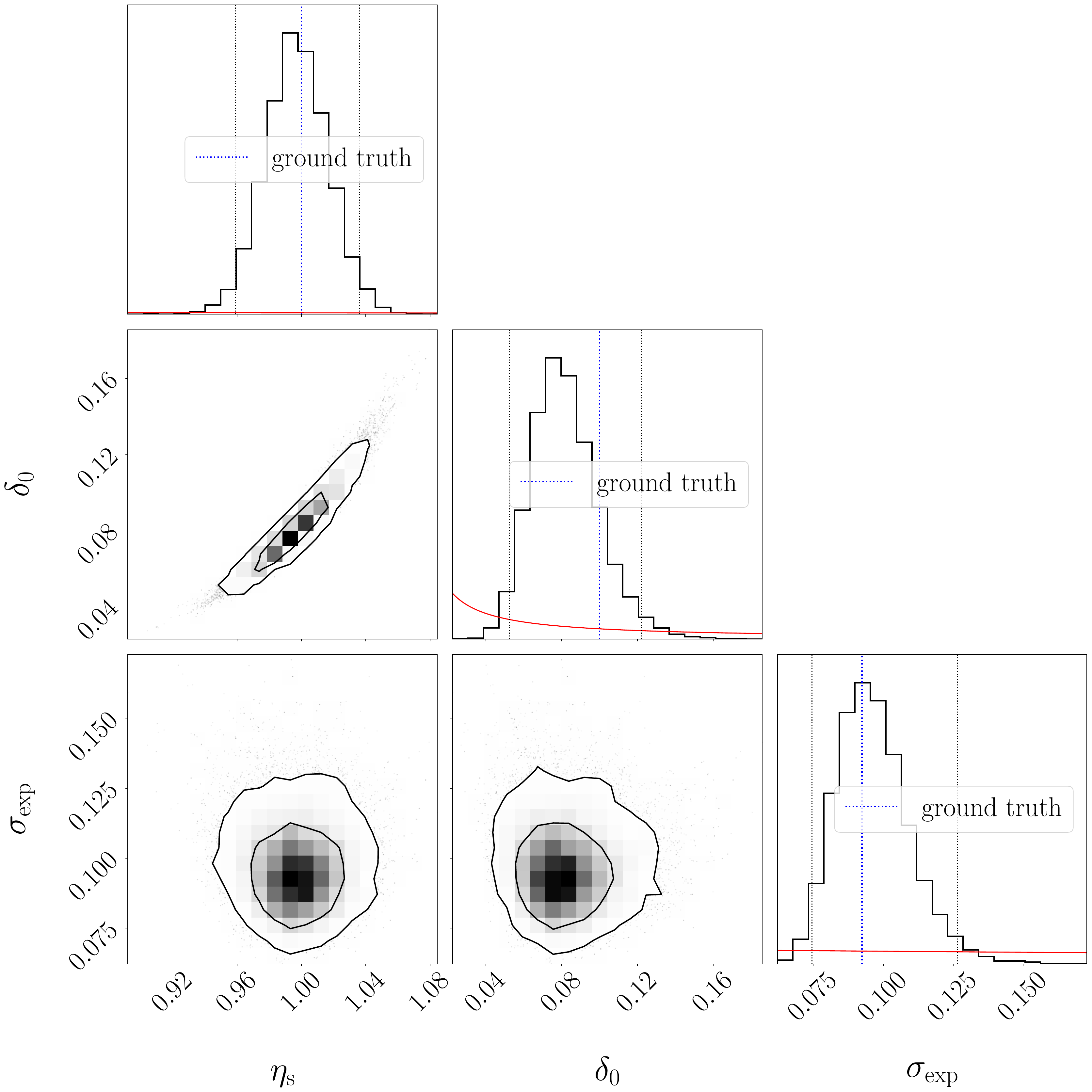}
\caption{Corner plot for the bounded-domain Newtonian case with $\delta_0/a = 0.1$ and $\theta = 45^\circ$. Diagonal panels show marginal posterior distributions, while off-diagonal panels show pairwise joint posterior distributions. Dashed lines indicate 95\% credible intervals, and contours enclose the 68\% and 95\% credible regions.} \label{newtonian:case3-corner-angle-45} \end{figure}

We now examine whether $\eta_{\mathrm{s}}$ and $\delta_0$ can be jointly
identified from a bounded-domain trajectory generated with oblique
forcing, $F=12\pi$ at $\theta=45^\circ$, $\delta_0/a=0.1$ (representative trajectories can be seen in 
Figure~\ref{fig:newtonian_displacement_3x3}), and a synthetic noise standard
deviation of $2\%$ of $\Delta X_{\max}$. We select 30 uniformly spaced points between $t = 0$ and $t = 5.8$. As before, we assign both $\eta_{\mathrm{s}}$ and $\delta_0$ broad log-uniform priors, and set the exponential prior on $\sigma_{\mathrm{exp}}$ so that its mean equals
$10\%$ of the maximum displacement as
listed in Table~\ref{tab:wall_eta_delta}. The inference is performed using only the $x$-displacement. The results of the inference are presented in Figure~\ref{newtonian:case3-corner-angle-45}, which shows the marginal posteriors and
the $\eta_{\mathrm{s}}$-$\delta_0$ correlation. These results clearly indicate that the oblique forcing allows for the identification of $\eta_{\mathrm{s}}$ and $\delta_0$ jointly, even when an uninformative, broad prior is employed. To extend this analysis to other values of $\delta_0$, Table~\ref{delta-newtonian} reports posterior means and standard
deviations for $\eta_{\mathrm{s}}$ and $\delta_0$ at three true gap
distances, with $\eta_{\mathrm{s}}=1$ and $2\%$ noise. The results indicate that at intermediate values of $\delta_0=0.01$, the inference is most accurate. We attribute this to weak nonlinear effects being present for larger $\delta_0$ (the trajectory tends to a straight line for large $\delta_0$) and smaller displacements
at small values of $\delta_0$ (the trajectory tends to zero displacement for small $\delta_0$). See also Figure \ref{fig:newtonian_displacement_3x3}.

\begin{table}[!ht]
\centering
\begin{tabular}{c cc cc cc}
\toprule
True $\delta_0$
& $\text{mean}({\eta}_{\subs})$
& $\stdof{{\eta}_{\subs}}$
& $\text{mean}(\delta_0)$
& $\stdof{\delta_0}$
& $\text{mean}({\sigma}_{\subexp})$
& $\stdof{{\sigma}_{\subexp}}$ \\
\midrule
$0.001$
& $1.03$ & $4.72\times 10^{-2}$
& $1.53 \times 10^{-3}$ & $7.10 \times 10^{-4}$
& $5.70\times 10^{-2}$ & $8.12\times 10^{-3}$ \\
$0.01$
& $1.01$ & $2.70\times 10^{-2}$
& $1.00 \times 10^{-2}$ & $2.73 \times 10^{-3}$
& $7.66\times 10^{-2}$ & $1.09\times 10^{-2}$ \\
$0.1$
& $0.99$ & $1.96\times 10^{-2}$
& $8.21 \times 10^{-2}$ & $1.77 \times 10^{-2}$ 
& $9.65\times 10^{-2}$ & $1.32\times 10^{-2}$ \\
\bottomrule
\end{tabular}
\caption{Posterior mean and standard deviation for ($\eta_{\mathrm{s}}$,
$\delta_0$), and the experimental-noise standard deviation
$\sigma_{\mathrm{exp}}$ at three initial gap distances.}
\label{delta-newtonian}
\end{table}

\begin{figure}[!ht]
\centering
\begin{subfigure}[b]{0.32\linewidth}
  \centering
  \includegraphics[width=\linewidth]{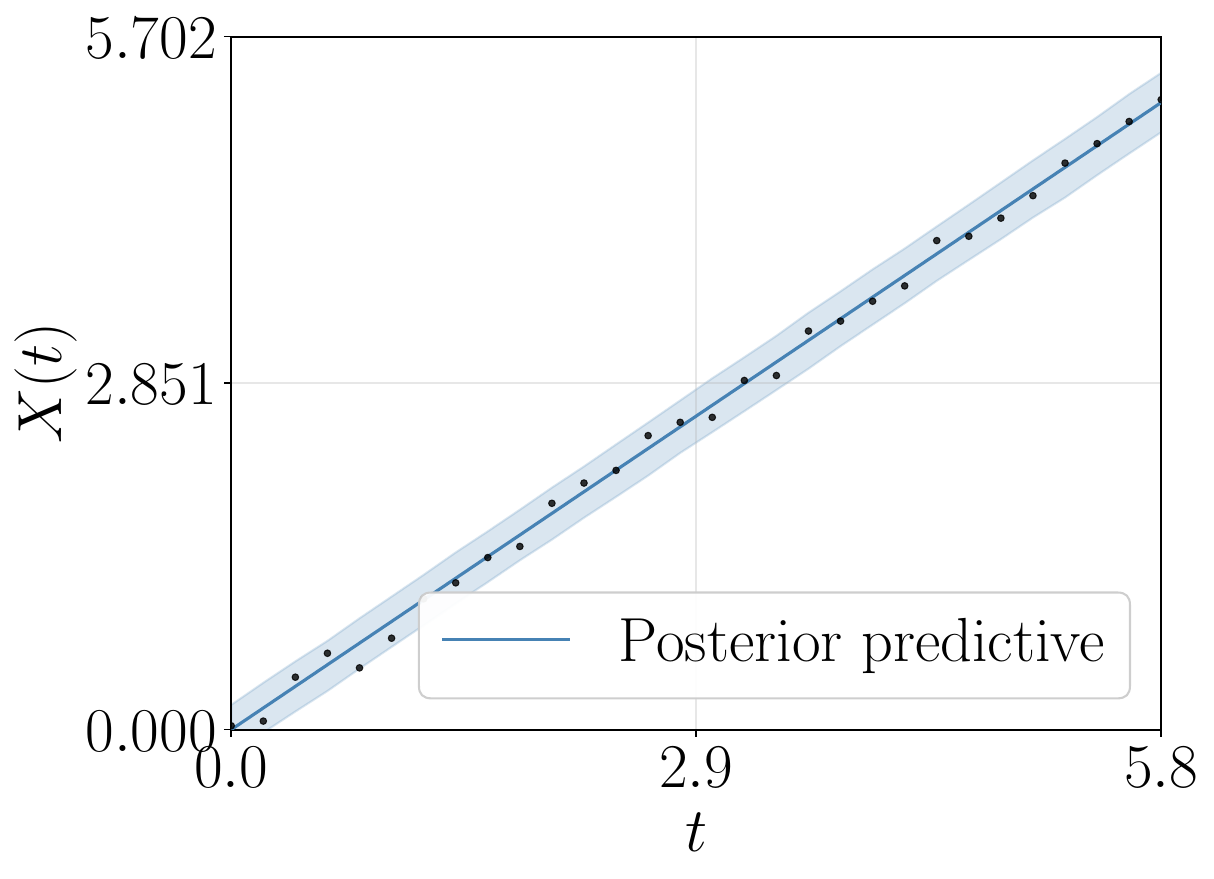}
  \caption{$\delta_0/a = 0.1, \theta = 0^{\circ}$.}
  \label{fig:newtonian:case2-kappa-01-b}
\end{subfigure}
\begin{subfigure}[b]{0.32\linewidth}
  \centering
  \includegraphics[width=\linewidth]{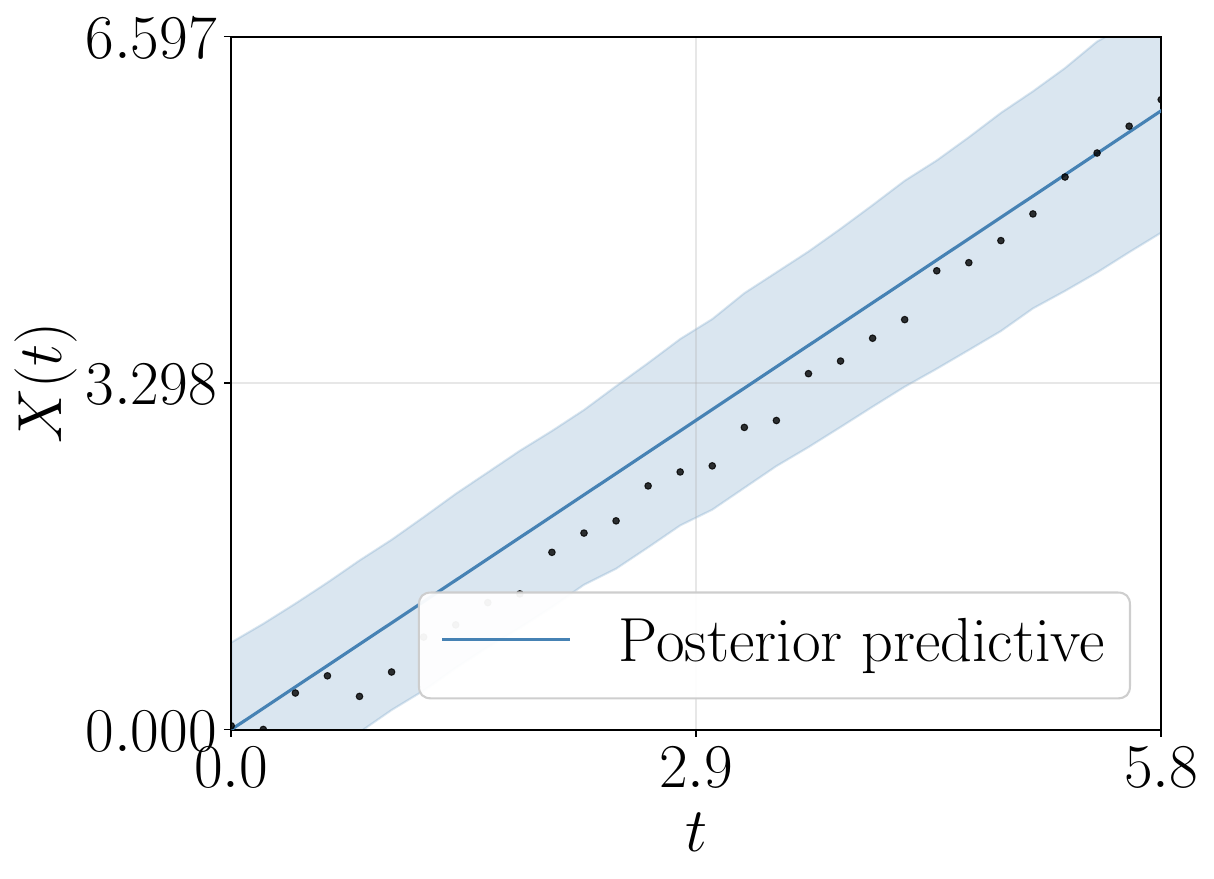}
  \caption{$\delta_0/a = 0.1, \theta = 45^{\circ}$.}

\end{subfigure}
\begin{subfigure}[b]{0.32\linewidth}
  \centering
  \includegraphics[width=\linewidth]{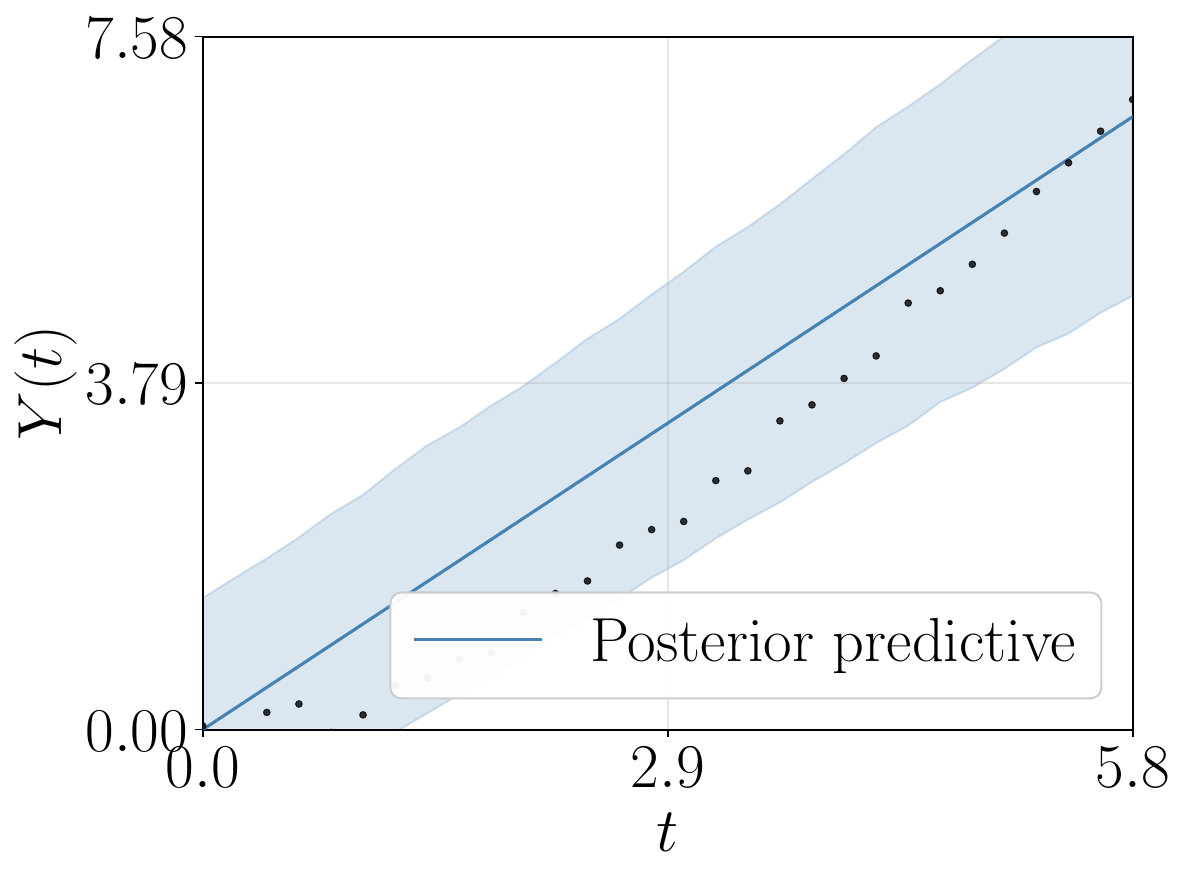}
  \caption{$\delta_0/a = 0.1, \theta = 90^{\circ}$.}
  \label{fig:newtonian:case2-kappa-001b}
\end{subfigure}
\caption{Posterior predictive bands obtained when the unbounded
Newtonian model is fit to bounded-domain FOM data at
$\delta_0/a = 0.1$. The correlation length is fixed at $\ell_{\mathrm{bias}}=6$.}
\label{fig:newtonian_model_error}
\end{figure}

The preceding results use the wall-corrected forward model. To isolate
the effect of model bias, we now use the unbounded SAM for the particle close to a wall under varying forcing directions. Since we anticipate a model bias, we also infer $\sigma_\text{bias}$ while setting $\ell_\text{bias}$ to a fixed value. 
We use the same prior for $\sigma_\text{bias}$ as for $\sigma_{\subexp}$ (Table \ref{tab:wall_eta_delta}).
The posterior predictive fits for $\ell_\text{bias}=6$ are presented in 
Figure~\ref{fig:newtonian_model_error}. For wall-parallel forcing ($\theta=0^\circ$), the
posterior predictive still follows the data closely, but only by
inferring biased parameters that compensate for the omitted wall drag. At
$\theta=45^\circ$, this compensation becomes less effective, and the band
widens; for wall-normal forcing ($\theta=90^\circ$), where the wall
correction is strongest, it widens further, and the mismatch becomes most
apparent. The posterior predictive thus reveals the growing structural
error as the motion becomes more wall-normal.

\begin{figure}[!ht] \centering \includegraphics[width=0.66\linewidth]{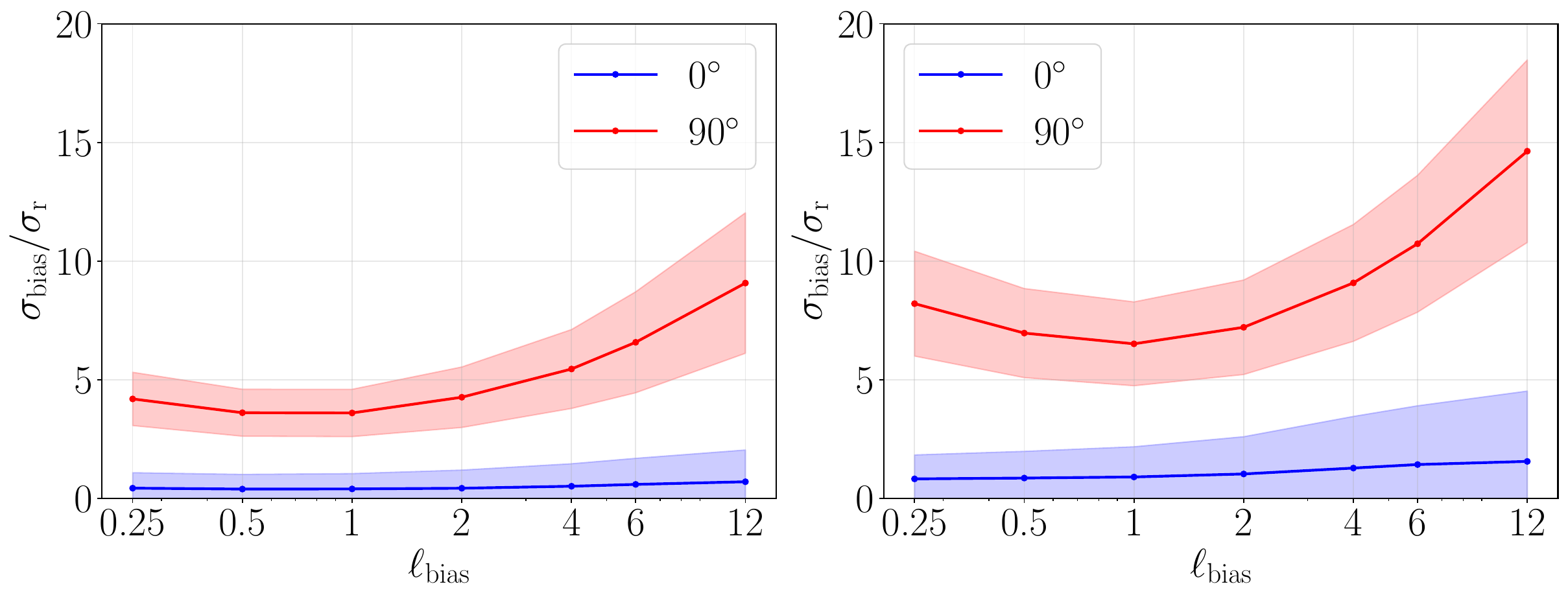} 
\caption{Normalized model bias, $\sigma_{\text{bias}}/\sigma_{\mathrm r}$, as a function of the correlation length $\ell_{\text{bias}}$, with $\sigma_{\mathrm r}$ denoting the realized noise standard deviation. Results are shown for $\delta_0/a=0.1$ (left) and $\delta_0/a=0.001$ (right). The solid lines represent the posterior means, while the shaded regions indicate the approximate 95\% credible intervals.}
\label{fig:l_sensitivity} 
\end{figure}

To study the effect of $\ell_\text{bias}$, 
Figure~\ref{fig:l_sensitivity} shows the normalized model-bias amplitude
$\sigma_{\mathrm{bias}}/\sigma_\text{r}$ as a function of the prescribed correlation
length $\ell_{\mathrm{bias}}$, similar to \cite{rinkens2026bayesian}. Here $\sigma_{\text{r}}$ is the realized standard deviation of the synthetic observational
noise added to the displacement data. We use an unbounded model to fit the bounded-domain data. Hence, $\sigma_{\mathrm{bias}}$ represents the wall-induced error that is not captured by the forward model. The smaller gap,
$\delta_0/a=0.001$, produces stronger wall effects than $\delta_0/a=0.1$ and
therefore requires a larger discrepancy amplitude, as seen by comparing
the two panels of Figure~\ref{fig:l_sensitivity}. This effect
is especially significant for the wall-normal ($90^\circ$) trajectory, for which
the unbounded model shows the strongest wall-resistance contribution. The upward trend for $\sigma_{\mathrm{bias}}/\sigma_{\mathrm{r}}$ with
$\ell_{\mathrm{bias}}$ reflects the known confounding between the
amplitude and correlation length of a Gaussian-process discrepancy,
for which only a single combination of the two is identifiable from
the data~\cite{Zhang2004,fuglstad2019constructing}.

\FloatBarrier

\subsubsection{Viscoelastic fluid}
\label{subsec:wall_ve}
\begin{figure}[!ht]
    \centering
    \begin{subfigure}[t]{0.33\textwidth}
        \centering
        \includegraphics[width=\linewidth]{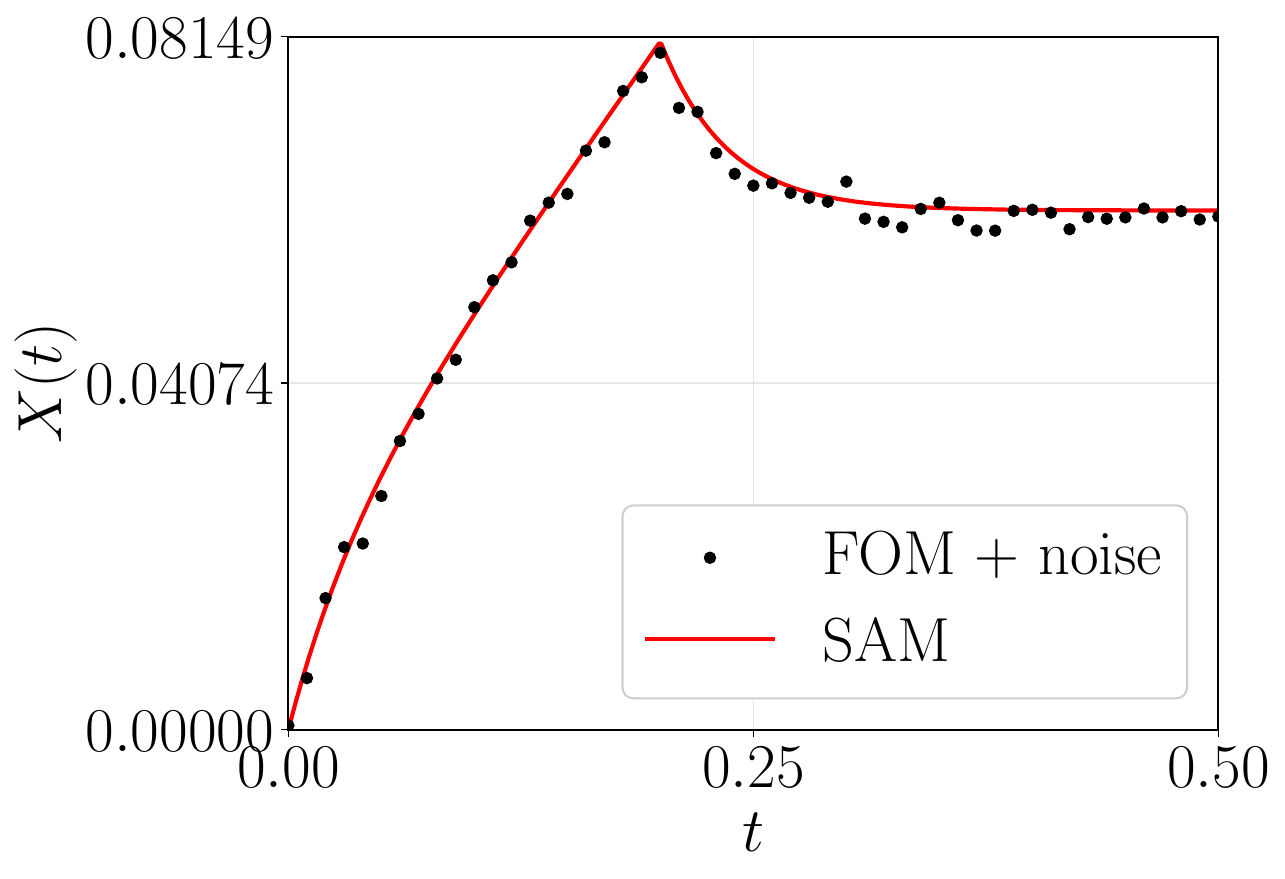}
        \caption{$x$-displacement}
        \label{fig:creep-45-unload-02}
    \end{subfigure}
    \begin{subfigure}[t]{0.33\textwidth}
        \centering
        \includegraphics[width=\linewidth]{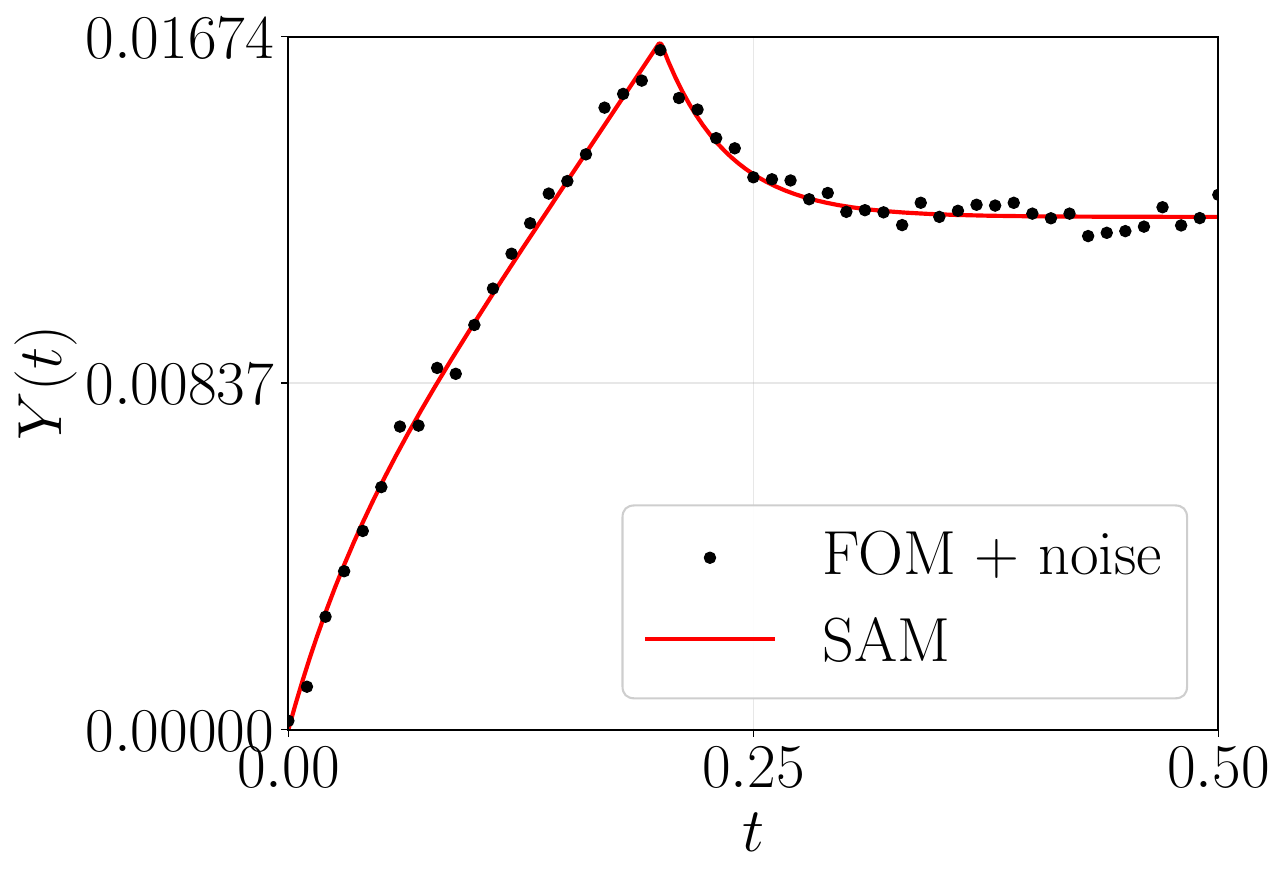}
        \caption{$y$-displacement}
        \label{fig:creep-45-unload-03}
    \end{subfigure}
    \caption{Full-order model (FOM) and Simplified analytical model (SAM)
    creep-recovery responses at $\theta = 45^\circ$: (a) the $x$-displacement and
    (b) the $y$-displacement.}
    \label{fig:creep-45-unload-time}
\end{figure}
\begin{table}[!ht]
\centering
\begin{tabular}{ll}
\toprule
Parameter & Prior \\
\midrule
$\hat{\eta}_{\mathrm{s}}$ & $\mathcal{U}(-2,\,2)$ \\
$\hat{\eta}_{\mathrm{p}}$ & $\mathcal{U}(-2,\,2)$ \\
$\hat{\lambda}$ & $\mathcal{U}(-2,\,2)$ \\
${\hat\delta_0}$ & $\mathcal{U}(-3,\,2)$ \\
$\sigma_{\mathrm{exp}}$           & $\mathcal{E}(125)$ \\
\bottomrule
\end{tabular}
\caption{Prior distributions used for the viscoelastic wall inference cases.}
\label{tab:wall_creep}
\end{table}
In this section, we investigate whether the wall separation $\delta_0$ can also be inferred from the creep-recovery data in a viscoelastic fluid. This bounded viscoelastic problem
combines the two identifiability mechanisms seen above: the transient shape of the
response fixes the material parameters
$(\eta_{\mathrm{s}},\,\eta_{\mathrm{p}},\,\lambda)$, while the oblique motion near
the wall separates the wall-parallel and wall-normal mobilities that carry
information about the geometric parameter $\delta_0$. At $\theta = 45^\circ$ the
particle moves in both $x$ and $y$, so a single experiment probes both
mechanisms. We generate the full-order data at $F = 8\pi$ and $\delta_0/a = 0.1$,
sample $N = 51$ equally spaced points, and add zero-mean Gaussian noise with a
standard deviation of $2\%$ of the maximum displacement. The SAM reproduces the full-order model (using the same parameters) in both components (Figure~\ref{fig:creep-45-unload-time}),
so we omit the model bias term, and we infer
$(\eta_{\mathrm{s}},\,\eta_{\mathrm{p}},\,\lambda,\,\delta_0)$ together with the
noise scale $\sigma_{\mathrm{exp}}$ using the priors in
Table~\ref{tab:wall_creep}.

\begin{figure}[!ht]
    \centering
    \includegraphics[width=0.9\linewidth]{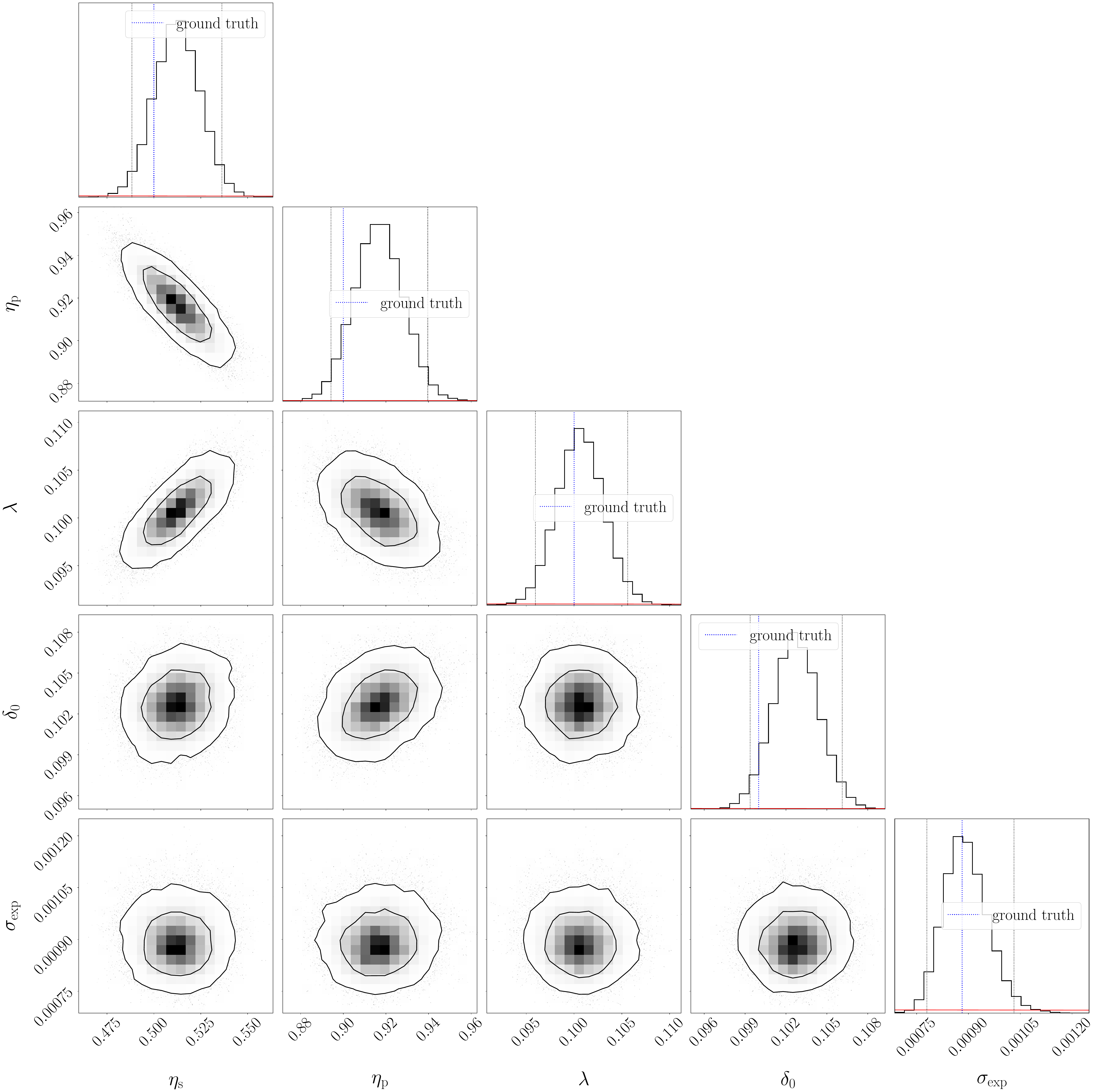}
    \caption{Marginal and joint posterior distributions of
$(\eta_{\mathrm{s}},\,\eta_{\mathrm{p}},\,\lambda,\,\delta_0)$ for the
bounded-domain creep-recovery case at $\delta_0 = 0.1$. Diagonal panels show the 1D marginal posteriors. Dashed vertical lines mark
 the 95\% credible
interval. Off-diagonal panels show the 2D joint posteriors, with contours
enclosing the 68\% and 95\% credible regions.}
    \label{creep:case2-corner-time3}
\end{figure}

Initial numerical experiments indicated that for this more complex problem, the
 $x$-displacement alone does not carry enough information to identify all five
unknowns. We solve this by inferring using the $x$-displacement and 
$y$-displacement simultaneously. Practically, this situation can arise when multiple cameras are used
to determine the location of the particle near a boundary that cannot be identified, \eg because it has similar optical properties to the bulk, or the gap is too small to resolve.
Although the standard deviation of the added noise differs between the $x$ and $y$ components, we infer a single value of $\sigma_{\mathrm{exp}}$ for simplicity.
Figure~\ref{creep:case2-corner-time3} shows the resulting
joint posterior for the physical parameters. The posterior updates from the prior and the ground truth values are contained within the $95\%$ credible interval, so all four physical parameters are identified, and the off-diagonal panels show the same correlations as in the isolated case.

\begin{table}[!ht] \centering \scriptsize \resizebox{\textwidth}{!}
{ \begin{tabular}{c cc cc cc cc cc} \toprule True $\delta_0$ & $\text{mean}({\eta}_{\mathrm{s}})$ & $\stdof{{\eta}_{\mathrm{s}}}$ & $\text{mean}({\eta}_{\mathrm{p}})$ & $\stdof{{\eta}_{\mathrm{p}}}$ & $\text{mean}(\lambda)$ & $\stdof{\lambda}$ & $\text{mean}(\delta_0)$ & $\stdof{\delta_0}$ & $\text{mean}({\sigma}_{\mathrm{exp}})$ & $\stdof{{\sigma}_{\mathrm{exp}}}$ \\ \midrule $0.005$ & $0.50$ & $1.52\times10^{-2}$ & $0.87$ & $1.95\times10^{-2}$ & $0.10$ & $2.57\times10^{-3}$ & $0.0049$ & $6.45\times10^{-4}$ & $5.25\times10^{-4}$ & $3.79\times10^{-5}$ \\  
$0.01$ & $0.50$ & $1.33\times10^{-2}$ & $0.88$ & $1.48\times10^{-2}$ & $0.099$ & $2.49\times10^{-3}$ & $0.0099$ & $7.38\times10^{-4}$ & $5.85\times10^{-4}$ & $3.09\times10^{-5}$ \\  
$0.1$ & $0.51$ & $1.21\times10^{-2}$ & $0.91$ & $1.15\times10^{-2}$ & $0.10$ & $2.44\times10^{-3}$ & $0.100$ & $1.70\times10^{-3}$ & $9.03\times10^{-4}$ & $6.47\times10^{-5}$ \\ \bottomrule \end{tabular} } \caption{Posterior mean and standard deviation for three true gap distances obtained from the MCMC analysis at $45^\circ$.} \label{tab:creep-delta} \end{table}

Having established that including both displacement components identifies all five
unknowns, we verify that this holds across wall separations.
Table~\ref{tab:creep-delta} reports the posterior means and standard deviations
for $(\eta_{\mathrm{s}},\,\eta_{\mathrm{p}},\,\lambda,\,\delta_0,\,\sigma_{\subexp})$ at
$\delta_0 \in \{0.005, 0.01, 0.1\}$. The means stay close to the true values at
every gap, confirming that the protocol recovers both the material parameters and
the wall separation over the full range of separations considered.

\FloatBarrier

\subsection{Particle-particle interactions}
\label{subsec:particle_particle}
In the previous sections, we considered two cases: (a) one where the particle is isolated and (b) one where the particle is close to a wall. However, in a suspension, the particle is not isolated, but it also interacts with its neighboring particles. We model this type of interaction with the particle in a periodic array (see Section~\ref{sec:hasimoto} and ~\ref{sec:ve_pp}) and examine it in two different material configurations.  In the first configuration, we consider a homogeneous material in which the modulus is spatially uniform, and the correction can be verified directly. In the second configuration, we consider a heterogeneous material in which the modulus varies locally in space. Here, we compare the local measurement of the modulus in space against a bulk measurement. 

In both configurations, the particle displacements are sampled at $51$ equally spaced points over
$t \in [0, 0.5]$. Likewise, we add zero-mean Gaussian noise with a standard deviation of $2\%$ of the maximum displacement. The priors are listed in
Table~\ref{tab:prior_particle_particle}.
\begin{table}[!ht]
\centering
\begin{tabular}{ll}
\toprule
Parameter & Prior \\
\midrule
$\hat{\eta}_{\mathrm{s}}$ & $\mathcal{U}(-2,\,2)$ \\
$\hat{\eta}_{\mathrm{p}}$ & $\mathcal{U}(-2,\,2)$ \\
$\hat{\lambda}$ & $\mathcal{U}(-2,\,2)$ \\
$\sigma_{\mathrm{exp}}$           & $\mathcal{E}(39)$ \\
\bottomrule
\end{tabular}
\caption{Prior distributions used for the particle-particle interactions.}
\label{tab:prior_particle_particle}
\end{table}

\subsubsection{Homogeneous material} 
\label{sec:pp_homogeneous}

In the homogeneous material, we consider the parameter set $(\eta_{\mathrm{s}},\eta_{\mathrm{p}},\lambda)=(0.5,0.9,0.1)$, which implies that 
the modulus is fixed as $G = 9$. Because we consider a spatially heterogeneous modulus in the next section, we focus on the inference of $G$.
When the Hasimoto correction is omitted, the finite size of the periodic box biases the
inferred modulus. Table~\ref{tab:hasimoto_homogeneous} shows that the inferred $G$ is strongly
overestimated for small domains when using the unbounded SAM: the inference returns $G = 19.65$ at $L = 5$, more than
twice the true value. The uncorrected error decays only slowly as the box grows, remaining
about $7\%$ at $L = 60$, because the periodic images interact through long-range hydrodynamic interactions. Table~\ref{tab:hasimoto_homogeneous} also shows that applying the Hasimoto mobility factor $Q(L)$
removes this contribution: the corrected modulus stays within roughly $2\%$ of $G_{\mathrm{true}}$ across every box size, showing no systematic dependence on $L$.
Even at $L = 5$, where the uncorrected estimate is off by more than $100\%$, the corrected
value $G = 9.16$ recovers the true modulus to about $2\%$. We apply the same
correction to the spatially varying material considered next.

\begin{table}[!ht]
\centering
\begin{tabular}{ccccccc}
\toprule
$L$  & $\text{mean}(G_\text{uc})$  & $\stdof{G_\text{uc}}$  & Error (\%)  & $\text{mean}(G_{\mathrm{H}})$  & $\stdof{G_{\mathrm{H}}}$  & Error (\%) \\
\midrule
5  & 19.65 & 0.956 & 118.3  & 9.16 & 0.438 & 1.78 \\
10 & 12.68 & 0.609 & 40.9 & 9.15 & 0.443 & 1.67 \\
20 & 10.65 & 0.519 & 18.3 & 9.14 & 0.434 & 1.56 \\
40 & 9.85  & 0.467 & 9.4  & 9.15 & 0.442 & 1.67 \\
60 & 9.61  & 0.475 & 6.8  & 9.14 & 0.451 & 1.56 \\
\bottomrule
\end{tabular}
\caption{Effect of domain size on the inferred modulus in a homogeneous material with
$G_{\mathrm{true}} = 9$. The uncorrected columns show the inference obtained without
applying the Hasimoto correction, while the corrected columns report the modulus after
multiplying by the Hasimoto mobility factor $Q(L)$. The errors are relative deviations
from $G_{\mathrm{true}}$.}
\label{tab:hasimoto_homogeneous}
\end{table}

\FloatBarrier

\subsubsection{Heterogeneous material}
\label{sec:heterogeneous_inference}

\begin{figure}[!ht]
    \centering
    \begin{subfigure}{0.33\linewidth}
        \centering
        \includegraphics[width=\linewidth]{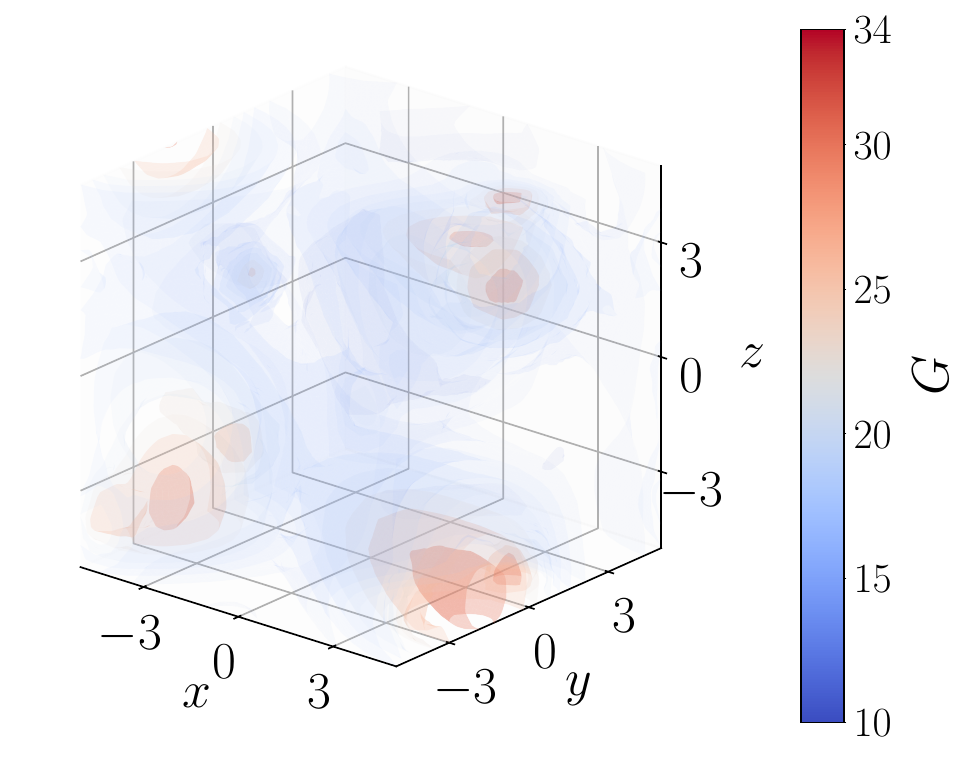}
        \caption{Three-dimensional iso-surfaces of the heterogeneous
                 modulus $G$.}
        \label{fig:G_field}
    \end{subfigure}
    \begin{subfigure}{\linewidth}
        \centering
        \includegraphics[width=0.7\linewidth]{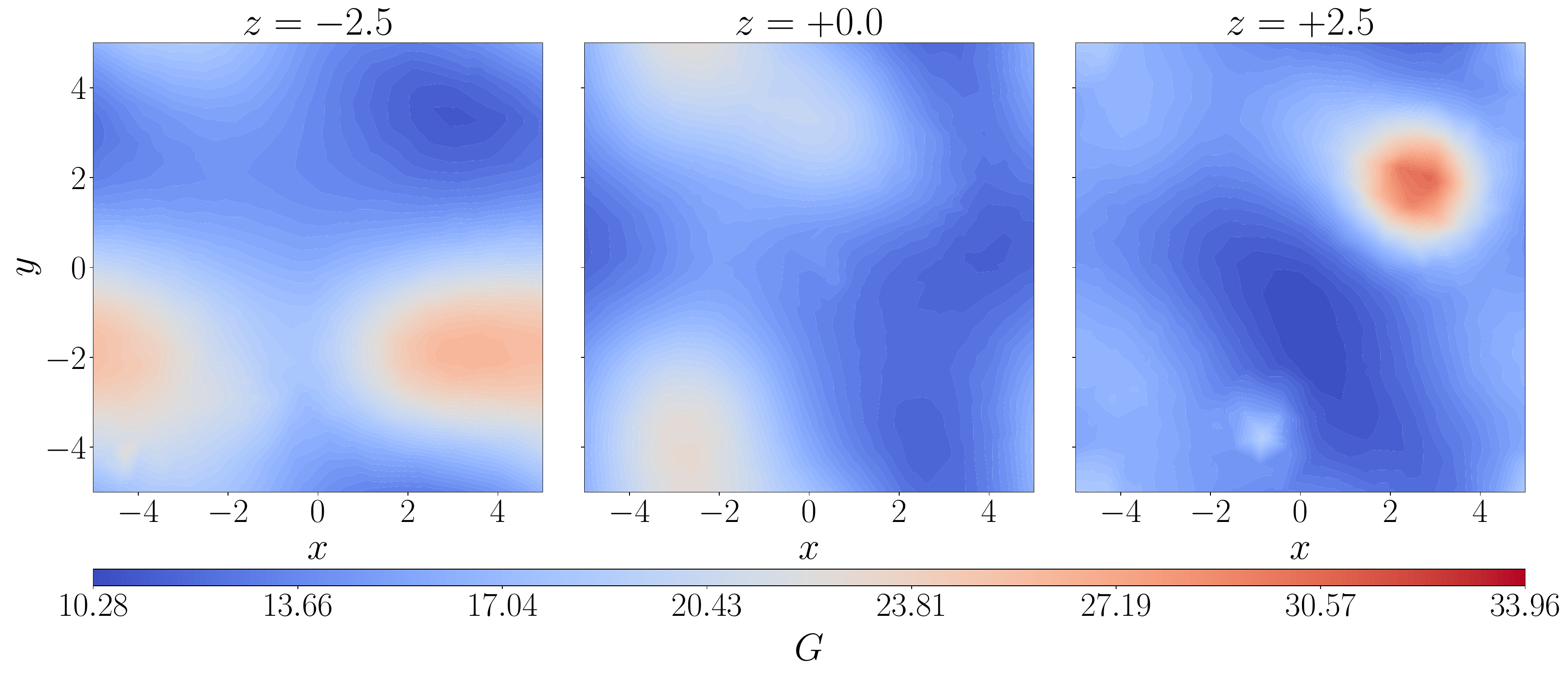}
        \caption{Planar slices of $G$ at $z = -2.5$, $z = 0$, and
                 $z = 2.5$.}
        \label{fig:g_slices}
    \end{subfigure}
    \caption{Visualization of the heterogeneous modulus field $G$. The
             iso-surfaces highlight the three-dimensional heterogeneity,
             while the slices show the spatial variation through the
             domain.}
    \label{fig:g_heterogeneous_medium}
\end{figure}

We now turn to a material in which the modulus varies in space,
$G = G(\vec{x})$. The construction of the heterogeneous microstructure in the
full-order model is done by introducing Gaussian "blobs" which locally increase $G$ and which are convected with the fluid flow (a full description is given in \ref{sec:blobs}). The microstructure is visualized in Figure~\ref{fig:g_heterogeneous_medium}.

Since the particle size is of the same order as the Gaussian blobs, the inferred modulus depends strongly on the particle location within the microstructure, and the particle
"feels" an averaged $G$ of the surrounding material. We
investigate this effect by computing the distribution of mean posterior values of the inferred modulus from many creep-recovery experiments with different particle locations.
The posterior of the $k$-th experiment is denoted by $G^{(k)}_{\mathrm{loc}}$, with mean 
$\bar{G}^{(k)}_{\mathrm{loc}}$.
In the creep-recovery inference, we assume that the probe interacts with its own periodic images as in the
homogeneous case, so the same finite-size correction of
Section~\ref{sec:hasimoto} removes the periodic-box artifact before the
particle response is converted into an apparent modulus. The obtained distribution of mean elastic modulus values is compared to an independent effective shear modulus $G_{\mathrm{eff}}$ from macroscopic shear (see \ref{sec:blobs}).

To sample the field without regenerating the particle mesh, the particle
is held fixed at the box center, and the entire periodic field is
translated rigidly by a random offset $\vec{s} = (s_x, s_y, s_z)$, with
each component drawn uniformly on $[0, L)$ and wrapped periodically. This
is equivalent to evaluating $G(\vec{x} + \vec{s})$ at the fixed particle
location. A new offset is drawn for each of $N_{\mathrm{run}} = 120$
independent runs, giving $120$ independent, Hasimoto-corrected local
posteriors $G_{\mathrm{loc}}^{(k)}$ of the shear modulus. 

\begin{figure}[!ht]
    \centering
    \includegraphics[width=0.33\linewidth]{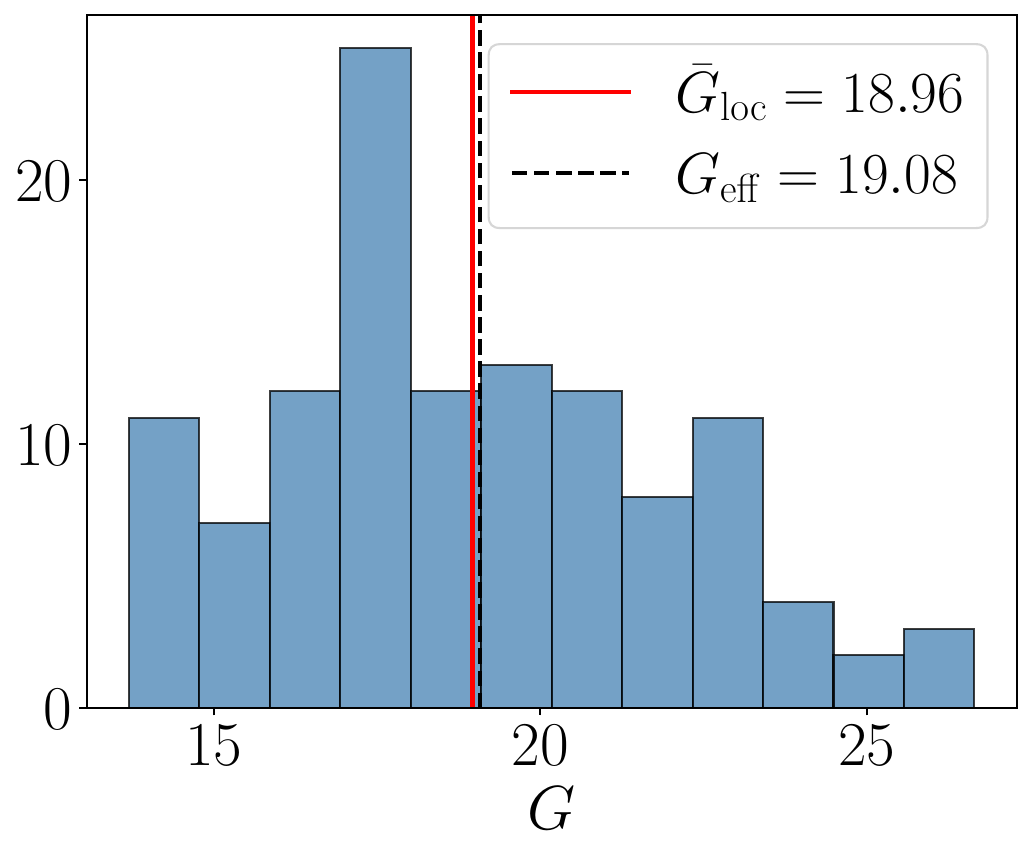}
    \caption{Histogram of the $120$ Hasimoto-corrected local estimates
    $\bar{G}^{(k)}_{\mathrm{loc}}$ (posterior means), with mean $18.96$ and
    standard deviation $3.02$ over the runs, shown with the effective modulus
    $G_{\mathrm{eff}} = 19.08$.}
    \label{fig:G_histogram}
\end{figure}

Figure~\ref{fig:G_histogram} shows the distribution of the local 
$\bar{G}^{(k)}_{\mathrm{loc}}$. 
The distribution ranges from approximately $14$ to $27$, with a standard deviation of
$3.0$, about five times the typical posterior standard deviation of an individual
estimate (approximately $0.6$), and thus reflects the heterogeneity of the material.
The distribution is, however, narrower than the range of
the modulus field itself ($10$ to $34$, Figure~\ref{fig:G_field}), because the
probe, being comparable in size to the blobs, responds to a weighted average of the
surrounding modulus. The mean of the local estimates, $\bar{G}_{\mathrm{loc}} = 18.96$, agrees with the
effective modulus of the same field, $G_{\mathrm{eff}} = 19.08$, as expected 
for uniformly sampled probe locations.

\FloatBarrier

\section{Conclusions}
\label{sec:conclusions}
Active microrheology becomes quantitatively reliable only when systematic effects of
the measurement environment are built into the forward model and the remaining
uncertainty is propagated into the inferred parameters. In this work, we have combined
simplified analytical models for Newtonian and linear viscoelastic fluids, including
wall and particle-particle hydrodynamic interactions, with Bayesian inference, and
assessed this approach on synthetic data from finite-element simulations.

Neglecting hydrodynamic interactions biases the inferred parameters, whereas including
them as mobility corrections removes this bias. Near a wall, the effect of viscosity
generally cannot be disentangled from the increased drag due to the wall when the probe
moves parallel to it. We show that oblique forcing breaks this degeneracy between
viscosity and wall distance, although the accuracy depends on how much the gap, and
hence the mobility, changes along the trajectory relative to the noise level. For
viscoelastic fluids, using both displacement components makes the data sufficiently
informative to identify all material parameters and the wall distance jointly from a
single creep-recovery experiment. For neighboring probes, the Hasimoto correction
recovers both the Newtonian and the viscoelastic creep-recovery response without
adjusting the constitutive parameters.

A good fit to a single trajectory is not a reliable indicator of model adequacy. In the
nonlinear viscoelastic regime, the linear model reproduces an individual trajectory
using biased parameters. Joint inference over multiple force levels, posterior
predictive checks, and an explicit model-bias term expose this inadequacy. Parameters
inferred in this regime should be regarded as "effective" only, since they do not 
capture the true underlying constitutive behavior.

When the probe size is comparable to the length scale of material
heterogeneity, the inferred modulus depends strongly on the probe location. 
The Bayesian framework presented here separates material heterogeneity from measurement
noise. For the scenario considered in this work, the resulting
distribution is narrower than the range of the modulus field itself, because the probe
responds to a weighted average of its surroundings, and its mean is consistent with the
effective modulus of the same field, as expected for uniformly sampled probe locations.

Future work includes more advanced forward models, such as nonlinear constitutive
models, wall slip, and coupling to other fields (\eg concentration or electromagnetic
fields). As the number of candidate models grows, Bayesian model selection becomes
essential to assess their relative plausibility given the data~\cite{freund2015quantitative, rinkens2026bayesian}.
Because both MCMC sampling and evidence computation require many forward evaluations,
such extensions will likely rely on reduced-order models or surrogates. Moreover, the
framework should be validated on experimental particle-tracking data, with particular
attention to the identification of experimental noise and uncertainty. Finally, the
results for heterogeneous materials suggest that repeated measurements could be used to
probe microstructural heterogeneity. However, it remains to be established how the 
distribution of local estimates depends on the microstructure and on the ratio of 
probe size to heterogeneity length scale.

\FloatBarrier 

\section*{Data and Code Availability}
The data generated by the finite-element model and source code for the Bayesian
inference used in this work are openly
available at \url{https://github.com/parajal/bayesian_microrheology}, including
the datasets, the Bayesian inference framework, and the scripts used to produce
all analyses and figures reported here.

\section*{Acknowledgments}
The authors thank M.A. Hulsen at the Eindhoven University of Technology (TU/e) for access to the TFEM software libraries.

\appendix

\section{Mesh and domain-size convergence studies}
\label{sec:mesh_conv}

Two sources of discretization error are assessed: (i) the spatial
element size (mesh convergence) and (ii) the size of the truncated
computational box (domain-size convergence). Both are quantified through
the particle displacement, which is the observable used throughout the
inference, and both are evaluated separately for the Newtonian and the
viscoelastic constitutive models. The studies are reported for each of
the three configurations introduced in Section~\ref{subsec:problem_def}:
the wall-bounded particle (\emph{Case~I}), the periodic array
(\emph{Case~II}), and the isolated particle (\emph{Case~III}).
Throughout, $h_{\mathrm{particle}}$ denotes the characteristic element
size near the particle surface and $h_{\mathrm{far}}$ the characteristic
size in the far field. Each mesh level halves both, and the reference
discretization identified below is adopted for all subsequent
simulations of that configuration.

\subsection{Case I: particle near a planar wall}
\label{sec:conv_wall}

\begin{figure}[!ht]
  \centering
  \includegraphics[width=0.31\linewidth]{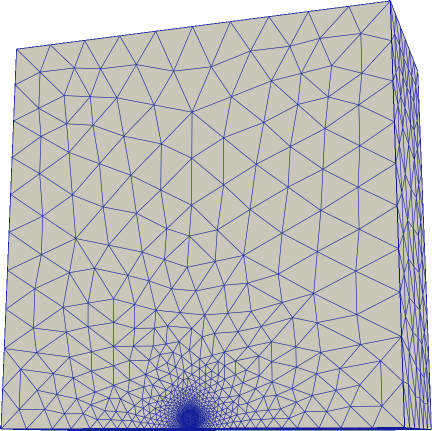}
  \caption{Mesh~M3 for the wall-bounded geometry (\emph{Case~I}) at
           $\delta/a = 0.1$.}
  \label{fig:mesh_3d}
\end{figure}

The wall-bounded configuration places two competing demands on the
discretization. The bounding box must be large enough that the finite
domain approximates a particle moving near a single wall in an otherwise
unbounded fluid, as assumed in Section~\ref{subsec:problem_def}, while
the mesh must resolve the thin gap between the particle and the wall.
Both are assessed at $\delta/a = 0.1$. The mesh hierarchy and the
domain-size levels are listed in Table~\ref{tab:3d-domain}, and the
reference mesh is shown in Figure~\ref{fig:mesh_3d}.

\begin{table}[!ht]
  \centering
  \begin{subtable}[t]{0.58\textwidth}
    \centering
    \begin{tabular}{lcccc}
      \toprule
      Mesh & $h_{\mathrm{particle}}$ & $h_{\mathrm{far}}$ & nodes & elements \\
      \midrule
      M1 & 0.8 & 48 &  1\,242 &     540 \\
      M2 & 0.4 & 24 &  3\,555 &  1\,827 \\
      M3 & 0.2 & 12 & 13\,918 &  8\,380 \\
      M4 & 0.1 &  6 & 85\,766 & 57\,644 \\
      \bottomrule
    \end{tabular}
    \caption{Mesh hierarchy on the reference domain D2.}
  \end{subtable}\hfill
  \begin{subtable}[t]{0.38\textwidth}
    \centering
    \begin{tabular}{cccc}
      \toprule
      Domain & $L$ & $W$ & $H$ \\
      \midrule
      D1 &  50 &  50 &  50 \\
      D2 & 100 & 100 & 100 \\
      D3 & 200 & 200 & 200 \\
      D4 & 400 & 400 & 400 \\
      \bottomrule
    \end{tabular}
    \caption{Domain-size levels.}
  \end{subtable}
  \caption{Spatial discretization and truncation levels for the
           wall-bounded geometry (\emph{Case~I}) at $\delta/a = 0.1$.}
  \label{tab:3d-domain}
\end{table}

\begin{figure}[!ht]
  \centering
  \begin{subfigure}[b]{0.31\linewidth}
    \centering
    \includegraphics[width=\linewidth]{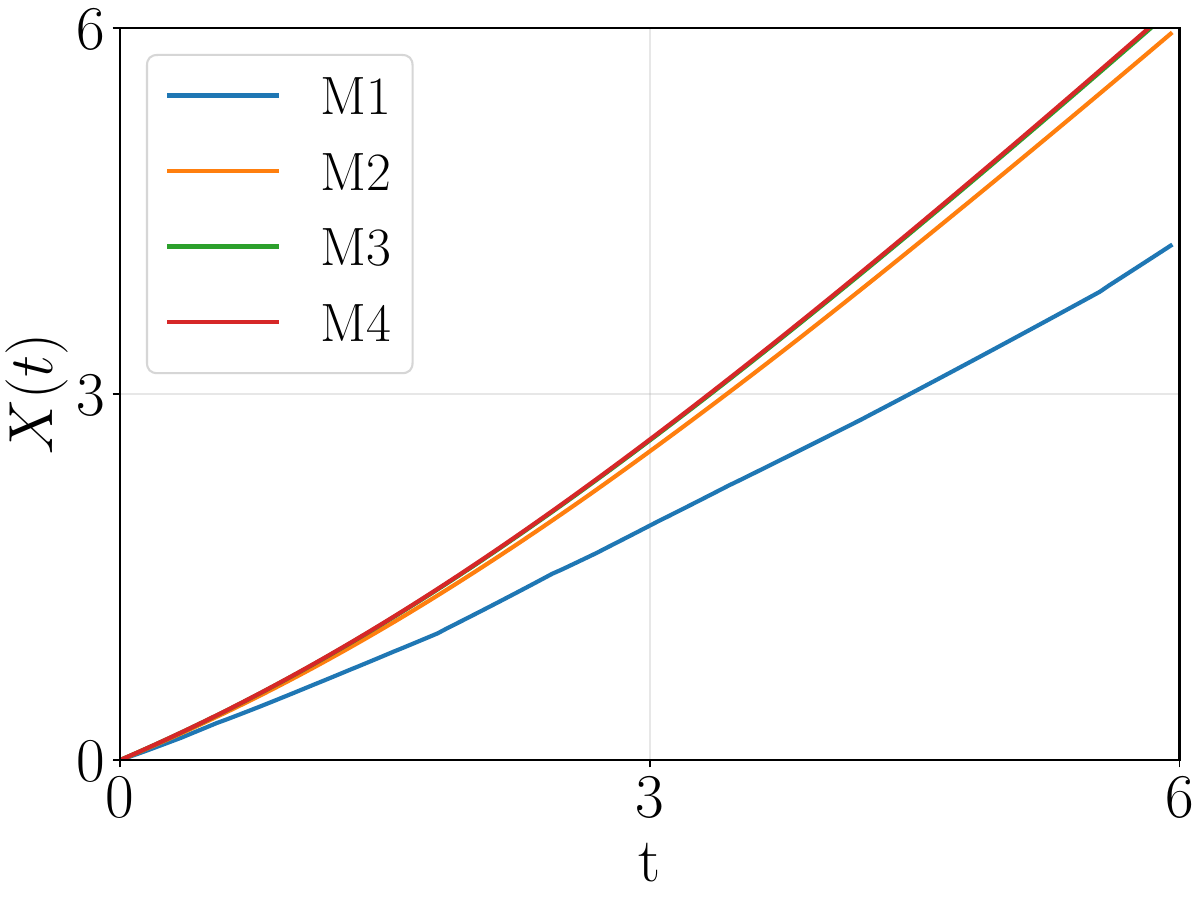}
    \caption{Mesh convergence.}
  \end{subfigure}\hspace{0.4cm}
  \begin{subfigure}[b]{0.31\linewidth}
    \centering
    \includegraphics[width=\linewidth]{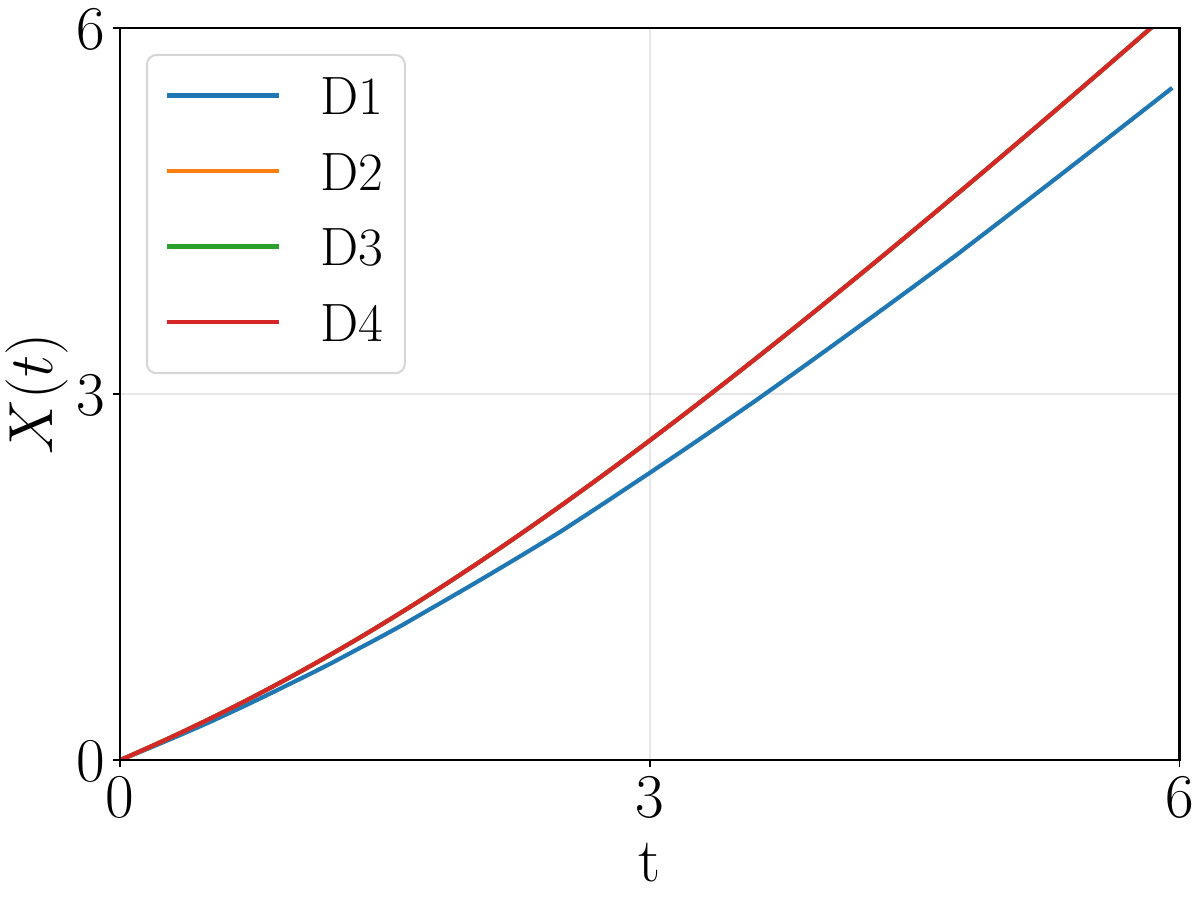}
    \caption{Domain-size convergence.}
  \end{subfigure}
  \caption{\emph{Case~I} ($\delta/a = 0.1$), Newtonian fluid: (a) effect
           of mesh refinement on domain D2 and (b) effect of domain size
           for the converged mesh.}
  \label{fig:conv_3d_newt}
\end{figure}

\begin{figure}[!ht]
  \centering
  \begin{subfigure}[b]{0.33\linewidth}
    \centering
    \includegraphics[width=\linewidth]{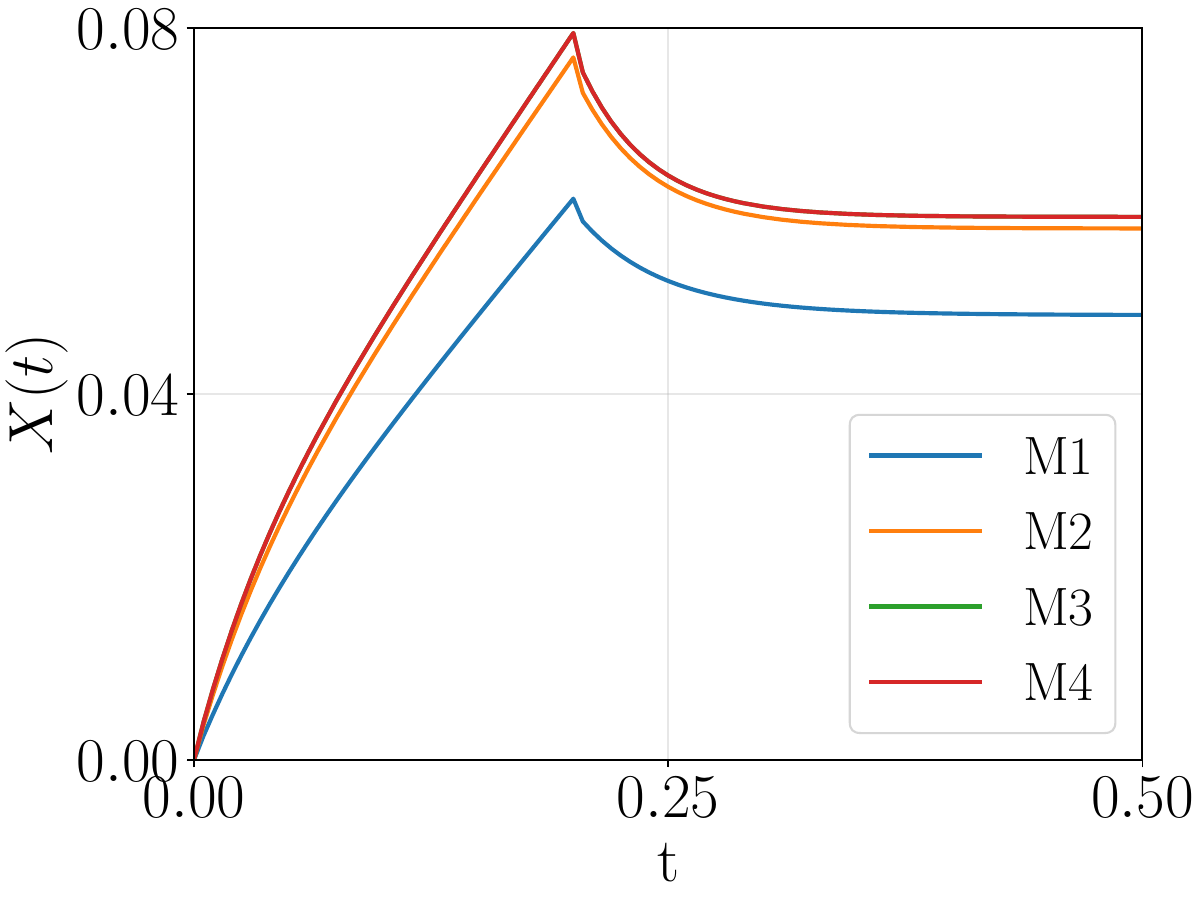}
    \caption{Mesh convergence.}
  \end{subfigure}\hspace{0.4cm}
  \begin{subfigure}[b]{0.33\linewidth}
    \centering
    \includegraphics[width=\linewidth]{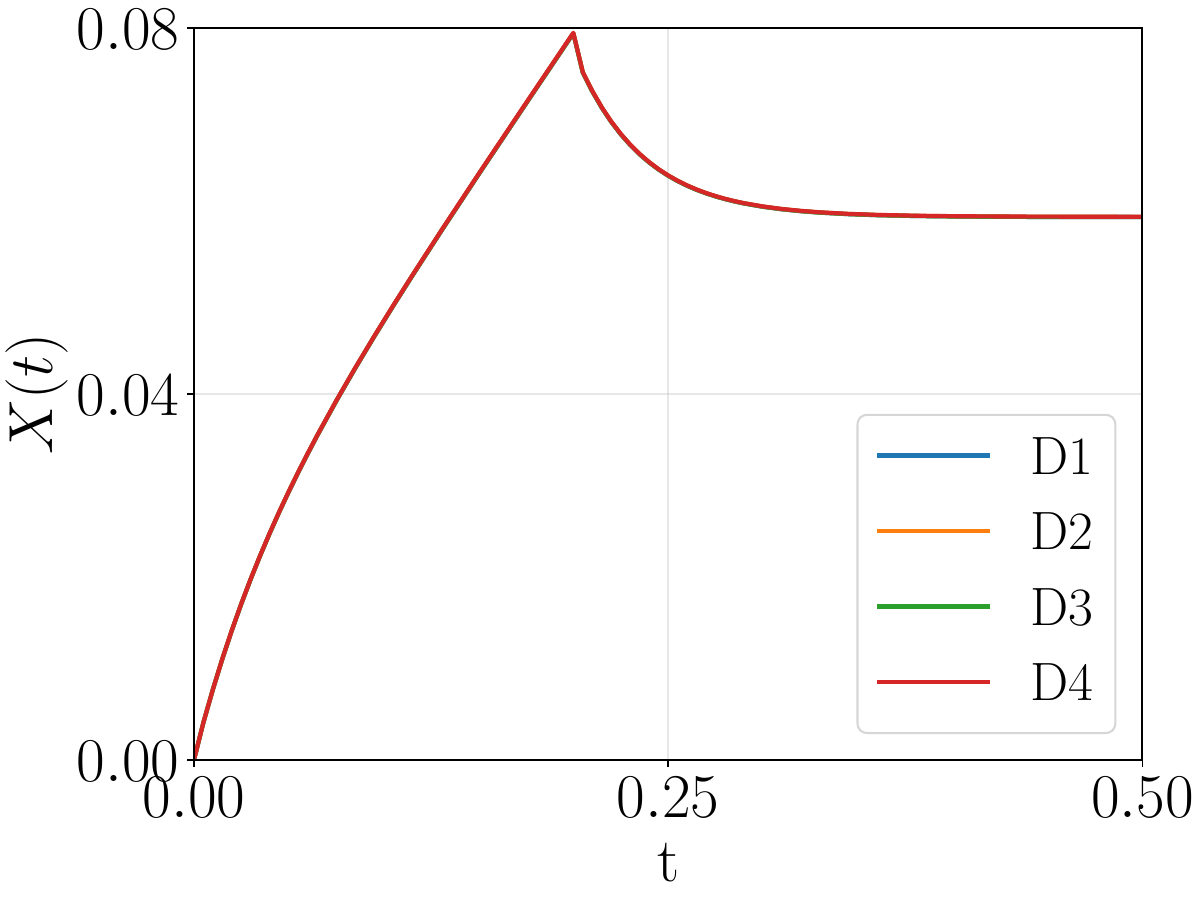}
    \caption{Domain-size convergence.}
  \end{subfigure}
  \caption{\emph{Case~I} ($\delta/a = 0.1$), viscoelastic fluid: (a)
           effect of mesh refinement on domain D2 and (b) effect of
           domain size for the converged mesh.}
  \label{fig:conv_3d_ve}
\end{figure}

The Newtonian results are shown in Figure~\ref{fig:conv_3d_newt} and the
viscoelastic creep-recovery results in Figure~\ref{fig:conv_3d_ve}. In
both cases, the displacement converges at mesh M3 on domain D2, and the
M3/D2 discretization is used for the wall-correction comparisons of
Section~\ref{sec:wall_corrections} and for the wall-effect inference of
Sections~\ref{subsec:wall_newtonian} and~\ref{subsec:wall_ve}. The
smaller gaps considered in those sections,
$\delta_0/a \in \{0.001,\,0.005,\,0.01\}$, are resolved by refining
$h_{\mathrm{particle}}$ in proportion to the gap so that the lubrication
layer is spanned by a comparable number of elements at every separation.

\FloatBarrier

\subsection{Case II: periodic array}
\label{sec:conv_periodic}

For the periodic configuration, mesh convergence is assessed at fixed $L = 10$. Three meshes are considered, listed
in Table~\ref{tab:periodic-mesh}.

\begin{table}[!ht]
  \centering
  \begin{tabular}{lcccc}
    \toprule
    Mesh & $h_{\mathrm{particle}}$ & $h_{\mathrm{far}}$ & nodes & elements \\
    \midrule
    M1 & 0.32 & 4 &  1\,860 &  1\,008 \\
    M2 & 0.16 & 2 & 12\,128 &  7\,899 \\
    M3 & 0.08 & 1 & 88\,485 & 62\,185 \\
    \bottomrule
  \end{tabular}
  \caption{Mesh hierarchy for the triperiodic geometry (\emph{Case~II})
           at box size $L = 10$.}
  \label{tab:periodic-mesh}
\end{table}

\begin{figure}[!ht]
  \centering
  \includegraphics[width=0.33\linewidth]{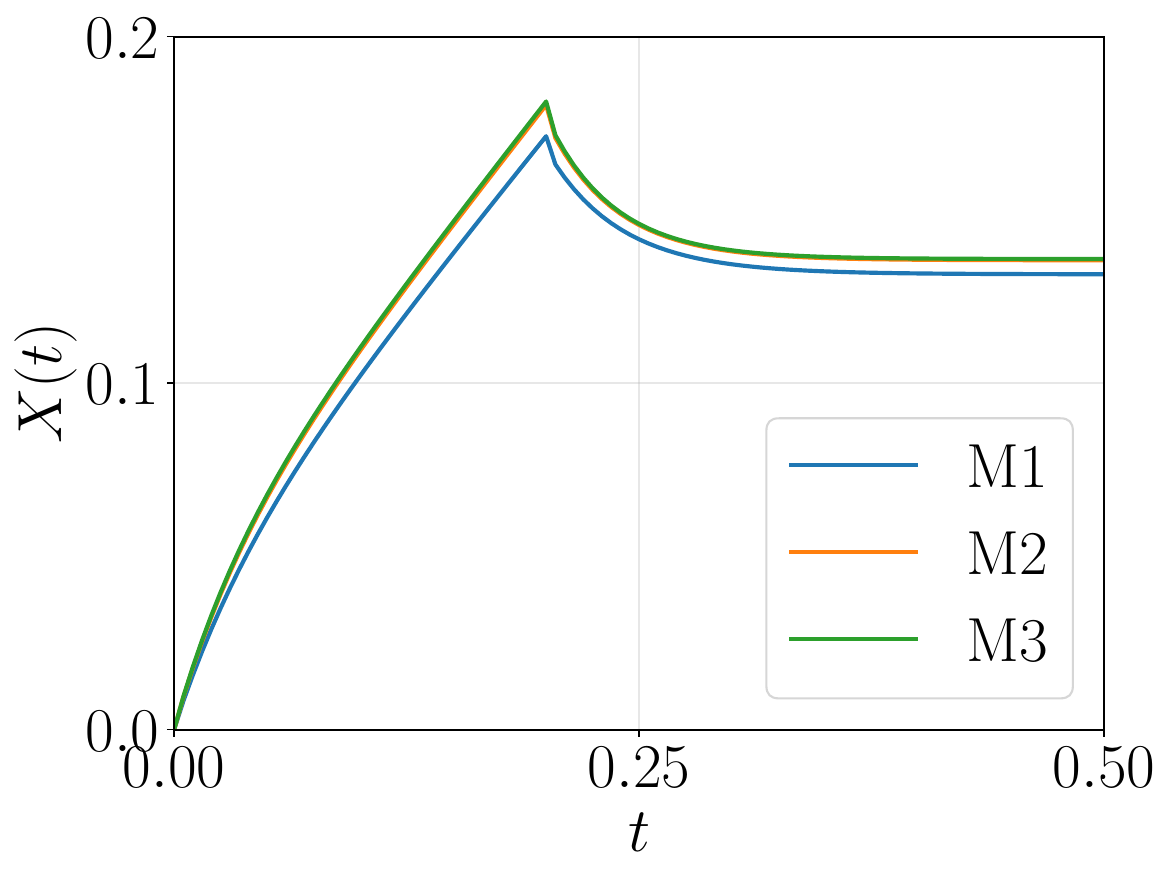}
  \caption{\emph{Case~II}, effect of mesh refinement at fixed box size
           $L = 10$ on the viscoelastic creep-recovery response.}
  \label{fig:conv_periodic}
\end{figure}

As shown in Figure~\ref{fig:conv_periodic}, the particle displacement converges on mesh~M2. To ensure sufficient spatial resolution across all domain sizes, however, mesh~M3 is used for all simulations in Section~\ref{subsec:particle_particle} with $L \in \{5,10,20,40,60\}$, as well as for the heterogeneous-field study in Section~\ref{sec:heterogeneous_inference}.

\subsection{Case III: isolated particle}
\label{sec:conv_isolated}

\begin{figure}[!ht]
  \centering
  \includegraphics[width=0.4\linewidth]{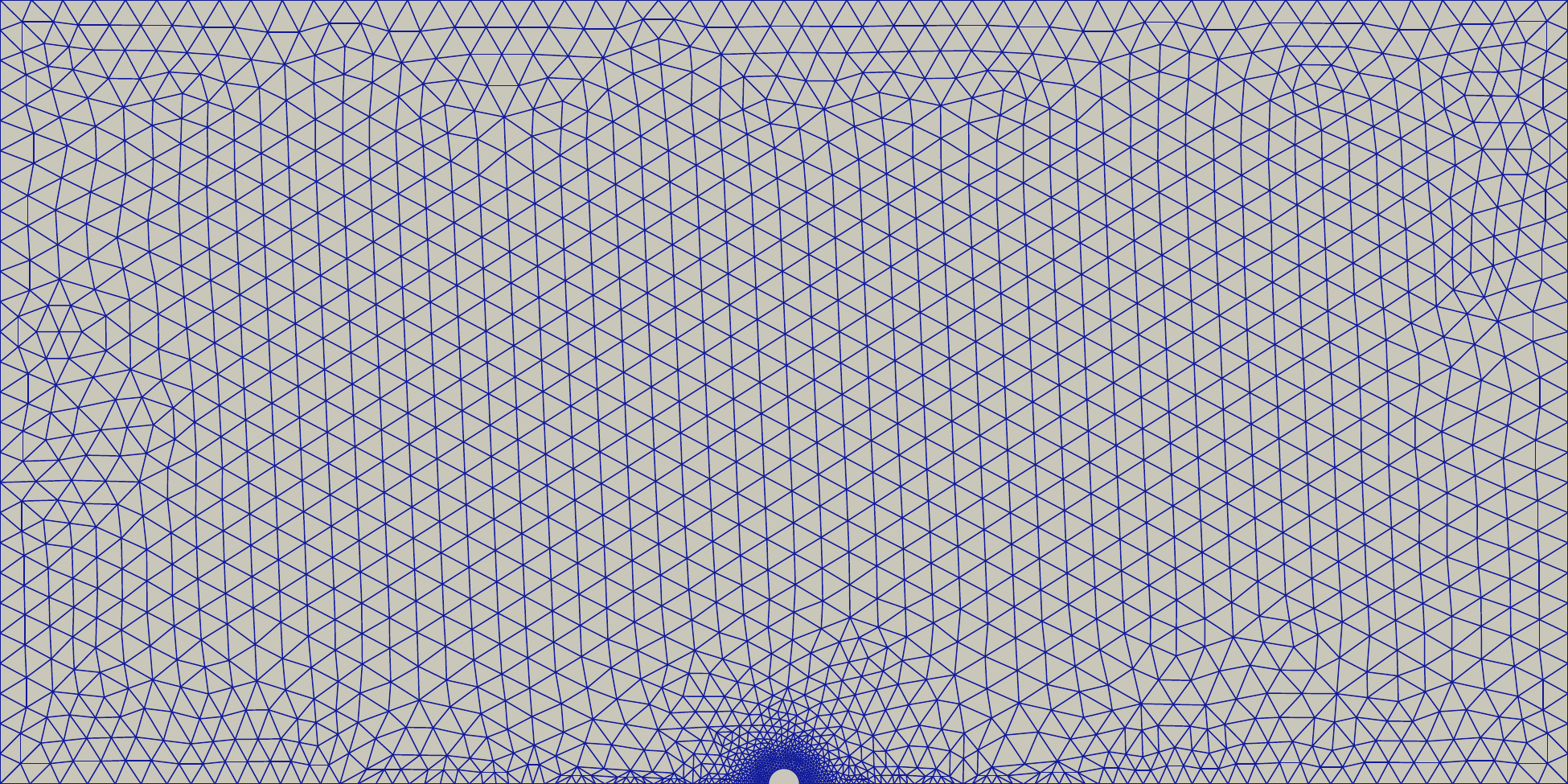}
  \caption{Mesh $M2$ for the axisymmetric isolated-particle geometry
           (\emph{Case~III}).}
  \label{fig:mesh_2d}
\end{figure}

The isolated case is solved on the two-dimensional meridional $(z,r)$
domain described in~\ref{sec:axisymmetric}. The unbounded problem is approximated by a closed cylinder of length $L$ and radius $H$. The mesh hierarchy and the domain-size levels are listed in
Table~\ref{tab:axi-domain}, and mesh~M2 is shown in
Figure~\ref{fig:mesh_2d}.

\begin{table}[!ht]
  \centering
  \begin{subtable}[t]{0.58\textwidth}
    \centering
    \begin{tabular}{lcccc}
      \toprule
      Mesh & $h_{\mathrm{particle}}$ & $h_{\mathrm{far}}$ & nodes & elements \\
      \midrule
      M1 & 1.6 & 16 &     586 &      261 \\
      M2 & 0.8 &  8 &  1\,894 &      887 \\
      M3 & 0.4 &  4 &  6\,785 &   3\,274 \\
      M4 & 0.2 &  2 & 25\,651 &  12\,590 \\
      \bottomrule
    \end{tabular}
    \caption{Mesh hierarchy on the reference domain D2.}
  \end{subtable}\hfill
  \begin{subtable}[t]{0.38\textwidth}
    \centering
    \begin{tabular}{ccc}
      \toprule
      Domain & $H$ & $L$ \\
      \midrule
      D1 &  50 & 100 \\
      D2 & 100 & 200 \\
      D3 & 200 & 400 \\
      D4 & 400 & 800 \\
      \bottomrule
    \end{tabular}
    \caption{Domain-size levels.}
  \end{subtable}
  \caption{Spatial discretization and truncation levels for the
           axisymmetric geometry (\emph{Case~III}).}
  \label{tab:axi-domain}
\end{table}

\begin{figure}[!ht]
  \centering
  \begin{subfigure}[b]{0.33\linewidth}
    \centering
    \includegraphics[width=\linewidth]{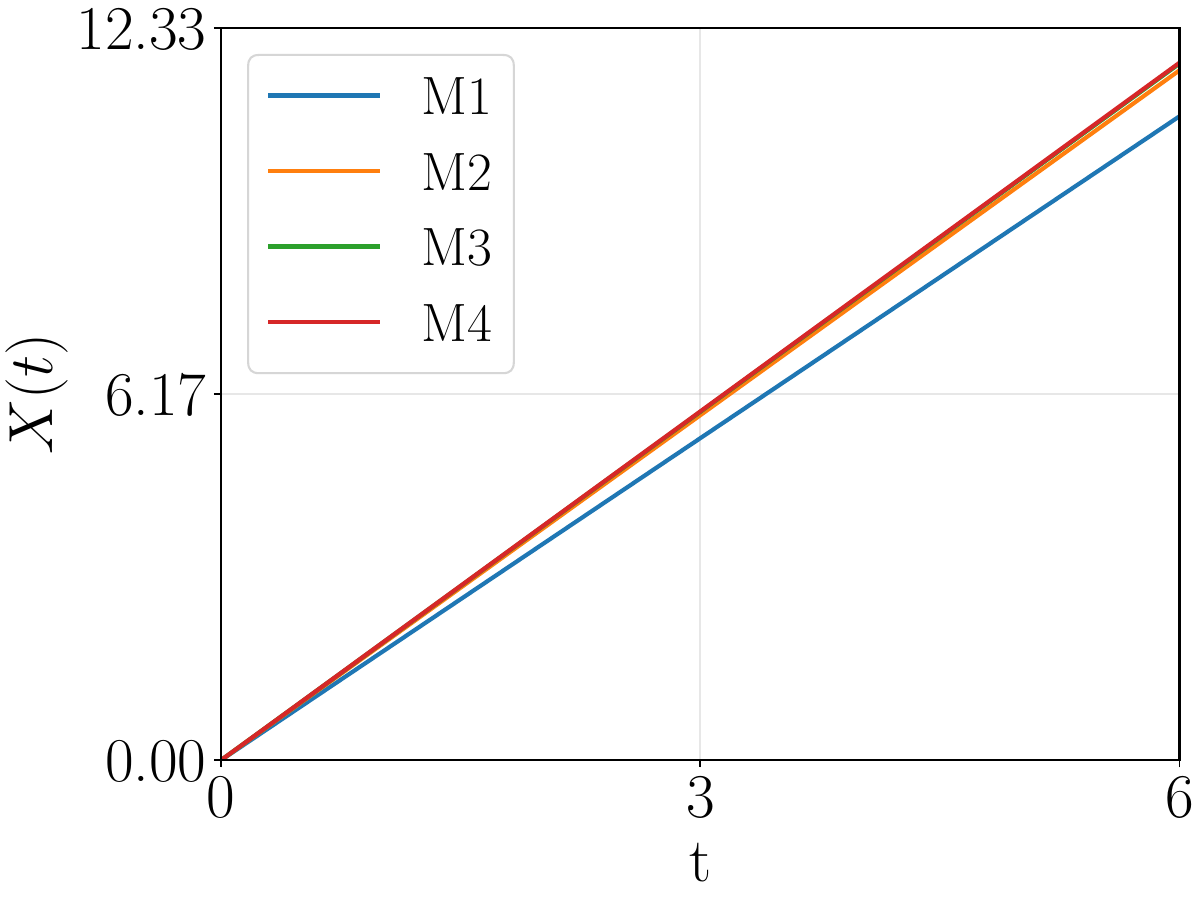}
    \caption{Mesh convergence.}
  \end{subfigure}\hspace{0.4cm}
  \begin{subfigure}[b]{0.33\linewidth}
    \centering
    \includegraphics[width=\linewidth]{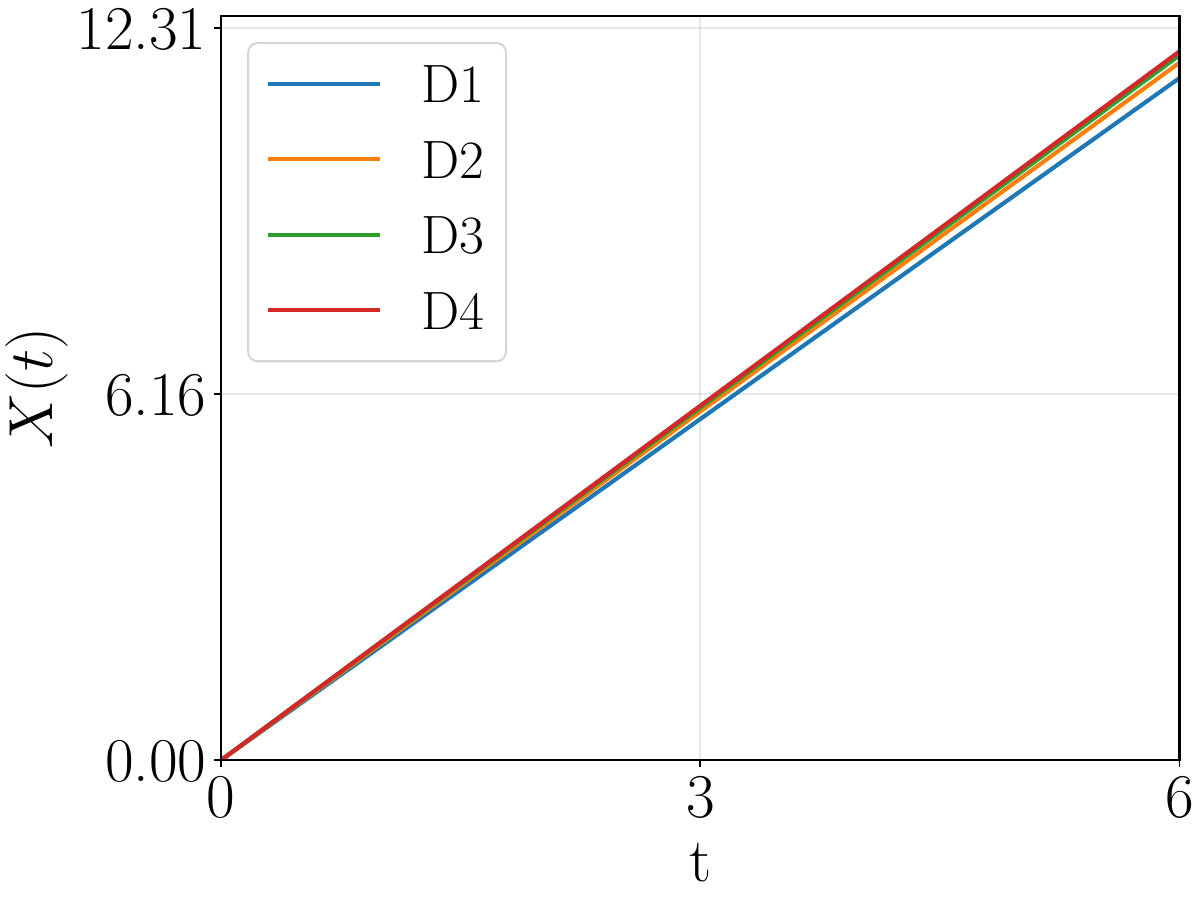}
    \caption{Domain-size convergence.}
  \end{subfigure}
  \caption{\emph{Case~III}, Newtonian fluid: (a) effect of mesh
           refinement on domain D2 and (b) effect of domain size for the
           converged mesh.}
  \label{fig:conv_axi_newt}
\end{figure}

\begin{figure}[!ht]
  \centering
  \begin{subfigure}[b]{0.33\linewidth}
    \centering
    \includegraphics[width=\linewidth]{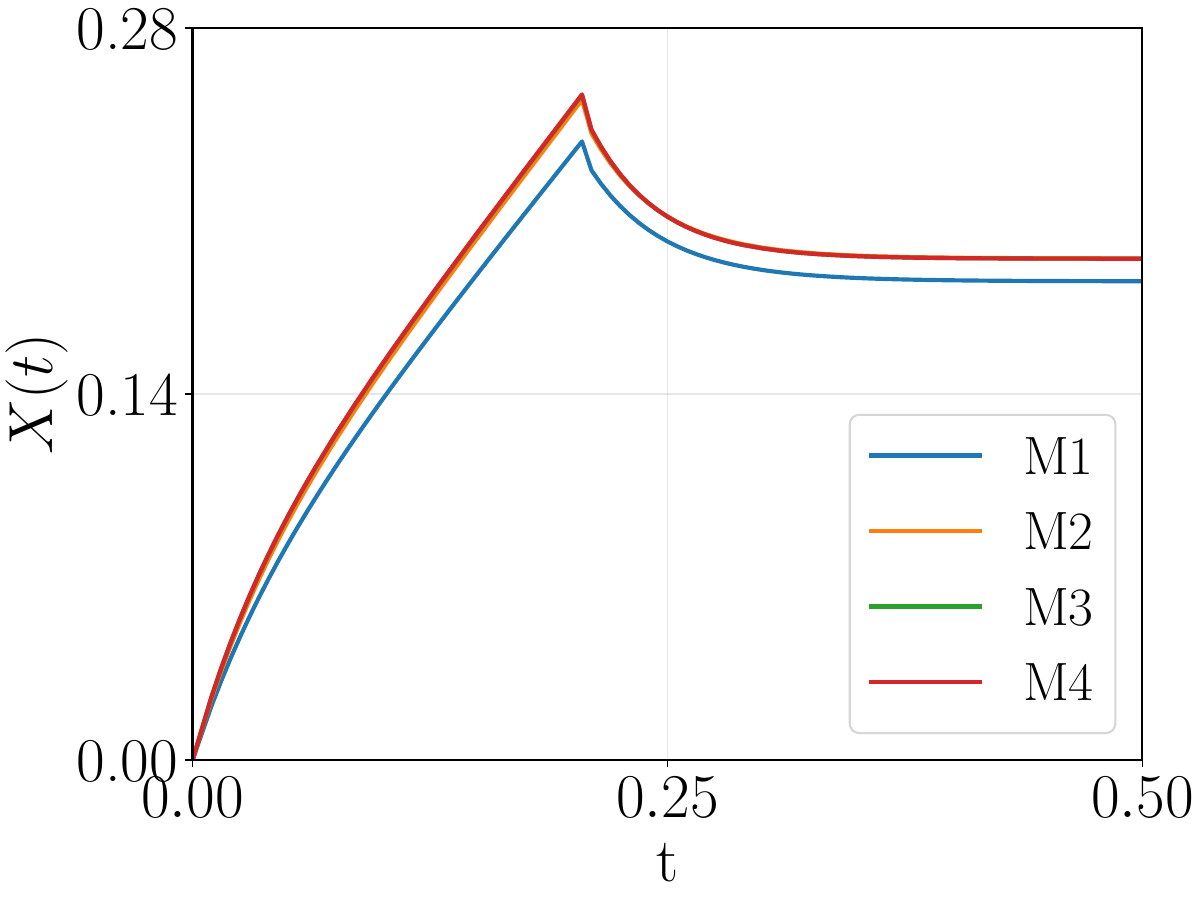}
    \caption{Mesh convergence.}
  \end{subfigure}\hspace{0.4cm}
  \begin{subfigure}[b]{0.33\linewidth}
    \centering
    \includegraphics[width=\linewidth]{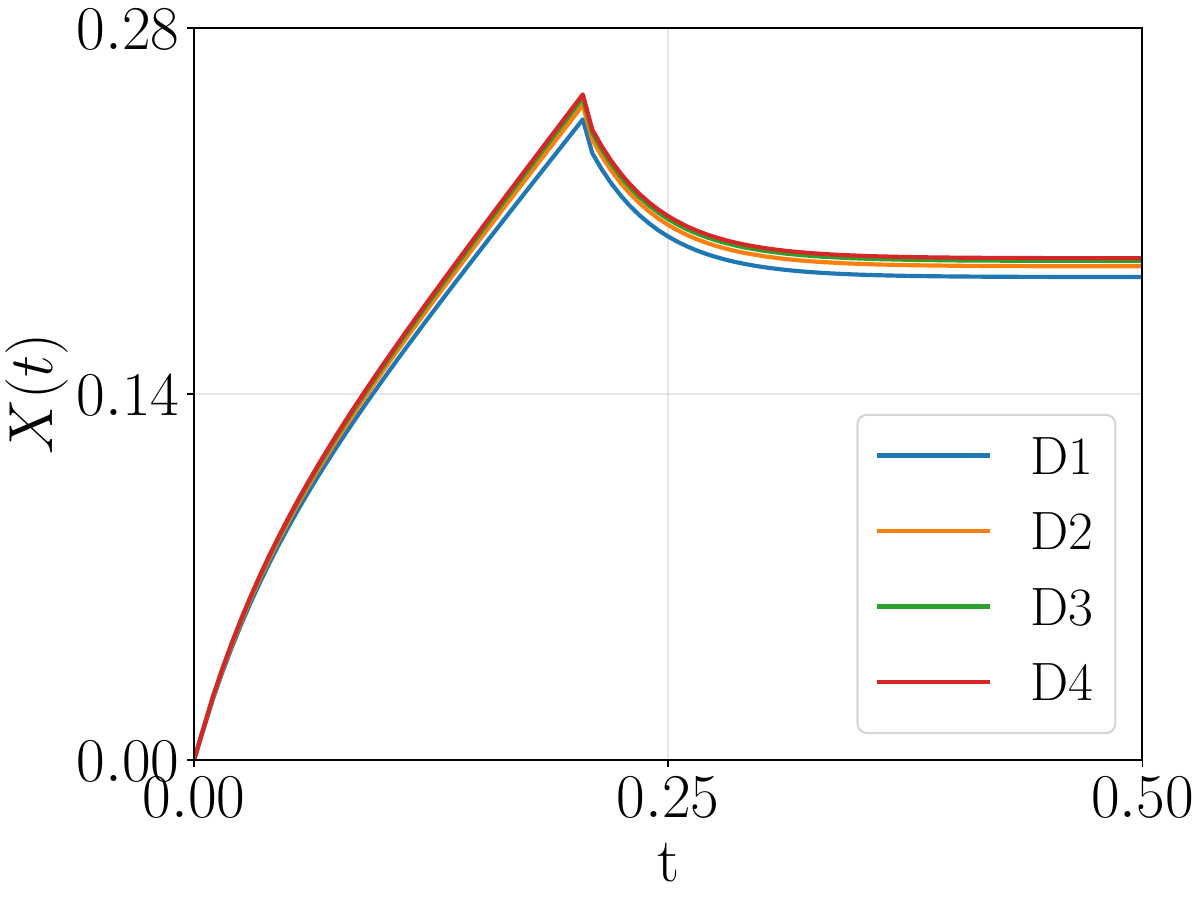}
    \caption{Domain-size convergence.}
  \end{subfigure}
  \caption{\emph{Case~III}, viscoelastic fluid: (a) effect of mesh
           refinement on domain D2 and (b) effect of domain size for the
           converged mesh.}
  \label{fig:conv_axi_ve}
\end{figure}

The Newtonian results are shown in Figure~\ref{fig:conv_axi_newt} and
the viscoelastic results in Figure~\ref{fig:conv_axi_ve}. The
displacement converges at mesh M2 on domain D2 for both constitutive
models. This discretization is used for the isolated-particle
experiments of Section~\ref{subsec:isolated}, including the
high-Weissenberg-number cases of Section~\ref{subsec:nonlinear}, for
which the same levels were checked at the largest applied force
$F = 1024\pi$.

\section{Axisymmetric problem: the isolated case}
\label{sec:axisymmetric}
\begin{figure}[!ht]
    \centering
    \includegraphics[width=0.5\linewidth]{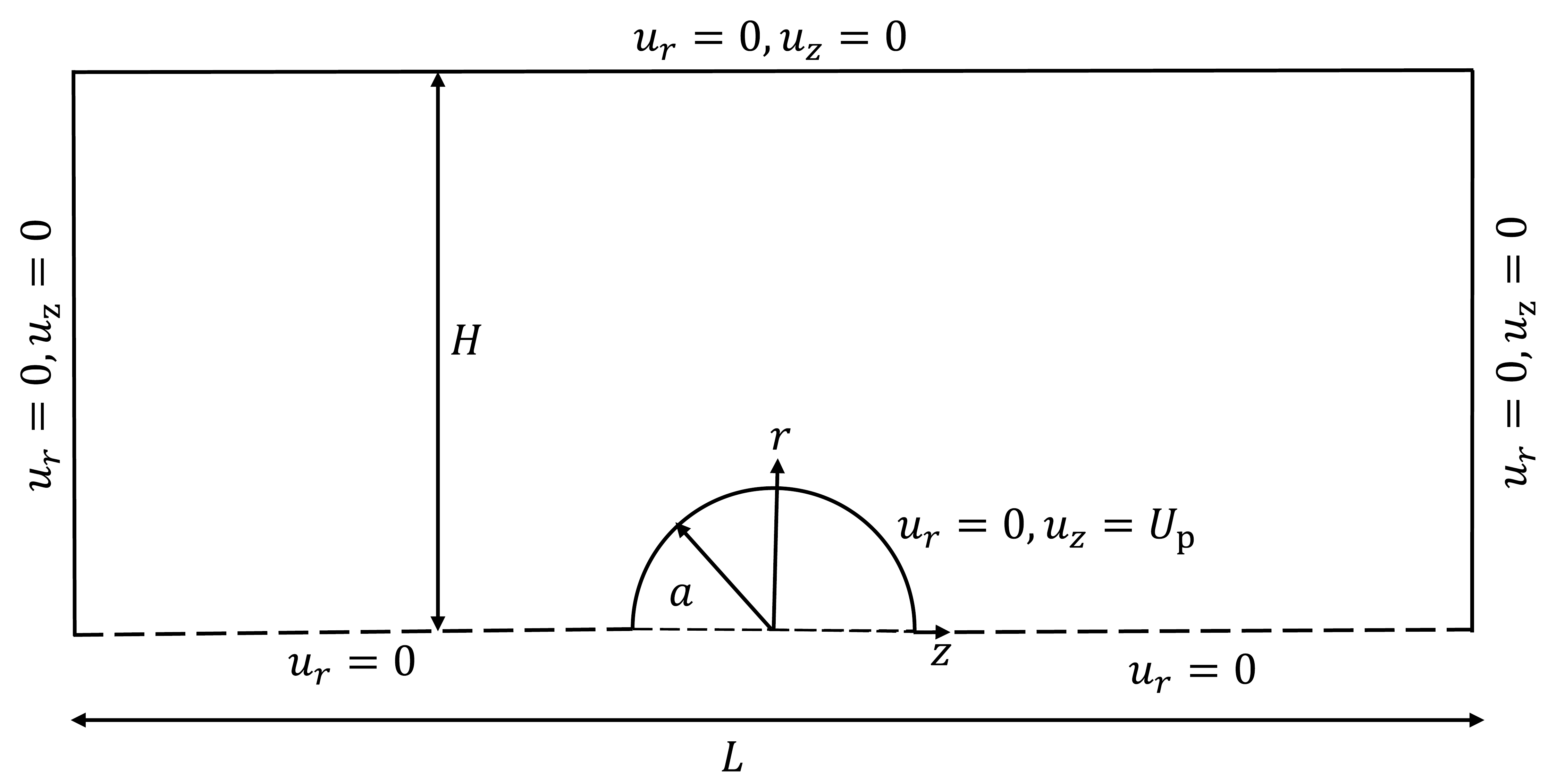}
    \caption{Schematic of the axisymmetric isolated-particle problem.}
    \label{fig:axi-schematic}
\end{figure}
The isolated case (\emph{Case~III}) represents a particle far from any wall or
neighboring image. Rather than enlarging the fully 3D domain, which is computationally expensive, we
exploit the symmetry of a sphere settling under an axial force and solve the problem in an axisymmetric formulation. The particle settles along the axis of a
closed cylindrical container, so the flow is invariant under rotation about that
axis, and the problem reduces to the two-dimensional meridional $(z,r)$ domain.
The cylinder has a length $L \gg a$ and a radius $H \gg a$,
which are large enough that the particle behaves as an effectively isolated body. The
force on the particle is prescribed, and its settling velocity is obtained as
part of the solution. Figure~\ref{fig:axi-schematic}
shows a schematic of this configuration.

Let $\Omega \subset \mathbb{R}^2$ denote the meridional $(z,r)$ domain, with boundary
\[
\partial\Omega = \Gamma_\mathrm{axis} \cup \Gamma_\mathrm{walls} \cup \Gamma_\mathrm{p},
\]
where $\Gamma_\mathrm{axis}$, $\Gamma_\mathrm{walls}$, and $\Gamma_\mathrm{p}$ denote
the axis of symmetry, the outer cylinder walls, and the particle surface,
respectively. On the axis of symmetry ($r=0$) and on the particle boundary, the
radial velocity component $u_r$ vanishes,
\begin{equation}
u_r = 0 \quad \text{on } \Gamma_\mathrm{axis} \cup \Gamma_\mathrm{p}.
\end{equation}
No-slip conditions are imposed on the two end walls and on the lateral cylinder wall,
\begin{equation}
\vec{u} = \vec{0} \quad \text{on } \Gamma_\mathrm{walls},
\end{equation}
while on the particle surface, the velocity is constrained to rigid-body translation,
\begin{equation}
\vec{u} = U_\mathrm{p}\,\vec{e}_z \quad \text{on } \Gamma_\mathrm{p},
\end{equation}
where $U_\mathrm{p}$ is the unknown settling velocity and $\vec{e}_z$ is the unit
vector in the $z$-direction. 

\FloatBarrier

\section{Modeling the heterogeneous microstructure}
\label{sec:blobs}
To probe a material whose
elasticity varies in space (Section~\ref{sec:heterogeneous_inference}), we equip
the full-order model with a transported microstructure. A scalar concentration
field $\phi(\vec{x},t)$ sets the local modulus of the fluid.

The field is initialized as a uniform background $\phi_{\mathrm{bg}}$ plus a
superposition of $N_b$ Gaussian blobs. Each blob $k$ is centered at $\vec{x}_k$
and carries its own peak $\phi_{\mathrm{blob},k}$ and width $w_k$,
\begin{equation}
\phi(\vec{x}) = \phi_{\mathrm{bg}}
+ \sum_{k=1}^{N_b} \phi_{\mathrm{blob},k}\,
\exp\!\left(-\frac{\lVert \vec{x}-\vec{x}_k\rVert^{2}}{2\,w_k^{2}}\right).
\label{eq:blob_field}
\end{equation}
The number of blobs $N_b$ is prescribed directly, the centers $\vec{x}_k$ are
placed by Latin-hypercube sampling (with a fixed seed) for even coverage and may
overlap, and the width and amplitude of each blob are drawn independently from
discrete sets. 

The field is transported by the flow, satisfying the advection equation
\begin{equation}
\frac{\partial \phi}{\partial t} + \vec{u}\cdot\nabla\phi = 0
\quad \text{in } \Omega \setminus P,
\label{eq:phi_advect}
\end{equation}
solved with a streamline-upwind/Petrov-Galerkin (SUPG) stabilized, second-order semi-implicit scheme consistent with the flow solver. The concentration enters the constitutive
model only through the local modulus, with the relaxation time $\lambda$ held fixed,
\begin{equation}
G(\phi) = \frac{\eta_{\mathrm{p}}}{\lambda}\,\phi,
\label{eq:G_of_phi}
\end{equation}
and $G$ is floored at a small positive value $G_{\min}$ to remain physical. Since
the polymeric stress is $\ten{\tau}_{\mathrm{p}} = G(\phi)\,(\ten{c}-\ten{I})$,
$G(\phi)$ is evaluated pointwise at every Gauss point from the convected field,
so the fluid's local modulus tracks the transported microstructure, and the
particle experiences a heterogeneous, evolving viscoelastic environment.

Applying the linear map \eqref{eq:G_of_phi} to the field \eqref{eq:blob_field}
gives the modulus directly as a background plus Gaussian peaks,
\begin{equation}
G(\vec{x}) = G_0 + \sum_{i=1}^{N_b} A_i\,
\exp\!\left(-\frac{\lVert \vec{x} - \vec{c}_i \rVert_2^2}{2\,w_i^2}\right),
\qquad
G_0 = \frac{\eta_{\mathrm{p}}}{\lambda}\,\phi_{\mathrm{bg}}, \quad
A_i = \frac{\eta_{\mathrm{p}}}{\lambda}\,\phi_{\mathrm{blob},i},
\label{eq:G_field}
\end{equation}
where each blob width $w_i$ equals its radius $R_i$. In the cases considered here
we use $N_b = 20$ blobs in a periodic box $L_x = L_y = L_z = 10$, with radii and
amplitudes drawn from $R_i \in \{0.5, 1.0, 1.5, 2.0\}$ and
$A_i \in \{2, 5, 10, 20\}$. With $\phi_{\mathrm{bg}} = 1$ the background reproduces
the homogeneous reference modulus $G_0 = 9$. 
The remaining parameters in this problem are $(\eta_{\mathrm{s}},\lambda)=(0.5,0.1)$ 
and $\eta_{\mathrm{p},0} = G_0\lambda = 0.9$. Because
the blobs are narrow compared with the box ($w_i \ll L$), the periodic images of
each Gaussian contribute negligibly at the box center, so $G_0$ is recovered as
the background modulus away from the peaks. 

As a reference, the same field is probed in shear, with no particle
present. A homogeneous simple shear is imposed through
triperiodic boundary conditions: a velocity jump
$\Delta u = u_x(\mathrm{top}) - u_x(\mathrm{bottom})$ across the periodic
$y$-faces sets a macroscopic shear rate $\dot\gamma = \Delta u / L_y$,
and the volume-averaged polymeric shear stress $\langle \tau_\text{p}^{xy} \rangle$ is
tracked in time. Starting from a stress-free state, the response at time $T$, where $T \ll \lambda$, probes the purely
elastic, pre-relaxation response of the material. The effective shear modulus is
estimated from \begin{equation}
G_{\mathrm{eff}}
= \frac{\langle \tau_\text{p}^{xy} \rangle}{\gamma},
\label{eq:G_bulk_secant}
\end{equation}
where $\gamma = \dot{\gamma} T$ is the shear strain at time $T$. Following this approach, we obtain a value of $G_{\mathrm{eff}}=19.08$ for the heterogeneous field considered here, using values of $T=0.001$ and $\dot{\gamma}=0.01$

\bibliographystyle{elsarticle-num}
\bibliography{references}

\end{document}